\documentclass[twocolumn]{aastex631}

\usepackage{amsmath}
\usepackage{multirow}

\begin{document}

\title{Wide post-common envelope binaries from Gaia: orbit validation and formation models}

\author[0000-0001-6970-1014]{Natsuko Yamaguchi}
\affiliation{Department of Astronomy, California Institute of Technology, 1200 E. California Blvd, Pasadena, CA, 91125, USA}

\author[0000-0002-6871-1752]{Kareem El-Badry}
\affiliation{Department of Astronomy, California Institute of Technology, 1200 E. California Blvd, Pasadena, CA, 91125, USA}

\author[0009-0004-7488-6542]{Natalie R. Rees}
\affiliation{University of Surrey, Stag Hill, University Campus, Guildford, GU2 7XH, United Kingdom}

\author[0000-0001-9298-8068]{Sahar Shahaf}
\affiliation{Department of Particle Physics and Astrophysics, Weizmann Institute of Science, Rehovot 7610001, Israel}

\author[0000-0002-3569-3391]{Tsevi Mazeh}
\affiliation{School of Physics and Astronomy, Tel Aviv University, Tel Aviv, 6997801, Israel}

\author[0000-0001-8006-6365]{René Andrae}
\affiliation{Max-Planck-Institut für Astronomie, Königstuhl 17, D-69117 Heidelberg, Germany}

\begin{abstract}
Astrometry from {\it Gaia} DR3 has enabled the discovery of a sample of 3000+ binaries containing white dwarfs (WD) and main-sequence (MS) stars in relatively wide orbits, with orbital periods $P_{\rm orb} = (100-1000)$ d. This population was not predicted by binary population synthesis models before {\it Gaia} and -- if the {\it Gaia} orbits are robust -- likely requires very efficient envelope ejection during common envelope evolution (CEE). To assess the reliability of the {\it Gaia} solutions, we measured multi-epoch radial velocities (RVs) of 31 WD+MS binary candidates with $P_{\rm orb} = (40-300)$ d and \texttt{AstroSpectroSB1} orbital solutions. We jointly fit the RVs and astrometry, allowing us to validate the {\it Gaia} solutions and tighten constraints on component masses. We find a high success rate for the {\it Gaia} solutions, with only 2 out of the 31 systems showing significant discrepancies between their {\it Gaia} orbital solutions and our RVs. Joint fitting of RVs and astrometry allows us to directly constrain the secondary-to-primary flux ratio $\mathcal{S}$, and we find $\mathcal{S}\lesssim 0.02$ for most objects, confirming the companions are indeed WDs. We tighten constraints on the binaries' eccentricities, finding a median $e\approx 0.1$. These eccentricities are much lower than those of normal MS+MS binaries at similar periods, but much higher than predicted for binaries formed via stable mass transfer. We present MESA single and binary evolution models to explore how the binaries may have formed. The orbits of most binaries in the sample can be produced through CEE that begins when the WD progenitor is an AGB star, corresponding to initial separations of $2-5$ au. Roughly 50\% of all post-common envelope binaries are predicted to have first interacted on the AGB, ending up in wide orbits like these systems.
\end{abstract}


\section{Introduction} \label{sec:intro}

Close binaries containing a main sequence (MS) star and a white dwarf (WD) are the end products of binary interactions that must have occurred when the WD progenitor was a giant. These systems are therefore important tools to constrain the physics of mass transfer (MT) processes. 

The most uncertain mode of binary interaction is common envelope evolution (CEE), in which a companion is engulfed by the envelope of an evolved star and spirals inward on a short timescale. 
If enough orbital energy is liberated in this process, the envelope is unbound and ejected, leaving behind a post-common envelope binary (PCEB) consisting of the donor's core and the companion star in a close orbit. Otherwise, the result is a stellar merger. CEE is critical in the formation of a wide variety of exotic systems. 

In the past few decades, several surveys have been successful in the search for PCEBs \citep{Rebassa-Mansergas2007MNRAS, Parsons2016MNRAS} and the population they have discovered has played an important role in calibrating CEE models \citep[e.g.][]{Zorotovic2010A&A}. However, in part due to selection biases, the majority of these known PCEBs have short orbital periods of less than a day. Their formation can be relatively well-explained by simply energy conservation arguments (the ``$\alpha$-formalism"; e.g. \citealt{Livio1988, Tout1997, DeMarco2011MNRAS}), where a fraction of the orbital energy lost during the binary's inspiral is enough to overcome the gravitational binding energy of the giant donor's envelope. 

A handful of binaries have been discovered which challenge these simplified CEE models \citep{Wonnacott1993, Kruse2014Sci, Kawahara2018AJ, Yamaguchi2024MNRAS}. These have longer orbital periods of months to a few years, making them too close to have avoided interaction when the WD progenitor was a giant, but much wider than naively expected for orbital energy losses to have been sufficient for successful envelope ejection. If these are in fact PCEBs, they may indicate that additional energy sources are needed to overcome the envelope binding energy and several such sources have been put forward and debated in the literature (e.g. recombination energy \citealt{1968IAUS...34..396P, Webbink2008ASSL, Ivanova2018ApJL, Belloni2024arXiv, 2024arXiv240117510B}; jets from the accretor \citealt{Sabach2017MNRAS, MorenoMendez2017MNRAS}). 

The third data release from the {\it Gaia} mission \citep{GaiaCollaboration2023} provided orbital solutions for over 169,000 astrometric binaries \citep{GaiaCollaboration2023A&A}. From these, \citet{Shahaf2024MNRAS} identified several thousand MS + WD binary candidates, most of which have orbital periods that range between of $\sim 100 - 1000\,$d and WD masses higher than predicted for binaries formed by stable mass transfer at these periods. This sample tells us that such systems are relatively common and that they were largely missed by previous searches, which were sensitive only to short periods. Therefore, this sample highlights missing pieces in our understanding of binary interactions, motivating follow-up observations to validate the {\it Gaia} solutions and put better constraints on their orbital and stellar parameters. 

In this paper, we present follow-up spectroscopic observations of 31 objects from the \citet{Shahaf2024MNRAS} sample of MS + WD binaries with orbital periods less than 300 days. Section \ref{sec:sample_selection} summarizes the original selection carried out by \citet{Shahaf2024MNRAS} and the properties of our sub-sample. In Section \ref{sec:follow_up}, we describe our spectroscopic observations to obtain radial velocities (RVs) and describe fits to the broadband spectral energy distributions (SEDs) to constrain the MS star mass. In Section \ref{sec:rv_fitting}, we jointly fit the RVs and the {\it Gaia} astrometry to constrain orbital parameters and flux ratios. We then discuss the resulting flux ratio distribution, inferred success rate of the {\it Gaia} astrometric solutions, and comparison of our systems to other populations in Section \ref{sec:discussion}. In Section \ref{sec:formation_channels}, we present MESA models testing formation channels through CEE as well as stable MT. We briefly discuss the completeness of our sample in Section \ref{sec:completeness}. Lastly, in Section \ref{sec:conclusion}, we summarize our main results and conclude. 

\section{Sample Selection} \label{sec:sample_selection}

We are carrying out an RV survey of WD + MS binary candidates selected by \citet{Shahaf2024MNRAS}. Their selection was based on the astrometric mass ratio function (AMRF) which was presented in \citet{Shahaf2019MNRAS} and defined in terms of observational parameters:
\begin{equation} \label{eqn:AMRF_1}
    \mathcal{A} = \frac{\alpha}{\varpi}\left(\frac{M_1}{M_{\odot}}\right)^{-1/3}\left(\frac{P}{\rm{yr}}\right)^{-2/3}
\end{equation} 
where $\alpha$ is the angular semi-major axis of the photocentric orbit, $\varpi$ is the parallax, $M_1$ is the mass of the primary (i.e. the more luminous star), and $P$ is the orbital period. 
This can also be written in terms of the mass ratio $q = M_2/M_1$ and the flux ratio $\mathcal{S} = F_2/F_1$, where $F_1$ and $F_2$ are the G-band fluxes of the primary and secondary: 
\begin{equation} \label{eqn:AMRF_2}
    \mathcal{A} = \frac{q}{\left(1+q\right)^{2/3}}\left(1 - \frac{\mathcal{S}\left(1+q\right)}{q\left(1+\mathcal{S}\right)}\right)
\end{equation}
Given a relation $\mathcal{S}(q)$ for different possible companions (e.g. a single MS star, a tight binary containing two MS stars, or a dark object), one can define regions in $\mathcal{A}$ - $M_1$ space in which each of these possibilities are most likely. In particular, above some critical AMRF for a given $M_1$, the system is very likely to host a dark companion/compact object \citep{Shahaf2023MNRAS}.

\citet{Shahaf2024MNRAS} selected $\sim$9000 astrometric binaries with AMRF values that are too large to be explained by a single luminous companion. These systems could either host WD companions, or be triples, in which the companion is a tight MS + MS binary. They then removed systems that fall above the main sequence in a color-magnitude diagram, which were suspected to be triples (For a detailed description, we refer readers to \citealt{Shahaf2024MNRAS}).

\begin{figure*}
    \centering
    \includegraphics[width=0.33\textwidth]{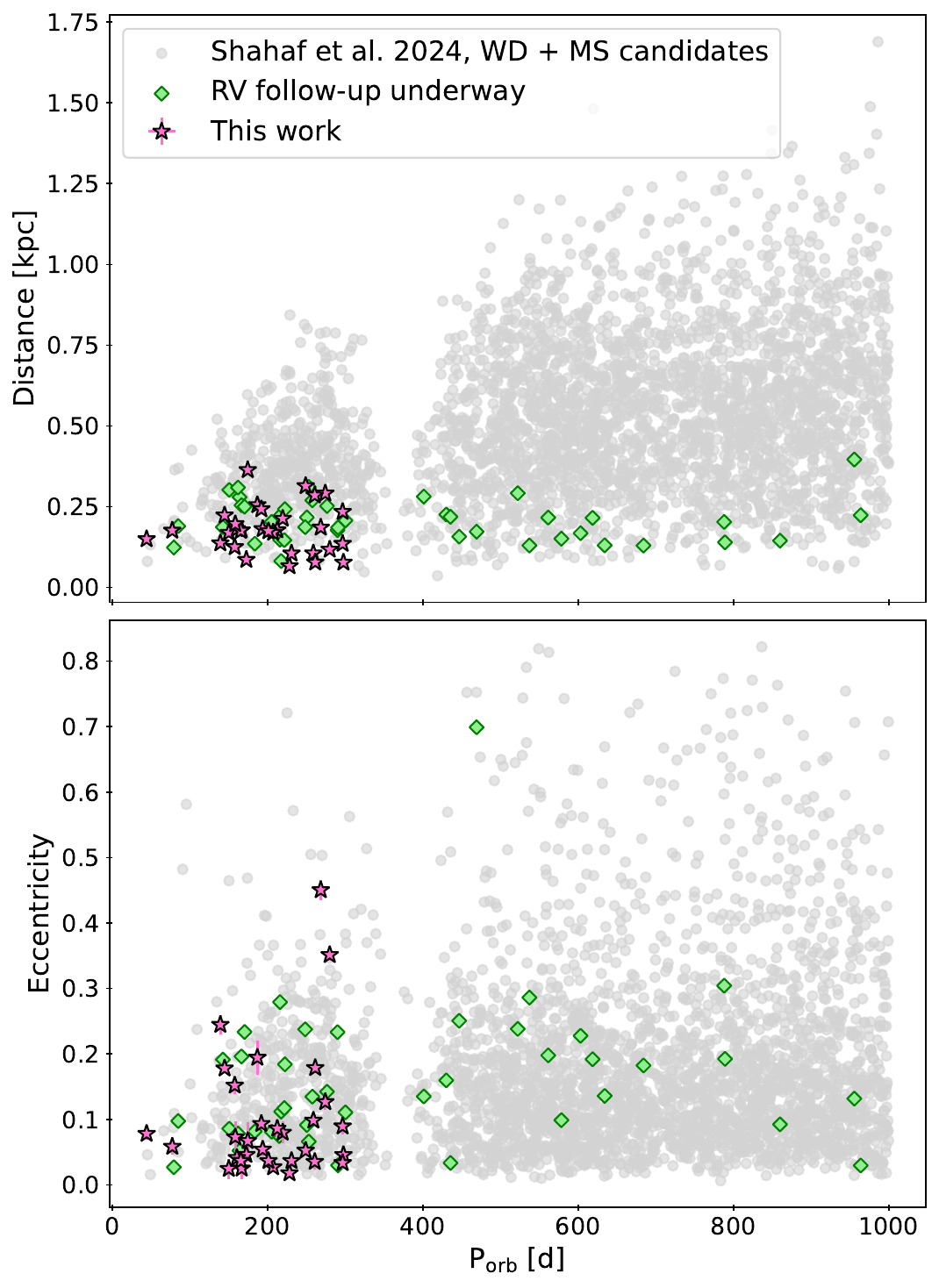}
     \includegraphics[width=0.475\textwidth]{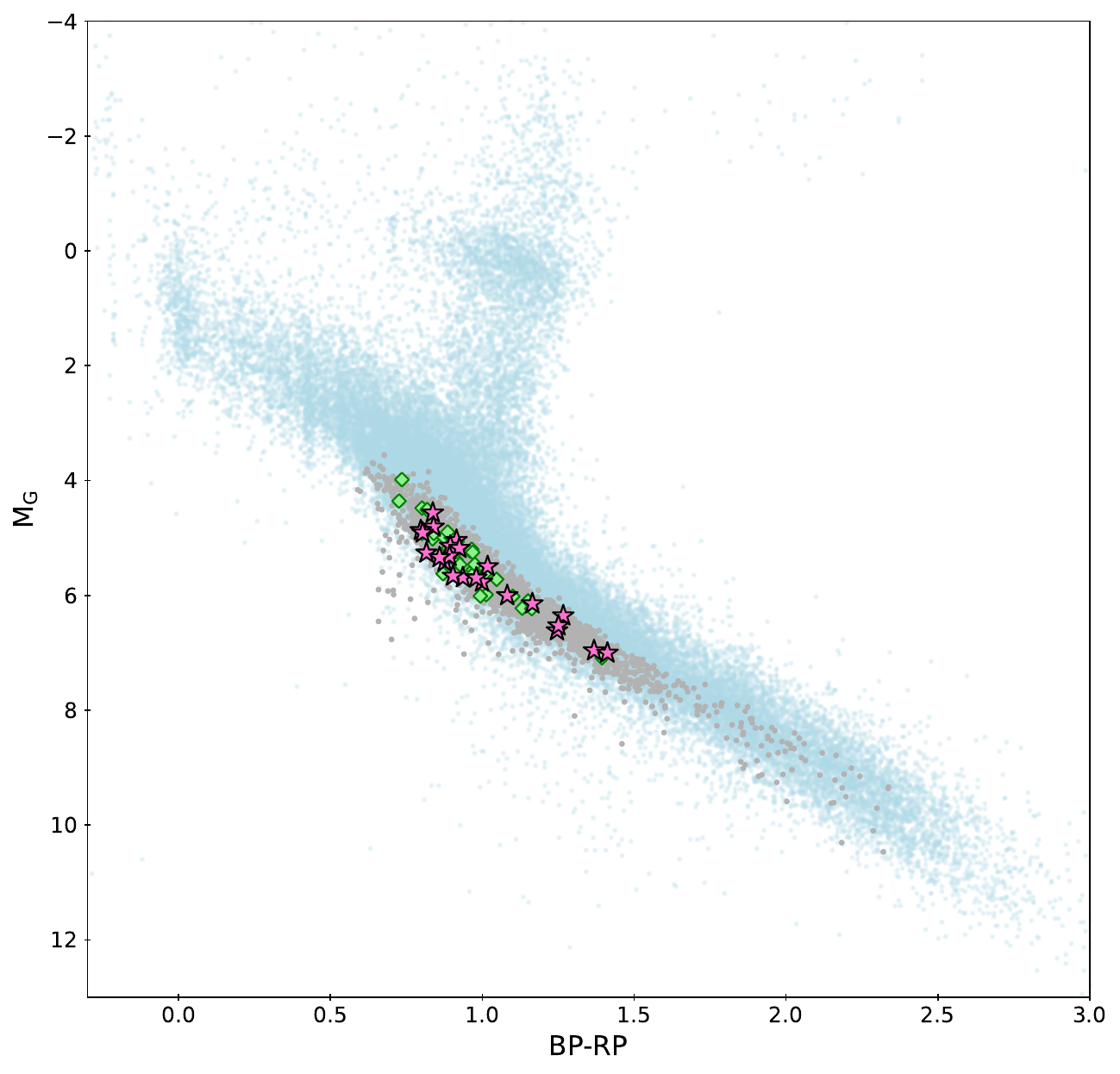} 
    \caption{\emph{Left}: Distance and eccentricity against orbital period of the full sample of MS + WD candidates from \citet{Shahaf2024MNRAS}. The 31 objects presented in this paper are marked with star markers. We have also highlighted $\sim 40$ more objects for which RV follow-up is ongoing with diamond markers. \emph{Right}: The same objects plotted on top of a {\it Gaia} (extinction-corrected) CMD, in blue.}
    \label{fig:sample_distribution}
\end{figure*}

For this paper, we selected 31 objects from this sample with orbital periods less than 300 days to obtain follow-up spectroscopy. Basic information for these are summarized in Table \ref{tab:basic_info}. These all have ``AstroSpectroSB1" solutions, meaning that both astrometry and RVs from Gaia were fit with a single orbital solution. The full \citet{Shahaf2024MNRAS} sample also contains objects with ``Orbital" solutions, which are fitted with just astrometry. Some of these are part of a larger sample for which we are also carrying out spectroscopic follow-up and which we will describe in future work. On the left panel of Figure \ref{fig:sample_distribution}, we show the distribution of binaries from the full sample across eccentricity and orbital period. We mark the 31 objects from this work with stars, and those from our larger sample (for which we have not yet obtained sufficient RVs for a well-constrained orbit) with diamonds. We see that these are roughly evenly sampled across orbital periods and cover a wide range of possible eccentricities. On the right panel, we plot the same objects on a {\it Gaia} extinction-corrected color-magnitude diagram (CMD)(for $\sim 45,000$ sources from \texttt{gaiadr3.gaia\_source} with \texttt{ag\_gspphot} values and \texttt{parallax\_over\_error} $>$ 5). The luminous stars are main-sequence G and K stars. They lie on the left half of the main sequence, because \citet{Shahaf2024MNRAS} discarded redder objects as possible triples. 

\begin{deluxetable*}{c c c c c c c c c}
\tablecaption{Basic information from {\it Gaia} DR3 of all objects studied in this work. The format for the name of each object is `J' for J2000 followed by the coordinates of the right ascension (RA) in hours and minutes, and declination (Dec) in degrees and arcminutes. $G$ is the G-band mean magnitude, GoF is the {\it Gaia} \texttt{goodness\_of\_fit}, $P_{\rm orb}$ is the orbital period reported by {\it Gaia}, and $\varpi$ is the parallax. Last column lists the number of RVs we measured for each object.}
\tablehead{
\colhead{Name} & \colhead{{\it Gaia} DR3 source ID} & 
\colhead{RA [deg]} & \colhead{Dec [deg]} & 
\colhead{$G$ [mag]} & \colhead{GoF} & 
\colhead{$P_{\rm orb}$ [d]} & \colhead{$\varpi [\rm mas]$} & 
\colhead{\# of RVs}
} 
\startdata
J2006+1234 & 1803080360967961856 & 301.73646 & 12.57663 & 12.65 & 3.33 & 219.56 & 4.69 $\pm$ 0.02 & 2 \\
 J0145-0208 & 2505477938150534016 & 26.42328 & -2.14026 & 10.06 & 2.57 & 259.50 & 9.48 $\pm$ 0.02 & 5 \\
 J0547-2018 & 2966114550741693440 & 86.79486 & -20.31002 & 12.45 & 4.80 & 186.98 & 3.93 $\pm$ 0.01 & 4 \\
 J0602-1747 & 2990664622462566528 & 90.71887 & -17.78789 & 11.42 & 0.88 & 297.65 & 4.26 $\pm$ 0.02 & 4 \\
 J1134-2456 & 3533798283973293056 & 173.59255 & -24.94907 & 12.99 & 2.08 & 167.21 & 5.62 $\pm$ 0.02 & 4 \\
 J2007-0840 & 4192464229290454912 & 301.99467 & -8.68059 & 12.84 & 2.10 & 212.15 & 5.65 $\pm$ 0.02 & 3 \\
 J1951-0305 & 4233885203132760320 & 297.98174 & -3.08986 & 10.60 & 1.36 & 158.03 & 8.01 $\pm$ 0.01 & 3 \\
 J2000-0047 & 4237613470960339456 & 300.18584 & 0.78755 & 12.20 & 1.53 & 150.41 & 5.92 $\pm$ 0.01 & 4 \\
 J1553-1117 & 4346259930053416064 & 238.27382 & -11.28546 & 10.41 & 3.63 & 173.33 & 11.74 $\pm$ 0.02 & 5 \\
 J1628+1408 & 4463782295536949632 & 247.04825 & 14.14072 & 12.42 & 3.63 & 261.44 & 3.51 $\pm$ 0.01 & 5 \\
 J1834+1525 & 4509912512048265088 & 278.54277 & 15.42166 & 11.59 & 1.66 & 195.27 & 5.55 $\pm$ 0.02 & 5 \\
 J0127-6350 & 4712016939794270720 & 21.98986 & -63.83941 & 12.67 & 3.76 & 249.43 & 3.18 $\pm$ 0.01 & 7 \\
 J0320-6129 & 4722462403337867520 & 50.05036 & -61.49707 & 11.47 & 3.36 & 207.75 & 5.89 $\pm$ 0.01 & 4 \\
 J0222-5032 & 4746089877426994048 & 35.63934 & -50.54031 & 12.32 & 3.19 & 160.85 & 5.53 $\pm$ 0.01 & 5 \\
 J0525-4059 & 4807112532250705280 & 81.43069 & -40.99854 & 10.76 & 3.54 & 298.07 & 12.98 $\pm$ 0.01 & 5 \\
 J0502-2803 & 4877335724279916032 & 75.54007 & -28.06128 & 12.61 & 4.13 & 174.66 & 2.75 $\pm$ 0.01 & 3 \\
 J0050-5739 & 4907589061595812224 & 12.62913 & -57.65659 & 11.74 & 2.11 & 77.57 & 5.69 $\pm$ 0.02 & 5 \\
 J0115-5039 & 4928000193592001920 & 18.97365 & -50.65588 & 11.08 & 1.80 & 201.41 & 5.74 $\pm$ 0.01 & 6 \\
 J0223-2354 & 5119937334246966656 & 35.77907 & -23.91649 & 12.63 & 2.59 & 191.71 & 4.14 $\pm$ 0.02 & 5 \\
 J0204-2602 & 5120822131869574784 & 31.15776 & -26.03718 & 13.10 & -2.50 & 44.48 & 6.70 $\pm$ 0.01 & 5 \\
 J0621-5218 & 5501078229653801600 & 95.45537 & -52.31318 & 12.37 & 4.07 & 165.72 & 5.71 $\pm$ 0.01 & 5 \\
 J1758-4658 & 5953913841603513984 & 269.66657 & -46.97284 & 11.49 & 3.34 & 228.66 & 15.16 $\pm$ 0.02 & 5 \\
 J1533-4345 & 6001340936385389440 & 233.30535 & -43.75493 & 12.52 & -0.32 & 144.49 & 4.51 $\pm$ 0.02 & 4 \\
 J1639-2827 & 6032398738252562688 & 249.75050 & -28.45857 & 10.78 & 2.17 & 139.53 & 7.32 $\pm$ 0.02 & 4 \\
 J2234-6856 & 6385252016957373824 & 338.65290 & -68.94279 & 13.00 & -0.90 & 297.38 & 7.42 $\pm$ 0.02 & 4 \\
 J2209-6024 & 6409382723774888448 & 332.30859 & -60.40201 & 12.69 & 4.50 & 158.63 & 5.10 $\pm$ 0.02 & 5 \\
 J1956-6409 & 6428833084470488704 & 299.14132 & -64.16665 & 9.71 & 3.79 & 280.57 & 8.53 $\pm$ 0.02 & 2 \\
 J2103-4747 & 6479954396567118976 & 315.91331 & -47.79355 & 11.84 & 2.25 & 269.34 & 5.39 $\pm$ 0.02 & 5 \\
 J1844-6103 & 6632738920293491584 & 281.18591 & -61.06593 & 11.68 & 2.93 & 231.09 & 9.53 $\pm$ 0.02 & 3 \\
 J1848-5709 & 6637389068504460288 & 282.01976 & -57.15752 & 10.18 & 1.92 & 261.41 & 12.94 $\pm$ 0.02 & 4 \\
 J1922-4624 & 6663317786070516736 & 290.57806 & -46.40878 & 12.68 & 2.74 & 272.68 & 3.51 $\pm$ 0.03 & 5 \\
\enddata
\end{deluxetable*} \label{tab:basic_info}

\section{Follow-up} \label{sec:follow_up}

\subsection{FEROS} \label{ssec:feros} 

We obtained 135 spectra with the Fiber-fed Extended Range Optical Spectrograph (FEROS; \citealt{Kaufer1999Msngr}) mounted on the MPG/ESO 2.2 m telescope at the La Silla Observatory. All observations used $1\times 1$ binning and the resulting spectra have resolution $R\approx 50,000$. Exposure times ranged from 600 to 1200 seconds, depending on the brightness of the target. The data were reduced with the CERES pipeline \citep{Brahm2017} which performs bias-subtraction, flat fielding, wavelength calibration, and optimal extraction. The pipeline measures and corrects for small shifts in the wavelength solution during the course a night via simultaneous observations of a ThAr lamp with a second fiber.

\subsection{GALAH and APOGEE RVs} \label{ssec:galah_apogee}

For a few objects, we also include RVs from the publicly available GALAH DR3 \citep{Buder2021MNRAS} and APOGEE DR17 \citep{Abdurro'uf2022ApJS} catalogues. From the GALAH DR3 Main and Value-Added calatalogues, we took the dates, RVs, and RV errors from the columns \texttt{heliocentricJD}, \texttt{rv\_galah}, and \texttt{e\_rv\_galah}, respectively. From the APOGEE DR17 allVisit file, we used the columns \texttt{JD}, \texttt{VHELIO}, and \texttt{VRELERR}. 

We obtained FEROS spectra at one epoch for five calibration stars with both GALAH and APOGEE RVs to obtain RV offsets between the three RV scales. We calculate median offsets from FEROS of $+0.19\,\mbox{km s}^{-1}$ and $-0.31\,\mbox{km s}^{-1}$ for GALAH and APOGEE RVs respectively, which we apply when fitting the RVs in Section \ref{sec:rv_fitting}. 

All RVs obtained for our objects are listed in Tables \ref{tab:all_rvs} to \ref{tab:all_rvs_3} in Appendix \ref{sec:all_rvs}. 

\subsection{SED fitting} \label{ssec:SED_fitting}

We constructed the broadband SED of each source using synthetic SDSS {\it ugriz} photometry from {\it Gaia} XP spectra \citep{GaiaCollaboration2023A&A_synphot}, 2MASS {\it JHK} photometry \citep{Skrutskie2006AJ}, and {\it WISE} $W_1W_2W_3$ photometry \citep{Wright2010AJ}. 

For declinations below $-30^{\circ}$, we obtained $E(B-V)$ values from the \citet{Lallement2022A&A} 3D dust map, and for declinations above $-30^{\circ}$, we used the Bayestar2019 \citep{Green2019ApJ} 3D dust map. We take $R_V = 3.1$ assuming a \citet{Cardelli1989ApJ} extinction law. We take $E(g-r)$ provided by the Bayestar2019 map to be approximately equal to $E(B-V)$ \citep{Schlafly2011ApJ} and take $A_0$ (extinction evaluated at $500\,$nm) provided by the \citet{Lallement2022A&A} map to be $A_V$. 

The SED fitting procedure is identical to that described in \citet{Yamaguchi2024MNRAS} to which we refer the readers for details, but we summarize it here. We use the \texttt{MINEsweeper} code \citep{Cargile2020ApJ} which was designed to jointly model photometry and spectra, but in this work, we only use the code to interpolate between MIST evolutionary models \citep{Dotter2016ApJS, Choi2016ApJ} and predict synthetic photometry. The free parameters are the parallax, mass $M_{\star}$, initial metallicity [Fe/H]$_{\rm init}$, and Equivalent Evolutionary Phase (EEP, a monotonic function of age; see \citealt{Dotter2016ApJS}). The code uses neural network interpolation to swiftly calculate the SEDs and photometry for a given set of parameters. We use \texttt{emcee}, a Python Markov chain Monte Carlo (MCMC) sampler \citep{Foreman-Mackey2013PASP}, to sample from the posterior. For the metallicities, we placed Gaussian constraints based on the values calculated using the {\it Gaia} XP very low-resolution spectra as described in \citet{Andrae2023ApJS}. However, we do not use the values as published in \citet{Andrae2023ApJS}, which used ``single-star" parallaxes from the \texttt{gaia\_source} table. Instead, we use values re-calculated using parallaxes from the \texttt{nss\_two\_body\_orbit} table which takes into account of the binary motion \citep{GaiaCollaboration2023A&A, Halbwachs2023A&A}. We also use a parallax prior based on the {\it Gaia} astrometric orbital solution.

We do not fit for the contribution from the WD companion. This is justified as these objects were selected on the basis of their AMRF values which suggested a dark companion. And for a typical WD companion with $T_{\rm eff} \sim 10,000\,$K, the expected flux contribution in the optical is small ($<$ 0.1 mag; see discussion in Appendix B of \citealt{Yamaguchi2024MNRAS}).  

The results from the SED fitting for two example objects are shown in Figure \ref{fig:sed_example}. The fits for all other objects can be found in Appendix \ref{sec:sed_fitting_all}. Note that about half of the objects have GALEX far/near ultraviolet (NUV/FUV) photometry which are plotted on top of the SEDs, but we did not use them in the SED fitting as hot WDs may contribute significantly in the UV. In fact, there are a few NUV and FUV observations where significant excess is present, one being J0947-2018 shown in Figure \ref{fig:sed_example}. 

The medians and standard deviations of the marginalized posterior distributions for each parameter for all 31 objects are given in Table \ref{tab:sed_params}. The masses range from $0.64$ to $1.09\,M_{\odot}$, with a median at $0.85\,M_{\odot}$. Note that although some of the mass uncertainties are formally very small, we adopt a minimum uncertainty of $0.03\,M_{\odot}$ in the later fitting to account for possible systematics in the stellar models and/or photometry (e.g. bias in u-band magnitudes from the {\it Gaia} XP spectra; \citealt{GaiaCollaboration2023A&A_synphot}). 

%Add a few sentences summarizing the results of the fitting. What are the typical masses, and how do they compare to those initially estimated by Shahaf et al? How evolved (or not) are most sources? How much extinction is typical? Maybe include a plot of M_SED vs M_Shahaf?

\begin{figure*}
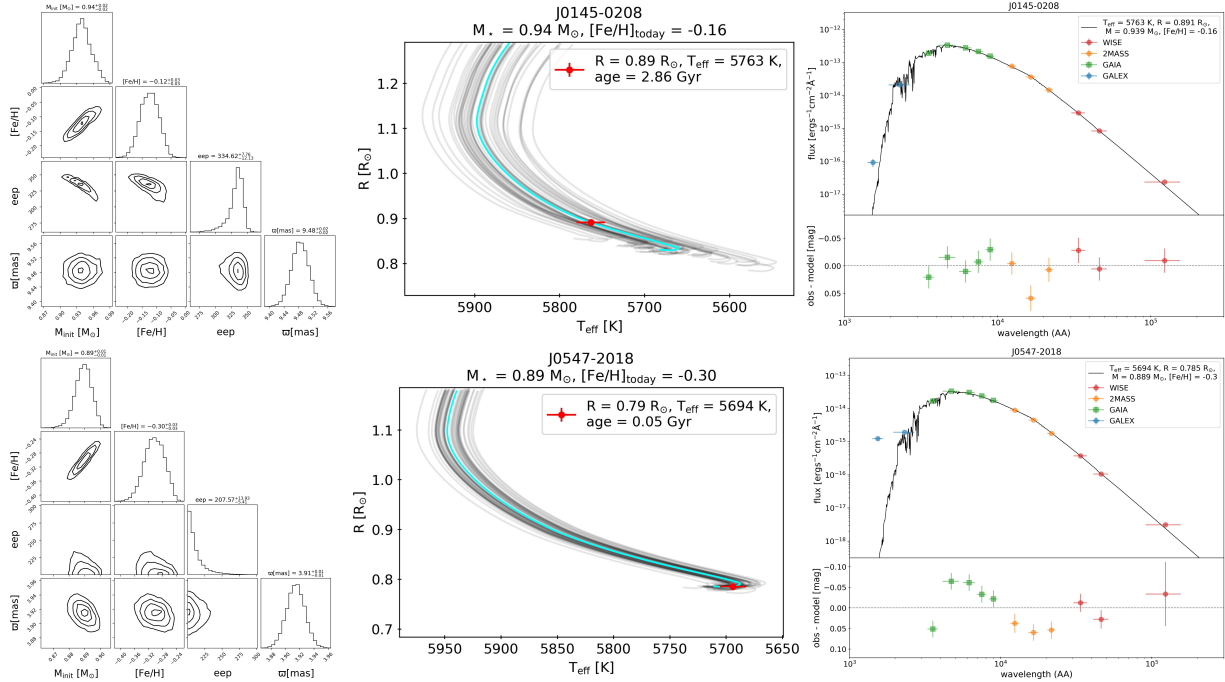

    \centering
    \includegraphics[width=0.9\textwidth]{Figures/2505477938150534016.jpeg}
    \includegraphics[width=0.9\textwidth]{Figures/2966114550741693440.jpeg}
    \caption{Results from SED fitting of two example objects,  J0145-0208 (top) and J0547-2018 (bottom). \textit{Left}: Corner plot showing the four fitted parameters. \textit{Center}: Isochrone predictions of the radius versus temperature from 50 random draws of the posterior. The marker indicates the best-fit current values from the SED fitting. \textit{Right}: Best-fit SED model plotted compared to the observed photometric points from WISE, 2MASS, {\it Gaia}, and GALEX along with the residuals. Note that GALEX points are plotted here, but were not used in the SED fitting. J0145-0208 does not show significant FUV excess, but J0547-2018 does.}
    \label{fig:sed_example}
\end{figure*}

\section{Joint Gaia + RV fitting} \label{sec:rv_fitting}

All of the objects in this sample have AstroSpectroSB1 solutions with bit indexes of 65535, meaning that the \textit{Gaia} astrometry and RVs were fitted with a single orbital model containing 15 parameters: \texttt{ra}, \texttt{dec}, \texttt{parallax}, \texttt{pmra}, \texttt{pmdec}, \texttt{a\_thiele\_innes}, \texttt{b\_thiele\_innes}, \texttt{f\_thiele\_innes}, \texttt{g\_thiele\_innes}, \texttt{c\_thiele\_innes}, \texttt{h\_thiele\_innes}, \texttt{center\_of\_mass\_velocity}, \texttt{eccentricity}, \texttt{period}, \texttt{t\_periastron} \citep{Halbwachs2023A&A, GaiaCollaboration2023A&A}. The Thiele-Innes elements A, B, F, G are transformation of the Campbell elements (photocenter semi-major axis, inclination, argument of periapsis, and the position angle of the ascending node) which describe the photocenteric orbit from astrometry. The C and H elements describe the motion of the primary as inferred from spectroscopic RVs.

For each object, we simultaneously fit the {\it Gaia} astrometric solution and the  follow-up RVs. The primary advantage of astrometry over just RVs in fitting orbits is that it constrains the inclination and thus the precise mass of the companion, as opposed to a lower limit. The benefit of having follow-up RVs is that they validate the {\it Gaia} solution and, given their small uncertainties, allow us to place much tighter constraints on the parameters. In addition, as we discuss below, RVs allows us to constrain the flux ratio of the components. 

We use the likelihood function as defined in \citet{El-Badry2023MNRAS}, which is the product (sum in log space) of two components. The first is the {\it Gaia} term. For each model with a vector of the 15 astrometric quantities $\theta_{\rm Gaia}$, the likelihood is calculated as
\begin{equation} \label{eqn:lnL_ast}
     \mathrm{ln} L_{\rm{Gaia}} = -\frac{1}{2}\left(\theta_{\rm{Gaia}}-\mu_{\rm{Gaia}}\right)^{\rm{T}}\Sigma_{\rm{Gaia}}^{-1}\left(\theta_{\rm{Gaia}}-\mu_{\rm{Gaia}}\right)
\end{equation} 
where $\mu_{\rm{Gaia}}$ is the vector of best-fit values from \textit{Gaia} and $\Sigma_{\rm{Gaia}}$ is the corresponding covariance matrix, which can be constructed with the correlation vector provided in the \texttt{gaiadr3.nss\_two\_body\_orbit} table. 

The second term comes from fitting the RVs. We use a standard Keplerian model and compare its predictions to our observations: 
\begin{equation} \label{eqn:lnL_RVs}
     \mathrm{ln} L_{\rm{RVs}} = - \frac{1}{2} \sum_i \frac{\left({\rm RV}_{\rm pred}\left(t_i\right) - {\rm RV}_i\right)^2}{\sigma^2_{{\rm RV}, i}}
\end{equation}
where ${\rm RV}_{\rm pred}\left(t_i\right)$ and ${\rm RV}_i$ are the predicted and measured RVs at a time $t_i$ and $\sigma_{{\rm RV}, i}$ is the error in ${\rm RV}_i$. 

Our model fits 15 parameters: ra, dec, parallax $\varpi$, pmra, pmdec, period $P_{\rm orb}$, eccentricity $e$, inclination $i$, argument of periapsis $\omega$, angle of the ascending node $\Omega$, periastron time $T_p$, center of mass velocity $\gamma$, luminous star mass $M_1$, WD mass $M_2$, and finally, the flux ratio $\mathcal{S}$. We place a Gaussian prior on $M_1$ using the best-fit value and uncertainty obtained from the SED fitting (with an uncertainty floor of $0.03\,M_{\odot}$; Section \ref{ssec:SED_fitting}). We also run a model using just the {\it Gaia} constraints (equation \ref{eqn:lnL_ast}) to assess the relative importance of the {\it Gaia} data and our RVs in constraining the orbit. 

The flux ratio $\mathcal{S}$ comes in because the angular semi-major axis of the photocenter $\alpha$ is related to that of the primary $\alpha_1$ by
\begin{equation}
    \alpha = \alpha_1 \left(1 - \frac{\mathcal{S} \left(1 + q\right)}{q \left(1 + \mathcal{S}\right)}  \right) 
\end{equation} \label{eqn:alpha}
By definition, $\mathcal{S}$ should be 0 or greater. It is zero in the case of a completely dark companion, where $\alpha = \alpha_1$. We emphasize that astrometry alone does not directly constrain this $\mathcal{S}$ -- RV measurements of the luminous primary are necessary. Since all objects presented in this work have combined astrometric \textit{and} single-lined spectroscopic solutions (AstroSpectroSB1), the {\it Gaia} solutions alone provide some constraint on this value (through the C and H elements) which will be tightened by the follow-up RVs. In the case of purely astrometric solutions (Orbital) however, follow-up RVs are necessary. 

As described in Section \ref{sec:sample_selection}, our objects were selected to have AMRF values and CMD positions indicative of dark companions. We therefore expect $\mathcal{S} \ll 1$ if the {\it Gaia} solutions are robust (this is true, even for those with Orbital solutions). This means that if a lower bound of zero is set, the posterior distributions will tend to run into this lower bound, and since $\mathcal{S}$ cannot be negative, this will lead to a bias toward higher $M_2$. Note also that if the {\it Gaia} solution is exactly correct, $\mathcal{S} = 0$ should give the minimum predicted RV variation (because in the absence of noise or systematic errors, the photocenter follows the primary). But if the {\it Gaia} solution is not exactly correct, it is possible for the true RV variation to be smaller than it predicts, which corresponds to negative $\mathcal{S}$. To avoid introducing a bias toward higher $M_2$, in the fitting, we allow $\mathcal{S}$ to range from -0.5 to 1 (none of the solutions run into the lower bound). 

\begin{figure*}
    \centering
    \includegraphics[width=0.85\textwidth]{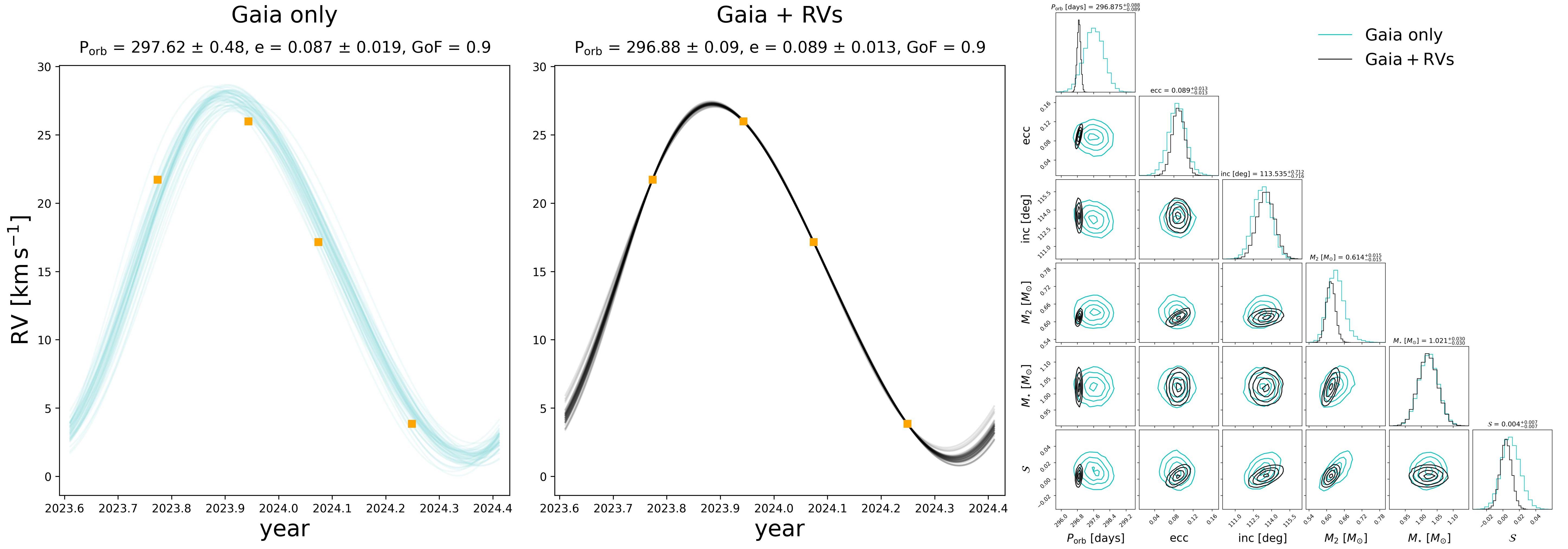}
    \includegraphics[width=0.85\textwidth]{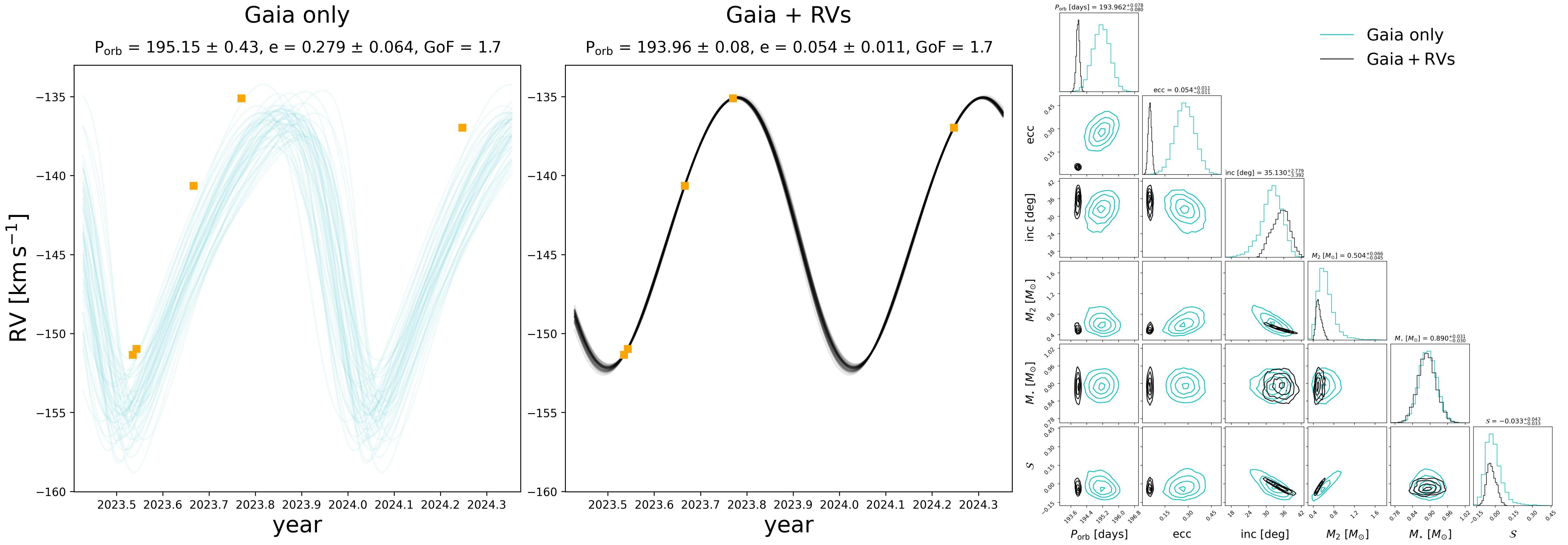}
    \caption{Results of joint {\it Gaia} + RV fitting for two example objects, J0602-1747 (\emph{top}) and J1834+1525 (\emph{bottom}). These represent cases with reliable and unreliable {\it Gaia} orbital solutions, respectively.  \emph{Left}: RV curves using orbital parameters from 50 random draws of the posterior resulting from the fitting of the {\it Gaia} solution only (cyan) and the joint fitting with our follow-up RVs (black). \emph{Right}: The corresponding corner plots showing  some key parameters for the same objects.}
    \label{fig:example_rv_curves}
\end{figure*}

The left panel of Figure \ref{fig:example_rv_curves} compares RV curves from 50 random draws of the posterior using just the {\it Gaia} solution (cyan) to those from jointly fitting the RVs and the Gaia solution (black), for two example objects, J0602-1747 and J1834+1525. These objects were chosen to represent a typical case with a reliable {\it Gaia} orbital solution that agrees with our measured RVs, and a case with an unreliable {\it Gaia} solution. Similar plots of the RV curves for all 31 objects can be found in Appendix \ref{sec:rv_fitting_all}. 
The right panel shows the corner plots for several key parameters. For J1951-0305, we see that the inclusion of our follow-up RVs tightens the constraints on the parameters (particularly $P_{\rm orb}$ and $M_2$) but the best-fit values from both fits are consistent with each other to within $1 \sigma$. Meanwhile, for J1834+1525, we see that there are significant disagreements between the solutions from the two fits. As we discuss further in Section \ref{ssec:gaia_success}, this is one of the few objects for which the {\it Gaia} solutions have underestimated errors, despite having \texttt{goodness\_of\_fit} values indicative of an unproblematic solution. 

% Table of fitted orbital parameters

\section{Results} \label{sec:discussion}

\subsection{Flux ratio} \label{ssec:flux_ratio}

Figure \ref{fig:S_hist_e_vs_S} shows a histogram of the distribution of $\mathcal{S}$. As expected, the majority of objects have $\mathcal{S} \lesssim 0$ which is indicative of a dark companion/WD. However, this is not conclusive, because depending on the masses of both components, a given flux ratio may or may not be large enough to be consistent with a triple scenario. 

\begin{figure*}
    \centering
    \includegraphics[width=0.8\textwidth]{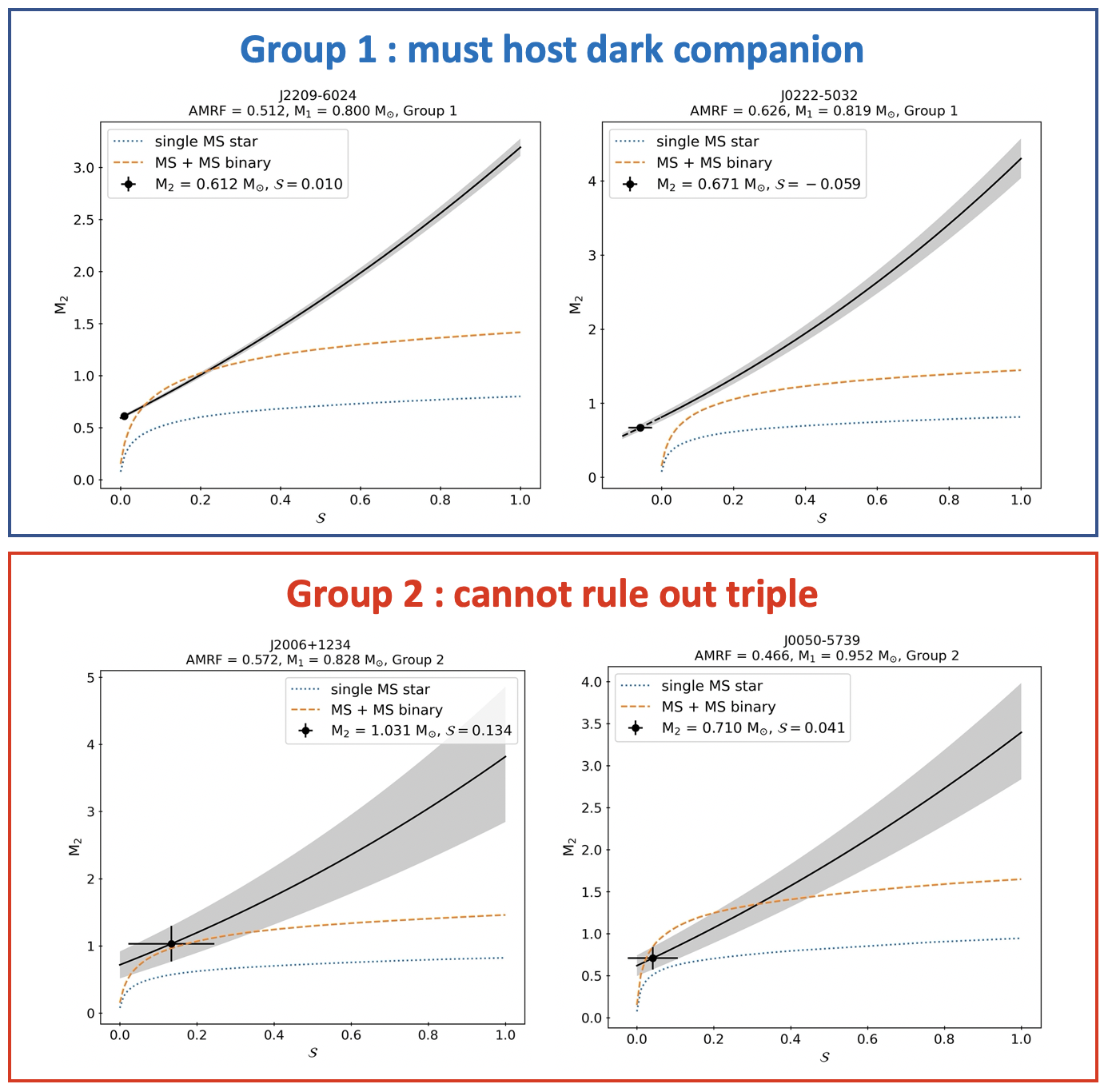}
    \caption{Secondary mass $M_2$ against the flux ratio $\mathcal{S}$ for several example objects. The black line is the expected relation obtained from the AMRF (Equation \ref{eqn:AMRF_1}) calculated using the best-fit orbital parameters derived in Section \ref{sec:rv_fitting};  black point marks the best-fit values of $M_2$ and $\mathcal{S}$. The blue solid line and orange dashed lines correspond to theoretical curves for single MS star and equal-mass inner binary companions. On the top row are ``group 1'' objects (as described in Section \ref{ssec:flux_ratio}) which must host dark companions: the constraints on $M_2$ and $\mathcal{S}$ are inconsistent with an inner MS+MS binary for any $\mathcal{S}$. Bottom row shows ``group 2'' objects, for which the flux ratio constraint cannot rule out an inner MS+MS binary.} %Each of the three objects shown belong to group 1, 2, and 3 (from left to right), as described in Section \ref{ssec:flux_ratio}. Groups 1 and 2 are binaries consistent with hosting a dark secondary, while group 3 are those consistent with hosting an inner binary of two MS stars. Note that none of our 31 objects are in group 3 so what is being plotted is that of a hypothetical system.}
    \label{fig:M2_v_S}
\end{figure*}

\begin{figure*}
    \centering
    \includegraphics[width=0.9\textwidth]{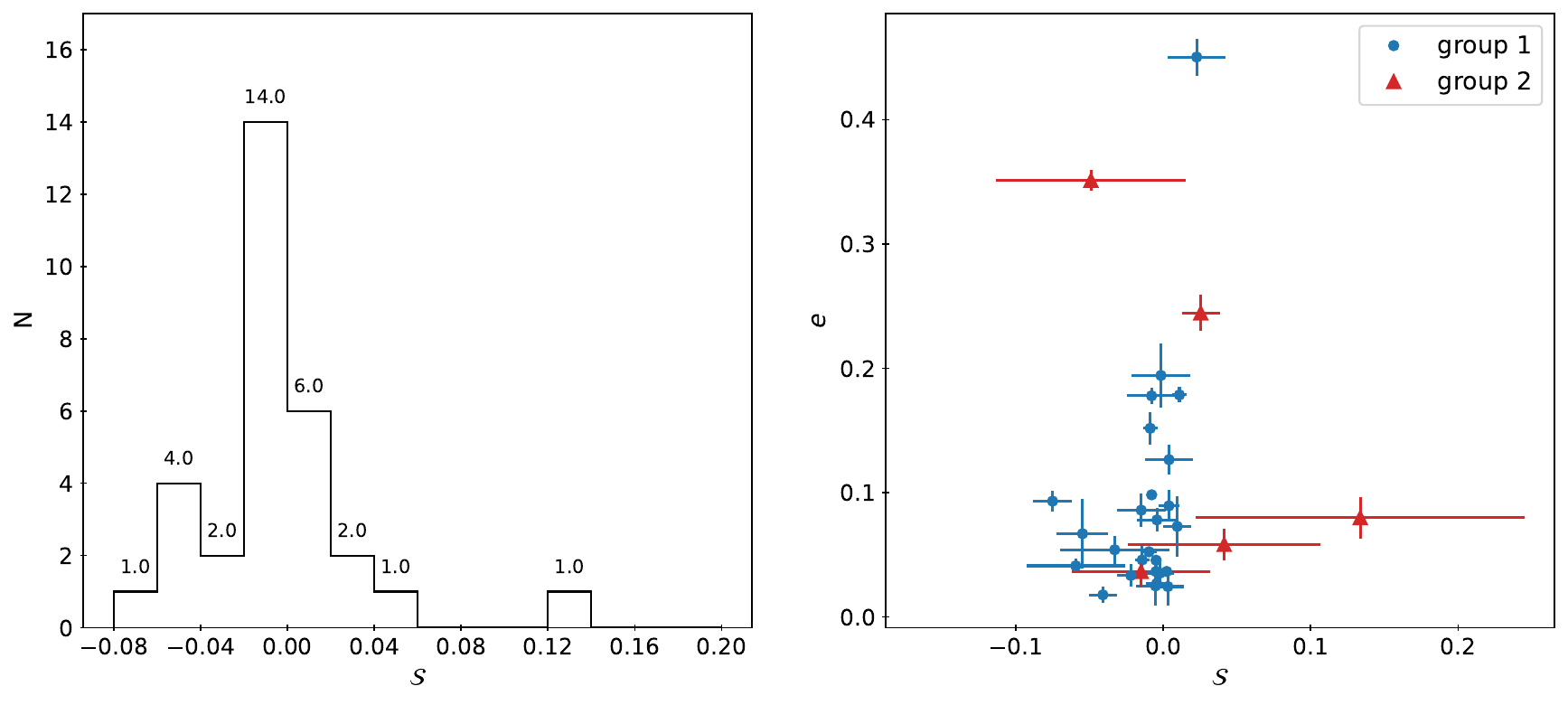}
    \caption{\emph{Left}: Histogram showing the distribution of the best-fit flux ratio $\mathcal{S}$ for the 31 objects in our sample. The majority of objects have $\mathcal{S} \sim 0$, which implies the less luminous secondary contributes little light to the system. \emph{Right}: Eccentricity $e$ against the flux ratio $\mathcal{S}$ of our sample resulting from the joint {\it Gaia} + RVs fits. The markers and colors of the points represents the groups that each object belongs to, as described in Section \ref{ssec:flux_ratio} and shown in Figure \ref{fig:M2_v_S}. All of the systems  are consistent with a dark WD secondary.}
    \label{fig:S_hist_e_vs_S}
\end{figure*}

Equating equations \ref{eqn:AMRF_1} and \ref{eqn:AMRF_2}, one can solve for the secondary mass $M_2$ as a function of the flux ratio $\mathcal{S}$, given all of the other parameters from the orbital fitting (Section \ref{sec:rv_fitting}). On Figure \ref{fig:M2_v_S}, we plot this curve as a black line for several exemplar objects. The best-fit values of $M_2$ and $\mathcal{S}$ must be a point that lies somewhere on this line (black marker). 

We use synthetic photometry of zero-age MS stars in the {\it Gaia} band from MIST to compute theoretical curves on this plot for a single MS companion (blue dotted) and an inner binary (orange dashed) of two equal-mass MS stars.  

Based on this plot, we can classify each object into one of two groups: 
\begin{itemize}
    \item Group 1: Inconsistent with being a triple system. The best-fit point (black marker) and its error bars (1-$\sigma$) do not cross the theoretical curve for an inner binary companion (orange dashed line). 
    \item Group 2: Cannot rule out a triple. All other cases -- the best-fit point is consistent with the orange dashed line, to within 1-$\sigma$.  Note that objects in group 2 could still host WD companions -- the data just do not rule out a triple. 
\end{itemize}

In Figure \ref{fig:M2_v_S}, we show examples of objects that belong to each of these two groups. In total, 26 out of the 31 objects are in group 1. This means the large majority of all objects are inconsistent with being in triples and are thus likely to host WDs. As for the remaining 5 objects, while we cannot rule out triples, we note that all of them are also consistent with hosting dark companions. 

Modeling of the SEDs with triple models could likely set more stringent limits on $\mathcal{S}$ for objects with only a few RVs, but here we only include constraints based on the joint fitting of the {\it Gaia} solutions and RVs. 

The right panel of Figure \ref{fig:S_hist_e_vs_S} plots the eccentricity against $\mathcal{S}$ for all objects, with the markers indicating their ``groups" as described above. We notice two key features: (1) all eccentricities are relatively low. This is expected for binaries hosting WD companions, but unexpected if the systems are primarily hierarchical triples (note also that those in group 2 tend to have larger uncertainties in $\mathcal{S}$); (2) even among the systems in group 1, most of the eccentricities are significantly different from 0. This indicates any tidal circularization process has not been very efficient. We compare these eccentricities to those of millisecond pulsar + WD binaries in Section \ref{ssec:P_v_ecc}.

\subsection{Success rate of Gaia solutions} \label{ssec:gaia_success}

\begin{figure*}
    \centering
    \includegraphics[width=0.9\textwidth]{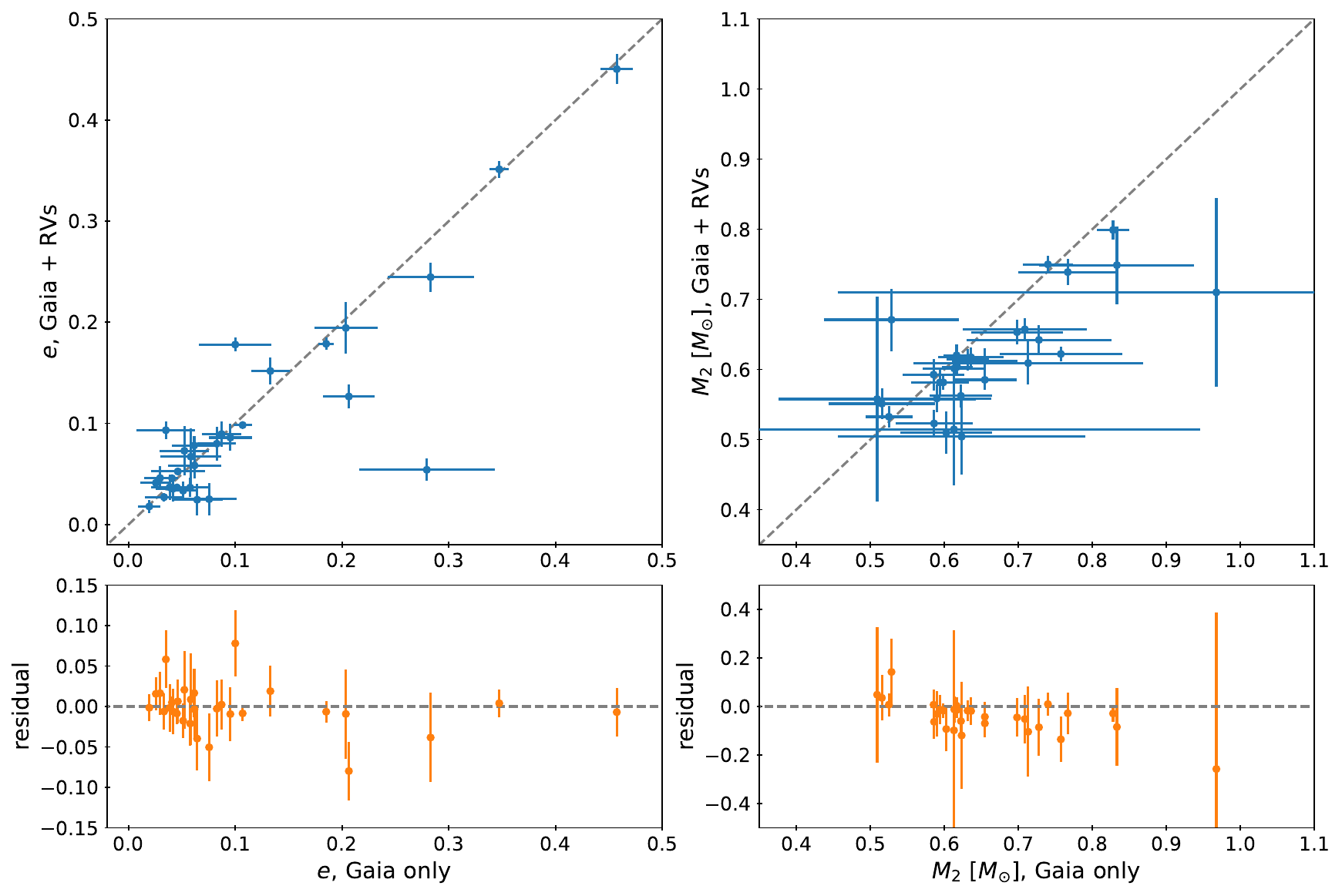}
    \caption{Best fit values of $e$ and $M_2$ resulting from joint {\it Gaia} + RVs fits, plotted against the values from the {\it Gaia} solution alone. The dashed grey line indicates where the values from the two fits are equal to each other. The lower panels show the residuals from this line. We see that in general, the inclusion of RVs lowers the measured $M_2$. (Note that in these plots, we have zoomed into regions where the majority of the points are located, so a few points have been cut out.)}
    \label{fig:M2_ecc_v_P}
\end{figure*}

One purpose of obtaining RV follow-up was to validate the {\it Gaia} astrometric solutions -- in this case, specifically those with AstroSpectroSB1 solutions. In Figure \ref{fig:M2_ecc_v_P}, we plot the best-fit values of $e$ and $M_2$ obtained from the fitting with and without the inclusion of RVs. If they are in agreement, the points would lie along the gray dashed line. We see that for the majority of objects, the change in the best-fit values are minimal, with the primary effect of the RVs being to reduce the uncertainties. However, there are two (five) cases with $> 2\sigma$ ($> 1\sigma$) discrepancy in $e$ between the two solutions. The two sources are J1834+1525 and J1922-4624. Figures \ref{fig:rv_astro_curves} and \ref{fig:rv_astro_curves_2} in Appendix \ref{sec:rv_fitting_all} show that the discrepancy between measured and predicted RVs is clearly visible. 
These objects have goodness-of-fit values of 1.7 and 2.7 respectively, which are not enhanced compared to the rest of the objects. These sources would therefore not have been identified to have spurious solutions without RV follow-up.

Thus, at least 2 out of the 31 objects in our sample ended up having unreliable {\it Gaia} solutions. While this suggests that most AstroSpectroSB1 solutions are reliable, there is a non-negligible amount which are not, so RV follow-up is important for objects that appear unusual based on their {\it Gaia}-only solutions. Note also that we pre-selected sources that did not have high goodness-of-fit values ($\lesssim 4.8$). The fraction of spurious solutions also appears to be higher for purely astrometric solutions, as we will describe in future work.

\subsection{Comparison to literature PCEBs}

\begin{figure*}
    \centering
    \includegraphics[width=0.95\textwidth]{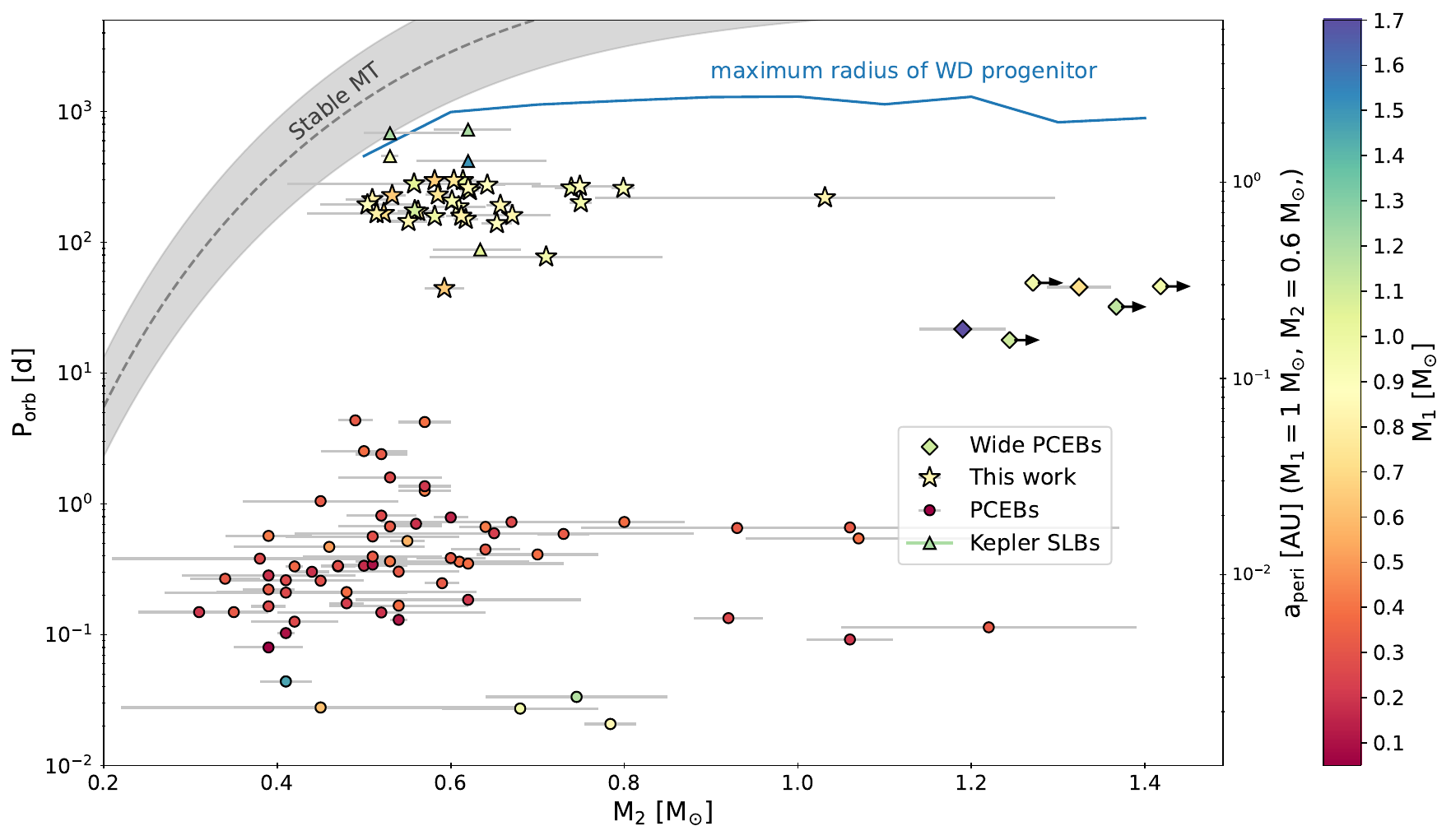}
    \caption{Orbital period, $P_{\rm orb}$, against WD mass, $M_2$. The colors of the points represent the masses of the luminous companion, $M_1$. On the right axis, we show the minimum orbital separation, $a_{\rm peri}$, corresponding to the period assuming $M_1 = 1\,M_{\odot}$ and $M_2 = 0.6\,M_{\odot}$. We plot the ``short-period PCEBs'' discovered with SDSS by \citet{Rebassa-Mansergas2007MNRAS} as well as those from \citet{Hernandez2021MNRAS, Hernandez2022MNRAS_vi, Hernandez2022MNRAS} with circle markers. ``Wide PCEBs'' which include IK Peg \citet{Wonnacott1993} and the systems from \citet{Yamaguchi2024MNRAS} are plotted with diamond markers. We also plot the self-lensing systems discovered with Kepler \citep{Kruse2014Sci, Kawahara2018AJ, Yamaguchi2024arXiv} with triangle markers. The solid line marks the maximum mass that is reached by the giant progenitor at each WD mass. The dashed line is the track along which binaries undergoing stable MT are expected to evolve \citet{Rappaport1995MNRAS}.}
    \label{fig:P_v_M2}
\end{figure*}

In Figure \ref{fig:P_v_M2}, we plot orbital period, $P_{\rm orb}$, against WD mass, $M_2$, for the objects in our sample as well as literature PCEBs. The colors of the points represent the mass of the luminous companion, $M_1$. We also mark the corresponding minimum orbital separation, $a_{\rm peri} = a (1 - e)$, on the right axis for a $0.6 M_{\odot}$ WD and a $1 M_{\odot}$ companion in a circular orbit (this is the default assumption that we make for any conversion between $P_{\rm orb}$ and $a_{\rm peri}$). Binaries formed through stable Roche lobe overflow are expected to evolve along the gray dashed line derived from the $P_{\rm orb}-M_2$ relation from \citet{Rappaport1995MNRAS}, with the shaded region showing the spread in $P_{\rm orb}$ by a factor of 2.4 (as a result of different giant masses and compositions). The blue solid line marks the maximum radius reached by an isolated WD progenitor. We obtained this using MIST evolutionary tracks for stars with a range of masses \citep{Dotter2016ApJS, Choi2016ApJ} which were converted to WD masses using the initial-final mass relation from \citet{Williams2009ApJ}. Any system lying below this line must have interacted at some point during its evolution. 

The best-studied sample of literature PCEBs comes from the Sloan Digital Sky Survey (SDSS) which identified many MS + WD binaries. From these, \citet{Rebassa-Mansergas2007MNRAS} discovered 37 new PCEBs. On Figure \ref{fig:P_v_M2}, we plot these, along with 25 previously known PCEBs, for a total 62 systems (compiled by \citealt{Zorotovic2010A&A}). We also plot the PCEBs from the ``white dwarf binary pathways survey" with more massive AFGK companions \citep{Hernandez2021MNRAS, Hernandez2022MNRAS_vi, Hernandez2022MNRAS}. All of these systems have orbital periods of $\lesssim 1\,$d, and their formation can be explained by simple energy conservation arguments, formally known as the $\alpha$-formalism. In the simplest case, this is the balance between a fraction $\alpha$ of the orbital energy loss and the gravitational binding energy of the donor's envelope. Various models of these systems have found that a value of $\alpha \sim 0.3$ can reproduce the observed population \citep{Zorotovic2010A&A, Davis2010MNRAS, Toonen2013A&A, Camacho2014A&A, Zorotovic2022, Scherbak2023MNRAS}. 

The above systems are labelled ``short-period PCEBs" in Figure \ref{fig:P_v_M2} to distinguish them from the ``wide PCEBs" that have been found over the years. These have much longer orbital periods of $\sim$ months, and the systems that have been characterized so far host unusually massive ($\gtrsim 1.2\,M_{\odot}$) WDs. These include IK Peg \citep{Wonnacott1993} as well as five systems from {\it Gaia} recently identified by \citet{Yamaguchi2024MNRAS}. Several self-lensing binaries (SLBs) discovered with Kepler are also plotted. These host less massive WDs and have even longer orbital periods from $\sim 100-700\,$d \citep{Kruse2014Sci, Kawahara2018AJ, Masuda2019ApJL, Yamaguchi2024arXiv}. All of these systems (with the possible exception of one SLB) lie firmly below both the blue solid line and the gray region, meaning that they must have interacted at some point but not via stable mass transfer, pointing towards CEE. However, the wide orbits of these systems imply less orbital shrinkage during the common envelope phase than is released in the formation of close PCEBs. This has to lead to works arguing that additional sources of energy are required to unbind the envelope \citep[e.g.][]{Davis2010MNRAS, Zorotovic2010A&A, Parsons2016MNRAS, Scherbak2023MNRAS, Yamaguchi2024MNRAS}, while others circumvent this with mass transfer from a donor on the TP-AGB (we explore this in Section \ref{ssec:cee}. See also \citealt{Belloni2024arXiv_b, 2024arXiv240117510B, Belloni2024arXiv}). 

The objects from this work are plotted with star markers. These have orbital periods ranging from $\sim 50 - 300\,$d, intermediate between the wide PCEBs and Kepler SLBs. They all lie in a region of this parameter space expected to be sparsely populated according to standard binary evolution models \citep{Shahaf2024MNRAS} -- having more massive WDs than expected at their orbital periods if they evolved through stable MT, while also having significantly longer orbital periods than expected for PCEBs. We discuss possible formation scenarios of these systems in Section \ref{sec:formation_channels}. 

\subsection{Period-eccentricity relation} \label{ssec:P_v_ecc}

\begin{figure}
    \centering
    \includegraphics[width=\columnwidth]{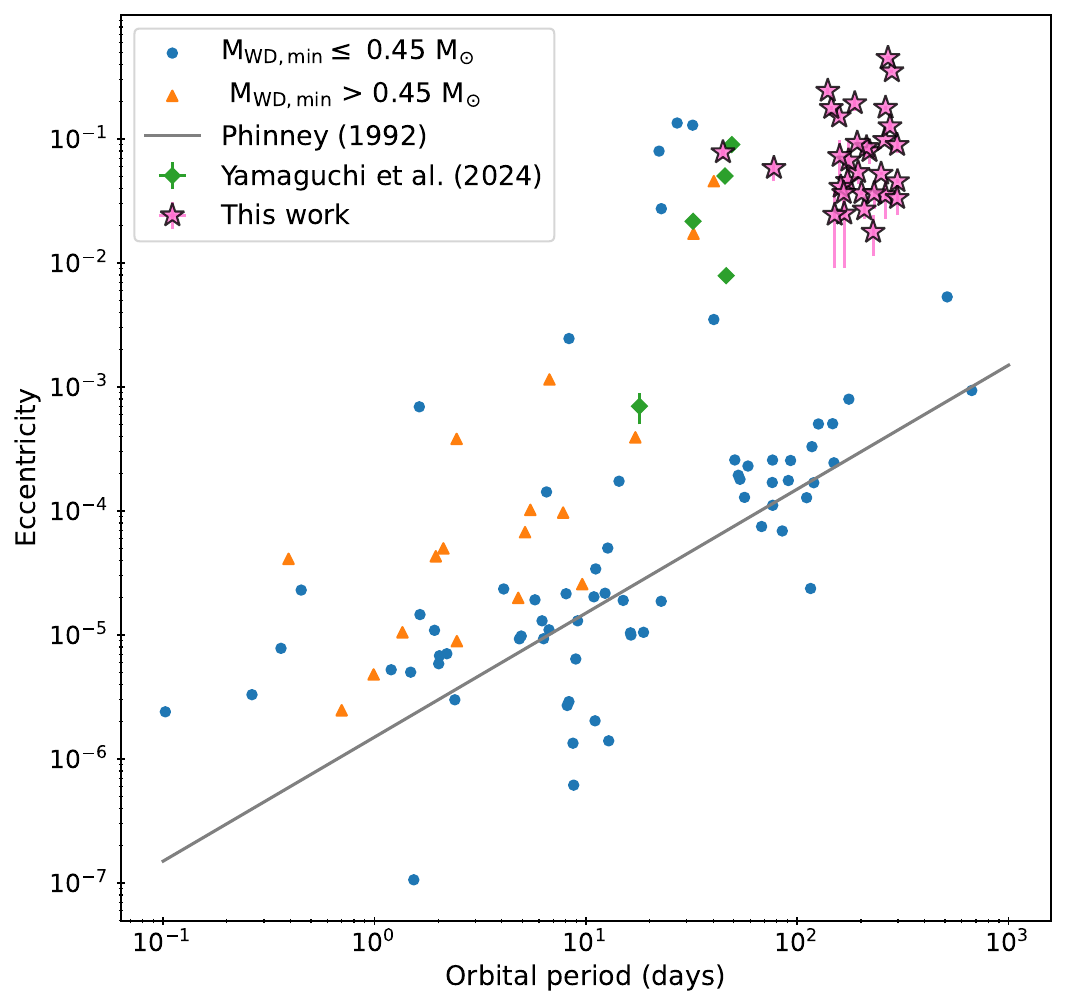}
    \caption{Period-eccentricity diagram of literature MSP + WD binaries from the ATNF pulsar catalog (\citealt{Manchester2005AJ}; compiled by \citealt{Hui2018ApJ}). The triangle and circle markers differentiate systems with WD masses above and below $0.45\,M_{\odot}$ respectively, which marks the approximate boundary between He and CO/ONeMg WDs. We also plot the objects from this work with star markers and the wide PCEBs from \citet{Yamaguchi2024MNRAS} with diamond markers. The solid line is a relation expected for He WD systems formed from stable MT from \citet{Phinney1992RSPTA}. We see that our objects have significantly higher eccentricities than predicted by this relation.}
    \label{fig:P_v_ecc}
\end{figure}

On Figure \ref{fig:P_v_ecc}, we plot the period against eccentricity of our objects, along with literature millisecond pulsars (MSPs) + WD binaries from the Australia Telescope National Facility (ATNF) pulsar catalogue (\citealt{Manchester2005AJ}; we plot the sample compiled by \citealt{Hui2018ApJ}). We distinguish systems with WD masses above and below 0.45 $M_{\odot}$. This is the approximate boundary between He WDs and more massive CO or ONeMg WDs, which correspond to systems thought to have formed from stable and unstable MT. The gray line is the theoretical relation expected for systems containing He WDs formed from stable MT derived in \citet{Phinney1992RSPTA}. \citet{Cohen2024arXiv} recently extended this relation to CO WDs with AGB progenitors and found an upper limit in the eccentricity of $\sim 3 \times 10^{-3}$, which is marked with a dashed horizontal line.

We see that our objects, plotted with pink stars, all have significantly higher eccentricities than the vast majority of the MSP + WD binaries at similar periods. We also plot the 5 systems from \citet{Yamaguchi2024MNRAS} with green diamonds, four of which lie at comparable eccentricities at shorter orbital periods. \textit{None} of these MS + WD binaries have eccentricities consistent with zero nor are they consistent with the \citet{Phinney1992RSPTA} relation. All of them also lie above the \citet{Cohen2024arXiv} limit. Since it is unlikely that all of these systems are in triples, this suggests that Kozai-Lidov oscillations are unlikely to be responsible for their eccentricities. Moreover, since many MSP + WD binaries -- which must have undergone stable MT in order to spin up the MSP -- do lie close to the \citet{Phinney1992RSPTA} relation, this suggests that these systems are not stable MT products. We note also that several recent 1D and 3D models have shown that initially eccentric systems that undergo CEE can retain eccentricities of $\sim 0.01 - 0.1$ post-inspiral, comparable to those of our systems \citep{Sand2020A&A, Glanz2021MNRAS, Bronner2024A&A}.

\section{Possible formation channels} \label{sec:formation_channels}

\subsection{Formation through CEE} \label{ssec:cee}

Using Modules for Experiments in Stellar Astrophysics (MESA; \citealt{Paxton2011ApJS, Paxton2013ApJS, Paxton2015ApJS, Paxton2018ApJS}) models of the WD progenitor, we can calculate the range of orbital periods that can result from CEE. We performed similar calculations in \citet{Yamaguchi2024MNRAS} and refer the reader to that work for more details. In short, we use a single star model of the WD progenitor (i.e. the donor) to trace its properties with time and use the $\alpha$-formalism to predict the final orbital separation of the PCEB as a function of the separation at the time when mass transfer began. We use MESA version 22.05.1. 

The $\alpha$-formalism is a statement of energy conservation between the binding energy of the star's envelope $E_{\rm bind}$ and a fraction $\alpha_{\rm CE}$ of the orbital energy released during CEE: 
\begin{equation} \label{eqn:alpha_formalism}
    E_{\rm bind} = \alpha_{\rm CE} \left(-\frac{GM_{\rm WD} M_{\star}}{2a_f} + \frac{GM_{i} M_{\star}}{2a_i}\right)
\end{equation} 
where $M_{\rm WD}$ is the WD mass, $M_{i}$ is the WD progenitor mass, $M_{\star}$ is the companion mass, $a_i$ is the initial separation at the onset of mass transfer (calculated using the \citealt{Eggleton1983ApJ} formula), and $a_f$ is the final separation at envelope ejection (i.e. the separation of the resultant PCEB). Given the short timescale of the CEE, we assume $M_{\star}$ remains constant. 

We define $E_{\rm bind}$ as the sum of the gravitational binding energy of the envelope $E_{\rm grav}$ and a fraction $\alpha_{\rm int}$ of its internal energy $E_{\rm int}$. The internal energy makes the total binding energy of the envelope less negative, meaning it is less bound and thus easier to eject \citep[see e.g.][]{1968IAUS...34..396P, Ivanova2013A&ARv}. This term was introduced to allow for the formation of wide PCEBs ($P_{\rm orb} \sim$ months - years) where the orbital shrinkage during the CEE may not be enough to overcome the gravitational binding energy \citep[e.g.][]{Scherbak2023MNRAS, Yamaguchi2024MNRAS}. MESA defines the internal energy as the sum of the ``thermodynamic" (thermal + radiation energy) and recombination energy \citep{Paxton2018ApJS}. It does not provide an easy way to extract the two components separately, so we estimate the specific thermal energy as $3P/(2\rho)$, and radiation energy as $3P/\rho$. The total binding energy of the envelope is:
\begin{equation}
    E_{\rm bind} = E_{\rm grav} + \alpha_{\rm int} E_{\rm int} = E_{\rm grav} + (\alpha_{\rm th} E_{\rm th} + \alpha_{\rm rec} E_{\rm rec})
\end{equation}
where we separately define the efficiencies $\alpha_{\rm th}$ and $\alpha_{\rm rec}$ corresponding to thermodynamic and recombination energies respectively. In the following calculations, we always take $\alpha_{\rm th} = 1$.  

For different combinations of $\alpha_{\rm CE}$ and $\alpha_{\rm rec}$, and taking observed values of $M_{\star}$, we can rearrange equation \ref{eqn:alpha_formalism} to solve for the final separation $a_f$ as a function of the donor model parameters when mass transfer starts. For all of our calculations, we use $M_{\star} = 0.85\,M_{\odot}$ which is the median mass of luminous companion in our systems. 

The WD masses for the objects in our sample range from around $0.5$ to $0.8\,M_{\odot}$ (excluding a single object J2006+1234 at $1\,M_{\odot}$), with a median at $\sim 0.6\,M_{\odot}$ which corresponds to progenitor masses of $1$ to $3\,M_{\odot}$, with a median at $\sim 1.5\,M_{\odot}$ \citep[e.g.][]{Williams2009ApJ, Cummings2018ApJ, El-Badry2018ApJL}. Therefore, we ran $1$, $1.5$, and $3\,M_{\odot}$ MESA models.

The inlists are based on those from the \texttt{1M\_pre\_ms\_to\_wd} test suite, though several changes and additions have been made following \citep{Rees2024MNRAS_2}. Unless otherwise stated, we use the wind prescription from \citet{Farmer2015ApJ}, setting \texttt{Reimers\_scaling\_factor} = 0.5 \citep{Reimers1975MSRSL, Reimers1977A&A} and \texttt{Blocker\_scaling\_factor} = 0.05 \citep{Bloecker1995A&A}. We make our inlists publicly available on Zenodo\footnote{https://zenodo.org/records/11114312}. 

Recently, \citet{Belloni2024arXiv_b, 2024arXiv240117510B, Belloni2024arXiv} emphasized the importance of mass transfer from a donor in the thermally pulsating AGB (TP-AGB) phase to produce wide PCEB orbits. This phase is difficult to model, with several instabilities that arise as the envelope mass is reduced by mass loss, causing it to expand and cool. These can terminate models prematurely or lead to prohibitively slow convergence, and are especially problematic for more massive stars which are the progenitors to ultra-massive WDs. In \citet{Yamaguchi2024MNRAS}, we terminated models at the onset of the TP-AGB phase for this reason. In this work, we run our models until the end of the TP-AGB, with help from inlists and custom MESA routines created by \citet{Rees2024MNRAS_2} which prevent instabilities and ensure convergence (we refer readers to their paper for details; their models are also discussed in \citealt{Rees2024MNRAS}). Note that in their work, they use the mass loss scheme of \citet{Vassiliadis1993ApJ} on the AGB, as opposed to that of \citet{Bloecker1995A&A}. We briefly discuss the effect of the chosen wind prescription in Section \ref{sssec:results_cee_mesa}. Models are terminated when most of the star's envelope has been stripped off (\texttt{envelope\_mass\_limit} = 0.01) or if the minimum timestep (\texttt{min\_timestep\_limit} = 1d-6) is reached. The latter only occurs for the $3\,M_{\odot}$ model, which has an envelope mass of $0.28\,M_{\odot}$ at termination. 

\subsubsection{Definition of the envelope in the calculation of binding energy}

In calculating the binding energy of the envelope, it has been standard practice in the literature to consider the \textit{entire} envelope, from the boundary of the donor's core to the photosphere \citep[e.g.][]{Marchant2021A&A, Scherbak2023MNRAS, Yamaguchi2024MNRAS, 2024arXiv240117510B}. However, in the formation of a wide PCEB, the companion never reaches the vicinity of the donor's core. It is thus not clear that the donor's orbital energy plays any significant role in ejecting material in the inner envelope. This means that rather than integrating energies from the core, it may be sufficient to integrate from the observed orbital separation to the surface. In practice, the star will likely expand when its outer envelope is stripped off, lifting more material from the inner envelope to the orbit of the secondary. However, in this case, most of the work to eject the inner envelope is done by the core, not the companion.

The binding energy of the outer envelope can be significantly less than that of the whole envelope. Moreover, the relative importance of the thermal and recombination energies varies inside the envelope, with recombination energy dominating in the outer layers and thermal energy dominating deeper in the interior. In the calculations below, we trace the binding energies of both (a) the entire envelope above the helium core, and (b) only the material beyond 0.7 AU, which is the typical final separation of the binaries in our sample.

\subsubsection{Results of CEE models} \label{sssec:results_cee_mesa}

\begin{figure*}
    \includegraphics[width=0.95\textwidth]{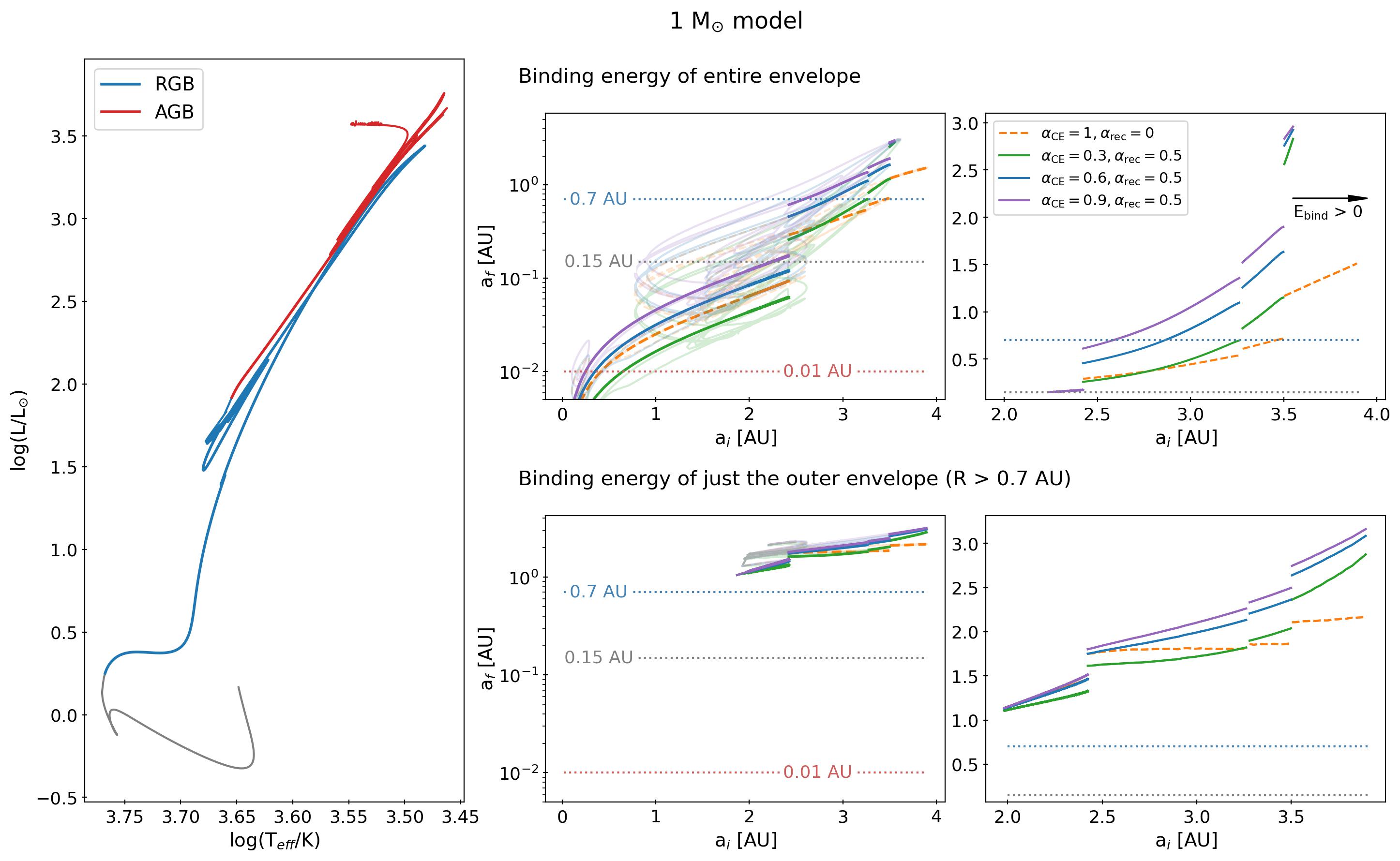}
    \caption{\textit{Left}: HR diagram of a MESA model of a $1.0\,M_{\odot}$ star. \textit{Middle}: Final separation, $a_f$, against the initial separation at the onset of mass transfer, $a_i$. The dashed line represents the case where all of the orbital energy loss goes into envelope ejection ($\alpha_{\rm CE} = 1$), but where no recombination energy is included ($\alpha_{\rm rec} = 0$). The three solid lines are cases for $\alpha_{\rm CE} = 0.3, 0.6$ and $0.9$, with the inclusion of 50\% of the available recombination energy ($\alpha_{\rm rec} = 0.5$). The horizontal dashed lines at 0.01, 0.15, and 0.7 AU represent the minimum separation below which the luminous primary would not fit in the orbit, the minimum separation of the ``wide PCEBs'', and the typical separation of the objects from this work. Black arrows indicate initial separations where the addition of recombination energy has resulted in the overall binding energy being positive, where the envelope is formally unbound. The upper and lower rows correspond to cases where the binding energy is calculated over the entire envelope vs. just above the final orbital separation of $0.7\,$AU, respectively. \textit{Right}: Zoom-in of the region above 0.15\,AU of the middle plot in the same row, isolating the regions with thicker line weight which can be reached by the donor in practice (i.e. where the radius is larger than at any previous point in the star's evolution). Wide orbits can form without the addition of any recombination energy if $\alpha_{\rm CE}$ is sufficiently large. This is true with both definitions of the envelope binding energy, though the range of initial separations over which they form significantly increased when just the outer envelope is considered.}
    \label{fig:mesa_af_1p0Msun}
\end{figure*}

\begin{figure*}
    \includegraphics[width=0.95\textwidth]{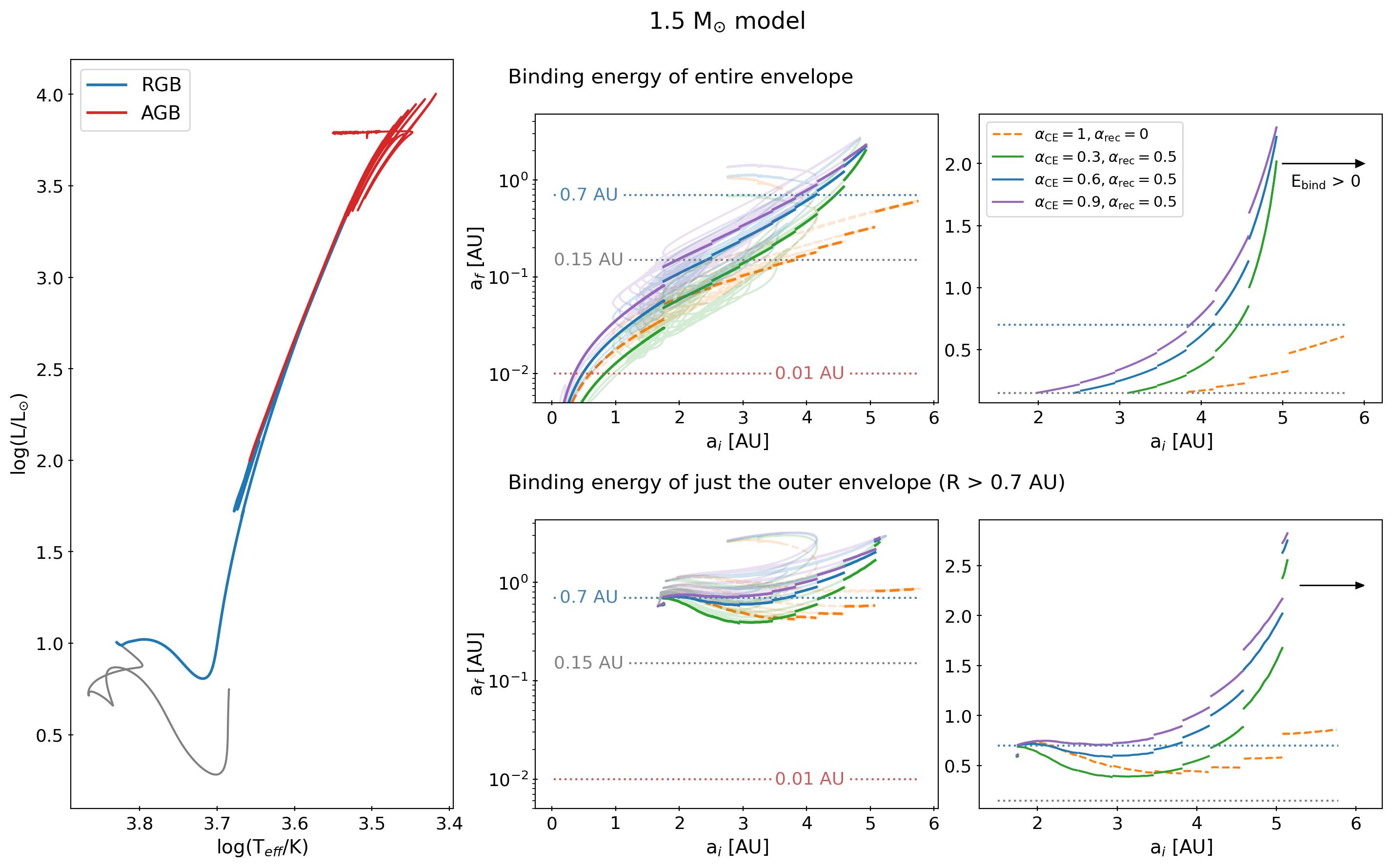}
    \caption{Same as Figure \ref{fig:mesa_af_1p0Msun}, but for a $1.5\,M_{\odot}$ model. For this model, we see that wide orbits like those of our systems ($a_f > 0.7\,$AU) do not form without recombination energy (orange dashed line) if the binding energy of the entire envelope is considered. Meanwhile, there is a small range of initial separations over which they can form if only the outer envelope is considered.}
    \label{fig:mesa_af_1p5Msun}
\end{figure*}

\begin{figure*}
    \includegraphics[width=0.95\textwidth]{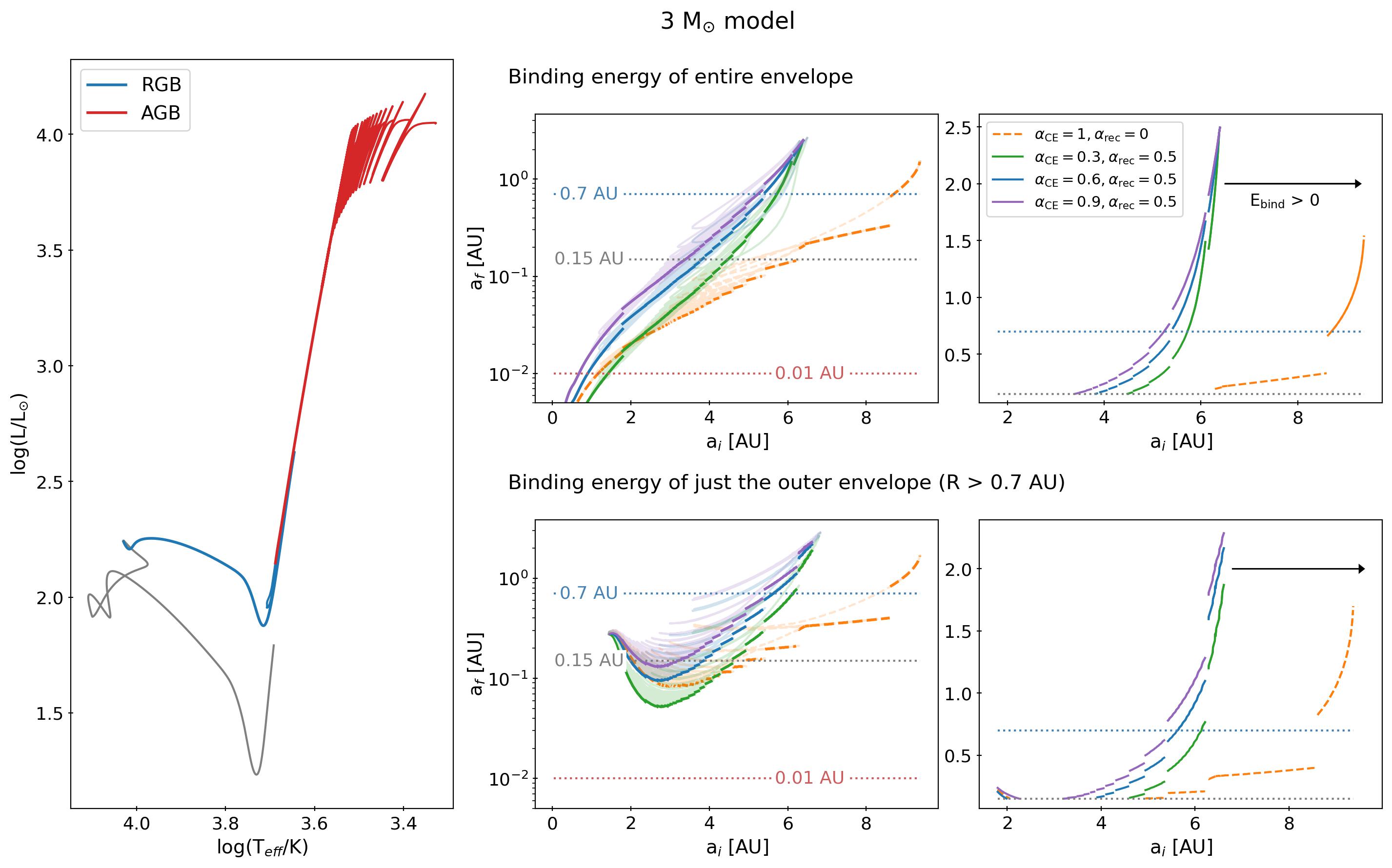}
    \caption{Same plots as Figure \ref{fig:mesa_af_1p0Msun} but for a $3\,M_{\odot}$ model.}
    \label{fig:mesa_af_3Msun}
\end{figure*}

Starting with the $1.0\,M_{\odot}$ model, the leftmost panel of Figure \ref{fig:mesa_af_1p0Msun} shows the evolutionary track of this star on a Hertzsprung-Russel (HR) diagram. On the middle columns, we plot $a_f$ against $a_i$ with and without the inclusion of recombination energy, and for various values of $\alpha_{\rm CE}$. We show the full tracks but in practice, only the sections with the thicker line weight is accessible, where $a_i$ (or $R$) is larger than it was at any earlier time. Otherwise, CEE would have been triggered prior to reaching that point in the donor's evolution. The red dashed lines at 0.01\,AU mark the approximate separation below which the MS companion would not fit in the orbit. The gray and blue dotted lines at 0.15 and 0.7\,AU respectively are the typical separations of previously discovered wide PCEBs including IK Peg and the systems from \citet{Yamaguchi2024MNRAS}, and the 31 objects from this work, respectively. The right columns zoom in on the parameter space that produces wide PCEBs. 

First, we examine the result of considering the binding energy of the entire envelope (upper row). We find that for a $1\,M_{\odot}$ donor, the orbits of wide PCEBs (both $\gtrsim 0.15\,$AU and $\gtrsim 0.7\,$AU) can form without the need to invoke any recombination energy ($\alpha_{\rm rec} = 0$; orange dashed line). This is true even for relatively modest values of $\alpha_{\rm CE} \sim 0.4$. As seen on the rightmost panel, even wider orbital separations can result with the inclusion of 50\% of the available recombination energy. 

Next, we move on to the same $1\,M_{\odot}$ model but considering the binding energy of the envelope above $0.7\,$AU (lower row). We see that the results change significantly. Small initial separations corresponding to the donor radii being less than $\sim 0.7\,$AU are, of course, excluded. We see that in this case, even larger separations of $\sim 2\,$ AU can be reached with $\alpha_{\rm rec} = 0$, for a wider range of initial separations. 

The situation changes slightly for the $1.5\,M_{\odot}$ case, shown in Figure \ref{fig:mesa_af_1p5Msun}. Once again considering the binding energy of the entire envelope first (upper row), we find that for $\alpha_{\rm rec} = 0$, the maximum separation predicted by our models is just below $\sim 0.7\,$AU for $\alpha_{\rm CE} = 1$. We note however that the exact results are dependent on model assumptions (e.g. using the  \citet{Vassiliadis1993ApJ} wind prescription results in a small range of $a_i$ that can produce $a_f > 0.7\,$AU). As expected, with $\alpha_{\rm rec} = 0.5$, wider orbits can form. For the case of the binding energy above $0.7\,$AU (lower row), there is a small range of initial separations resulting in $a_f \gtrsim 0.7\,$AU (and a much larger region for $a_f \gtrsim 0.15\,$AU), even with $\alpha_{\rm rec} = 0$. 

Lastly, for the $3\,M_{\odot}$ model (Figure \ref{fig:mesa_af_3Msun}), wide orbits ($a_f > 0.7\,$AU) can form without added recombination energy for $\alpha_{\rm CE} \gtrsim 0.4$, similar to the $1\,M_{\odot}$ model. This suggests that there is a non-monotonic relationship between the donor mass and post-common envelope orbital separation, which may be worth exploring with finer model grids. Unlike the $1\,M_{\odot}$ case, the consideration of just the outer envelope in the $3\,M_{\odot}$ model does not significantly change the resulting final orbits nor the range of initial separations over which wide orbits can form, which reflects the change in envelope structure between the two models.

In these calculations, we exclude regions where the total binding energy becomes positive due to the addition of recombination energy (indicated by the black arrows in the figures), where the $\alpha$-formalism is no longer appropriate. In these regions, the envelope is formally unbound and therefore, even wider orbits with little to no shrinkage may be produced. 

\begin{figure}
    \includegraphics[width=\columnwidth]{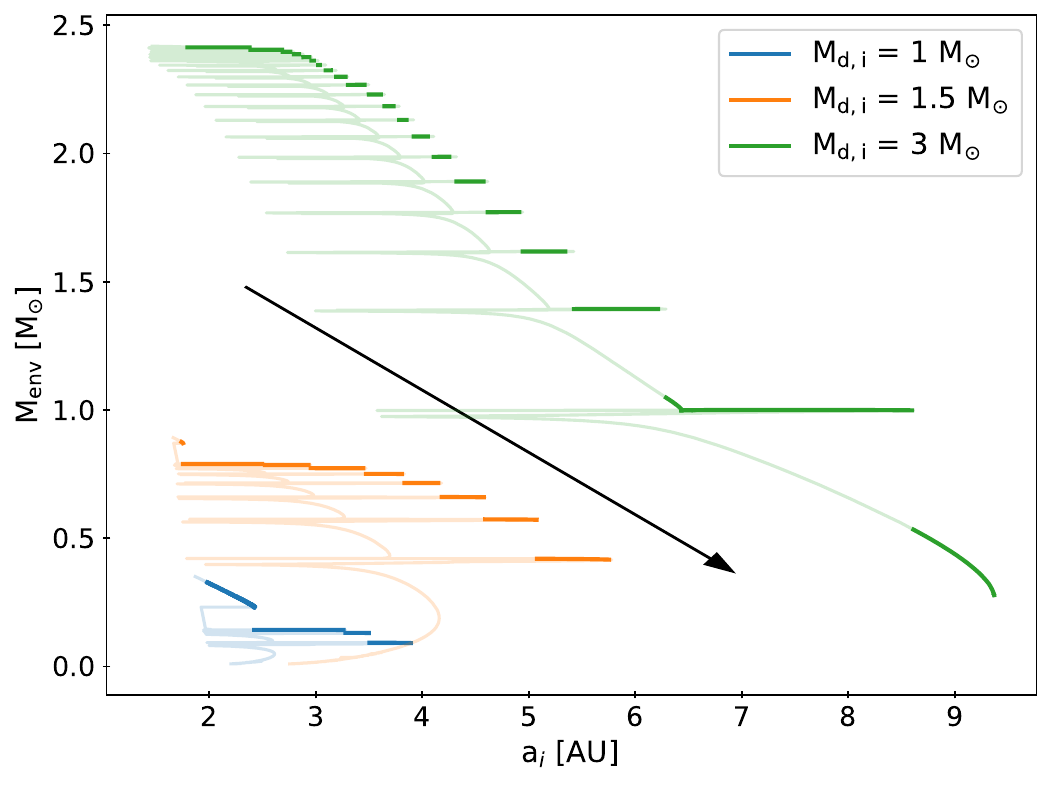}
    \caption{Envelope mass against separation at initial Roche lobe overflow for the 1, 1.5, and $3\,M_{\odot}$ models. As in Figure \ref{fig:mesa_af_1p0Msun}, regions where $a_i$ is larger than at any previous point have been plotted with a thicker line weight. The black arrow indicates the direction of evolution. We see that the range of initial separations needed to form the orbits of our systems in each of these cases (see Figures \ref{fig:mesa_af_1p0Msun}, \ref{fig:mesa_af_1p5Msun}, and \ref{fig:mesa_af_3Msun}) occur primarily during the thermal pulses.}
    \label{fig:Menv_v_ai}
\end{figure}

In Figure \ref{fig:Menv_v_ai}, we plot the envelope mass against initial orbital separation for the three models.  Comparing to Figures \ref{fig:mesa_af_1p0Msun}, \ref{fig:mesa_af_1p5Msun}, and \ref{fig:mesa_af_3Msun}, we see that wide orbits with $a_f \gtrsim 0.7\,$AU are formed predominantly when mass transfer begins during the thermal pulses. This corresponds to envelope masses below roughly $0.2$, $0.4$, and $0.8\,M_{\odot}$ for the $1, 1.5$, and $3\,M_{\odot}$ models respectively. In panel h of Figure \ref{fig:mesa_tpagb_1p5Msun}, we see that for the 1.5$\,M_{\odot}$ model, this point is reached during the last $10^5$ years. This highlights the importance of modeling the TP-AGB phase in its entirety, beyond the first thermal pulse and until a significant amount of the envelope is lost. 

The predicted core masses corresponding to wide final orbits range from $\sim 0.55 - 0.65\,M_{\odot}$ across the 1, 1.5, and 3$\,M_{\odot}$ models. These values are similar to the masses of most of the WDs in our sample.

\subsubsection{The importance of thermal pulses to forming wide PCEBs} \label{ssec:importance_of_TPs}

\citet{Belloni2024arXiv_b, 2024arXiv240117510B} ran a grid of MESA models to explain the formation of KOI-3278 \citep{Kruse2014Sci} and FN Sgr \citep{Brandi2005A&A} and concluded that such systems can form through CEE if the donor is on the TP-AGB without the need for any recombination energy. Despite some differences between their MESA models and ours (single star vs. binary models, various parameter choices), the final orbits are predicted using the $\alpha$-formalism and their results are in reasonably good agreement with ours.

\begin{figure*}
    \includegraphics[width=0.9\textwidth]{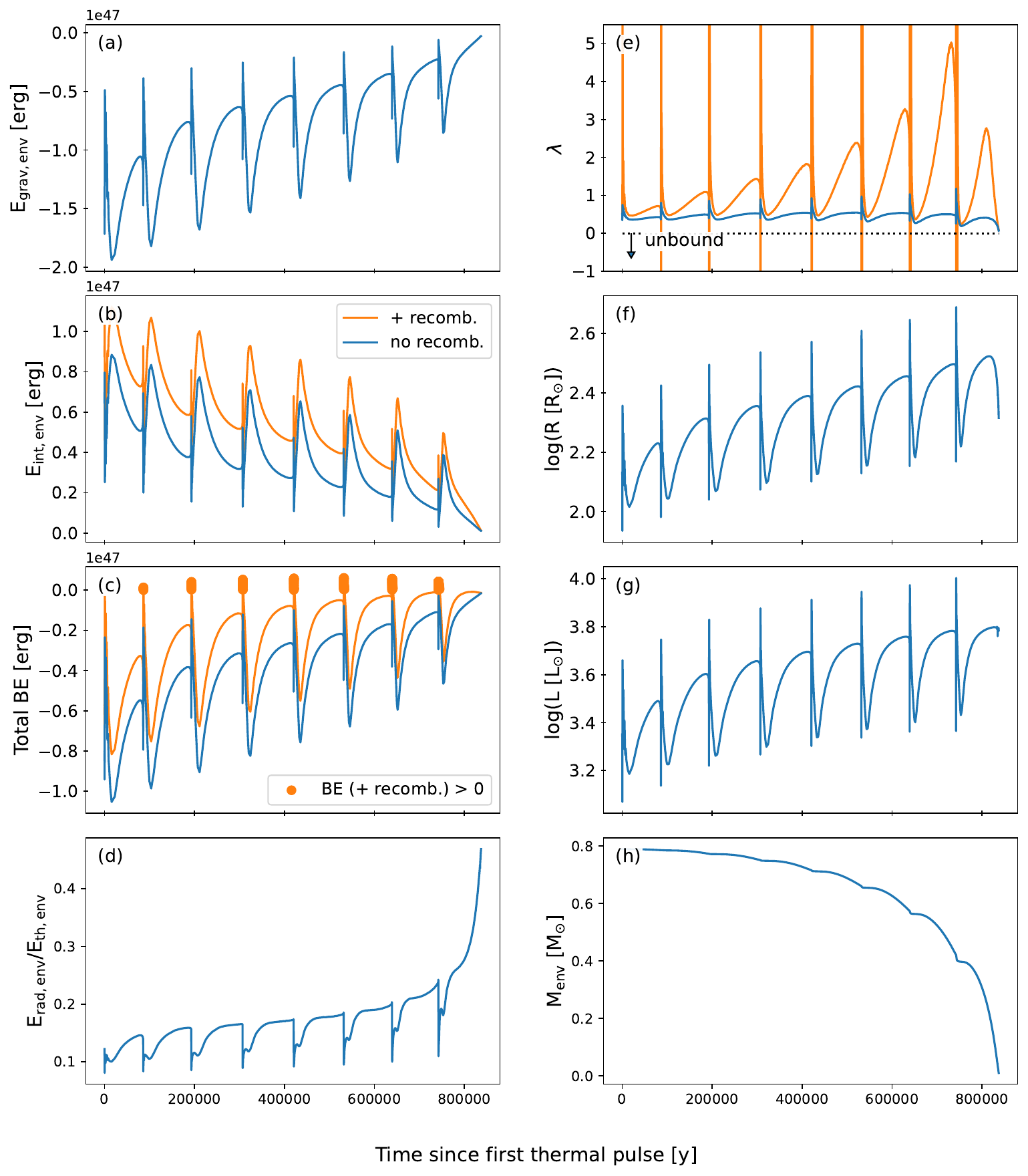}
    \caption{Stellar properties as a function of age for a $1.5\,M_{\odot}$ star on the TP-AGB. On the left column, we plot the gravitational binding energy (a), internal energy with and without recombination energy (b), total binding energy (c; $E_{\rm grav, env}$ + $E_{\rm int, env}$), and ratio of radiation to thermal energy (d). All energies are of the entire envelope. In panel (c), points where the total binding energy becomes positive (i.e. the envelope is formally unbound) are plotted with circle markers - this only occurs if the recombination energy is considered. On the right column, the structural binding energy parameter $\lambda$ (e; see text), radius (f), luminosity (g), and envelope mass (h).}
    \label{fig:mesa_tpagb_1p5Msun}
\end{figure*}

In Figure \ref{fig:mesa_tpagb_1p5Msun}, we plot several stellar properties against time for the $1.5\,M_{\odot}$ model on the TP-AGB (the energies plotted are of the entire envelope). From panels a - c, we see that as expected, the internal energy increases with the addition of recombination energy, and correspondingly, the total BE of the envelope becomes less negative. Unlike the models in \citet{Belloni2024arXiv_b}, we find that the total binding energy of our star without recombination energy approaches zero but always remains negative. Note also that the contribution of the radiation pressure to the total thermodynamic energy is sub-dominant to that of the gas pressure throughout the star's evolution until the final few thermal pulses (panel d). The predicted evolution is similar for the other models. 

In panel (e), we plot the structural
binding energy parameter, $\lambda$, defined as
\begin{equation}
    E_{\rm bind} = - \frac{G M_d M_{d, \rm env}}{\lambda R_d}
\end{equation}
where $M_d$ is the total mass of the donor, $M_{d, \rm env}$ is its envelope mass, and $R_d$ is its radius. \citet{deKool1990ApJ} first introduced $\lambda$ to account for differences in envelope structure in calculating the binding energy, $E_{\rm bind}$ (generalizing the approximation from  \citealt{Webbink1984ApJ}). Tabulated values of $\lambda$ have been widely used in the context of the $\alpha$-formalism to estimate $E_{\rm bind}$ without the use of detailed stellar models like MESA. Without including recombination energy and in between thermal pulses, we see that $\lambda$ takes typically assumed values of $\sim 0.5$. However, during pulses, the envelope is less tightly bound and it increases to $\sim 1$. With recombination energy included, $E_{\rm bind}$ can change signs, which causes $\lambda$ to blow up and then become negative as $E_{\rm bind}$ becomes positive during thermal pulses. Even between pulses, $\lambda$ can get as large as $\sim 5$. This illustrates the very large range of possible $\lambda$ values for TP-AGB stars, building on previous works on the dependence of $\lambda$ and a star's evolutionary phase \citep[e.g.][]{Dewi2000A&A, Tauris2001A&A, Xu2010ApJ}. 

On the AGB, CEE will almost always be initiated during a thermal pulse, since it is during thermal pulses that stars reach their largest radii (panel f) and will first encounter their companions (and have the least bound envelopes; panels c and e). Similar conclusions were made by \citet{Gonzalez-Bolivar2022MNRAS} who carried out 3D smoothed particle hydrodynamic simulations of CEE from an AGB star during a thermal pulse to a low mass companion. It is critical that future stellar model grids used in population synthesis codes capture the properties of AGB stars during the thermal pulses, not between them where they spend most of their time. 

We emphasize that while recombination is not {\it required} to explain wide PCEB orbits based on energy conservation arguments, it dominates the internal energy in the outer envelope of a TP-AGB donor (on average, $E_{\rm rec}/E_{\rm int} \sim 0.8$ integrated over $R > 0.7\,$AU). Therefore, recombination likely does play a role in the expansion and ejection of the envelope and its effect should not be neglected in making predictions about the orbits of PCEBs. Indeed, it has been shown to be important in more detailed models of the CE phase itself, both in 1D and 3D (e.g. \citealt{Sand2020A&A, Gonzalez-Bolivar2022MNRAS, Bronner2024A&A}). 

\subsubsection{Expected PCEB period distribution}

\begin{figure}
    \centering
    \includegraphics[width=\columnwidth]{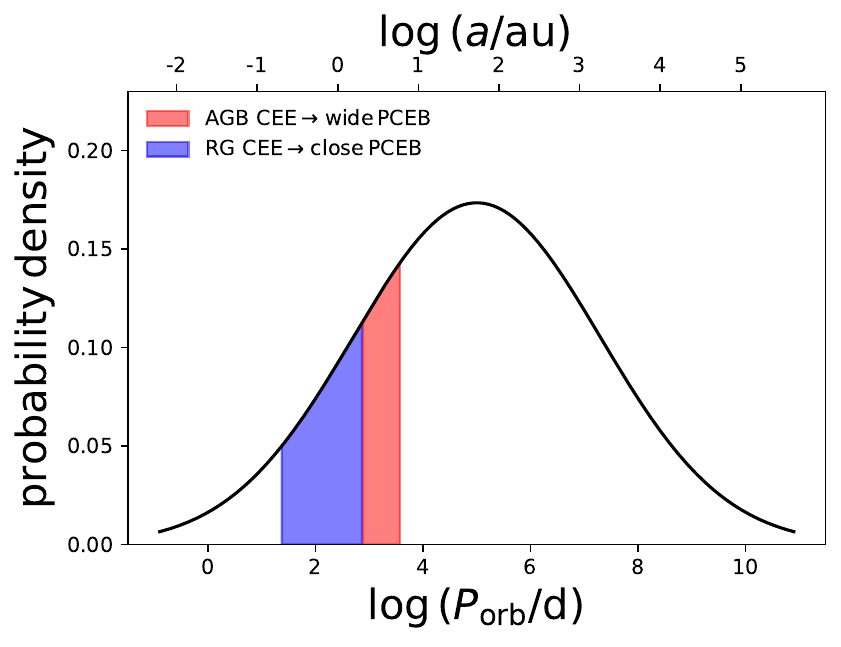}
    \caption{Period distribution of solar-type MS + MS binaries \citep[e.g.][]{Raghavan2010ApJS}. In blue is the region corresponding to systems where the first interaction will occur when one star becomes a red giant. If such binaries undergo CEE, these are expected to form close PCEBs. In red are systems where the first interaction will occur when one star is on the AGB, which is expected to form wide PCEBs. These systems each make up $\sim 10\%$ of the total population.}
    \label{fig:Porb_dist_MSMS}
\end{figure}

Figure \ref{fig:Porb_dist_MSMS} shows the distribution of the orbital period and separation of solar-type MS + MS binaries \citep{Raghavan2010ApJS}. Systems in the red region of the plot ($a_i\sim2 - 6\,$AU) will undergo the first common envelope event when the donor (i.e. the more massive MS star) becomes an AGB star. As we showed with our MESA models, these can form wide PCEBs ($a_f \gtrsim 0.15\,$AU). Those in tighter orbits ($a_i\sim0.2 - 2\,$AU), shaded in blue, will interact as the donor goes on the first giant branch and will likely form close PCEBs. Those in even smaller orbits will likely merge while those in orbits $\gtrsim 6\,$AU will likely never interact. The red and blue regions are comparable in area ($\sim 10\,$\%), suggesting that we expect similar number of PCEBs in close ($P_{\rm orb} \lesssim 1\,$d) and wide ($P_{\rm orb} \sim 10 - 1000$d) orbits. We emphasize that this is an over-simplification -- we see from Figure \ref{fig:P_v_M2} that there are several short-period PCEBs with massive WDs that likely formed from AGB progenitors. Thus, not all systems in the red region of Figure \ref{fig:Porb_dist_MSMS} will form wide orbits. One possibility is that CEE only leads to wide orbits if interaction begins during the most extended phase of a thermal pulse. Despite this caveat, our analysis suggests that a significant fraction of binaries that begin mass transfer when the donor is on the AGB produce wide PCEBs, and thus that wide and close PCEBs may form with comparable birth rates. 

\subsection{Formation through stable MT} \label{ssec:stable_mt}

Given the very wide orbits of our systems compared to previously identified PCEBs, we also consider the possibility of formation through stable mass transfer, as suggested by \citet{Hallakoun2023arXiv}. There are several different tests for this possibility which we describe briefly here. 

The first one, which has already been discussed, is whether or not systems fall on the theoretical $P_{\rm orb}-M_2$ relation along which systems undergoing stable Roche lobe overflow are expected to evolve. We plot this relation from \citet{Rappaport1995MNRAS} on Figure \ref{fig:P_v_M2}. We note that this specific relation was derived for giants with masses between 0.8 to 2$\,M_{\odot}$, which have degenerate cores on the RGB, and thus should be appropriate for the majority of our objects. The fact that none of our objects lie within the uncertainties of the relation suggests that their orbits are inconsistent with having formed from stable mass transfer.

A different commonly used criterion of mass transfer stability is whether mass transfer is expected to self-regulate or lead to a runaway. This boundary is often defined by a critical mass ratio, $q_c$. Using the initial mass - final mass relation for WDs, we can calculate that the initial mass ratios of our systems ($q = M_{\rm donor}/M_{\rm accretor}$) range from $\sim 1 - 7$, with a median value of $\sim 2-3$. Note that here we are taking the mass of the accretor to be equal to the mass of the luminous companion today, making these values of $q$ upper limits. Values of $q_c$ above which dynamically mass transfer occur vary between works, with typical values of $\sim 1$ to the more conservative values of $\sim 3 - 4$ \citep[e.g.][]{Hjellming1987ApJ, Chen2008MNRAS, Ge2010ApJ, 2015ApJ...812...40G, Temmink2023A&A}. Therefore, the single object with $q \sim 7$ lies firmly in the unstable regime. Meanwhile, the others have $q$ within the range of literature $q_c$ values. 

\begin{figure*}
    \includegraphics[width=0.95\textwidth]{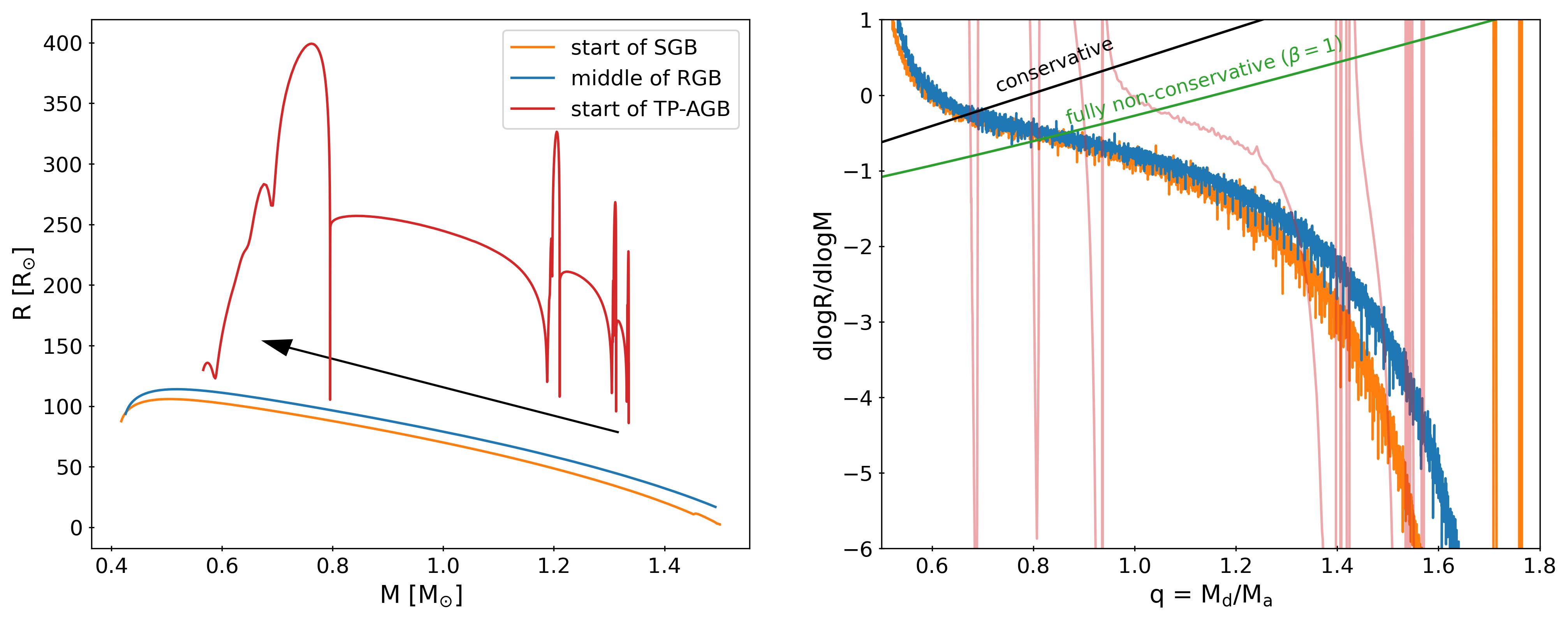}
    \caption{\emph{Left}: Change in the radius of an initially $1.5\,M_{\odot}$ donor as mass is stripped away from it at different points in its evolution. The black arrow indicates the direction of evolution. We see that up until the final stages of envelope stripping, its radius consistently increases as it loses mass (i.e. $\zeta_d < 0$). \emph{Right}: $\zeta_d$ against mass ratio, assuming $M_a = 0.85\,M_{\odot}$. We also plot curves above which there are conservative and fully non-conservative ($\beta = 1$) stable mass transfer \citep{Soberman1997A&A}. We see that our models lie firmly below these lines when $M_{\rm d} / M_{\rm a} > 1$, suggesting that mass transfer will be unstable at its onset. }
    \label{fig:mesa_enhanced_wind}
\end{figure*}

To contextualize these mass ratios, we perform another experiment which involves comparing the donor's response to mass loss, $\zeta_d = d\mathrm{log}R_d/d\mathrm{log}M_d$, to that of the Roche lobe radius, $\zeta_L = d\mathrm{log}R_L/d\mathrm{log}M_d$ \citep{Soberman1997A&A}. In the case where the donor expands faster than the Roche lobe, the mass transfer becomes increasingly more rapid, and thus unstable. To this end, we take the $1.5\,M_{\odot}$ model evolved regularly up to a certain point (Section \ref{ssec:cee}) and from there, remove its mass by increasing the wind strength (setting \texttt{Reimers\_scaling\_factor} = 5.0 and \texttt{Blocker\_scaling\_factor} = 0.5). This corresponds to mass loss rates ranging from $\sim 10^{-10}$ to $10^{-6} M_{\odot}\mbox{yr}^{-1}$. We start mass loss at the start of the sub-giant branch (SGB), roughly halfway up the RGB, and the start of the TP-AGB. The left panel of Figure \ref{fig:mesa_enhanced_wind} shows the change in the radius as a function of mass, from which we can get $\zeta_d$. We can then compare this to $\zeta_L$ which is just a function of the mass ratio $q$ and the parameter $\alpha$ (not to be confused with $\alpha_{\rm CE}$) which quantifies the mass lost as an isotropic wind from the accretor (along with several other parameters for other modes of MT; \citealt{Soberman1997A&A}). On the right panel, we plot $\zeta_d$ and $\zeta_L$ as a function of mass ratio where the mass of the accretor is once again taken to be $0.85\,M_{\odot}$ and assumed to be constant (i.e. fully non-conservative, $\beta = 1$). We see that for MT that begins on the SGB and RGB, $\zeta_d < 0$ and $\zeta_L > 0$  for most values of $q$ (i.e. the donor expands as the donor loses mass while the Roche lobe shrinks). Therefore, we expect mass transfer occurring at these mass ratios to become unstable and remain that way. The behaviour for MT beginning on the TP-AGB is complex due to the occurrence of thermal pulses, making this stability criterion not well suited for this case. We note also that the donor's response can vary depending on the mass loss rate \citep[e.g.][]{Passy2012ApJ, Ge2020ApJ} so the outcomes of this analysis do not necessarily encompass all possible scenarios.  

\begin{figure*}
    \includegraphics[width=0.95\textwidth]{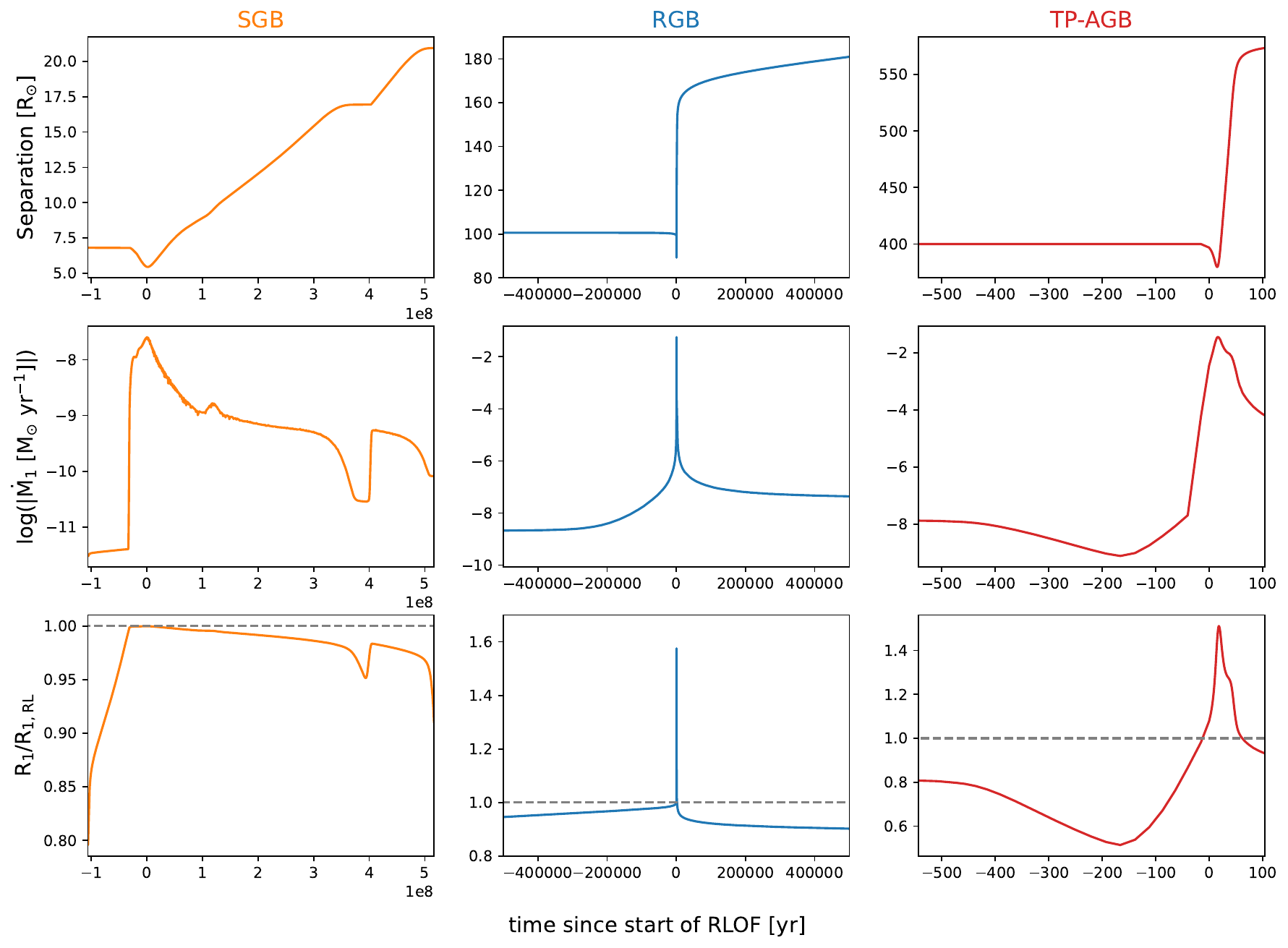}
    \caption{Orbital separation, mass loss rate, and Roche lobe filling factor as a function of time for binary MESA models where mass transfer from a $1.5\,M_{\odot}$ donor begins on the SGB, RGB and TP-AGB. The start of the Roche lobe overflow (RLOF) is defined to be the point where the filling factor, $R_1/R_L$, first reaches 1 (gray dashed line). We see that mass transfer from a donor on the SGB remains stable, while in the other two cases, instability seems to be triggered (mass loss rate reaches very high values, log$|\dot{M_1}| > -2$, and $R_1/R_L$ significantly exceeds 1).}
    \label{fig:mesa_binary}
\end{figure*}

Lastly, we run several binary models of a $1.5\,M_{\odot}$ donor and a $0.85\,M_{\odot}$ point-mass companion. We take an isolated model of the donor star, as described in Section \ref{ssec:cee}, load it into a binary model at a wide enough separation for there to be no initial interaction, and let it evolve until it overfills its Roche lobe and mass transfer begins. For the MT scheme, we use the optically thick overflow from \citet{Kolb1990A&A} and assume that it is fully non-conservative (\texttt{mass\_transfer\_beta} = 1; \citealt{rappaport_evolution_1982, Tauris2006csxs}). Models are terminated when the envelope is stripped off. In Figure \ref{fig:mesa_binary}, we plot the orbital separation, mass loss rate, and Roche lobe filling factor as a function of time since the start of mass transfer from a donor on the SGB, RGB, and TP-AGB (left, middle, and right columns, respectively). We see that for the SGB donor, the mass transfer remains stable -- there is an initial rise in the mass loss rate which peaks at $\sim 10^{-8}\,M_{\odot} \mbox{yr}^{-1}$ and then gradually declines, and the donor is contained within its Roche lobe throughout (i.e. filling factor $\lesssim 1$). Therefore, at these mass ratios, it is possible to have stable MT if the donor is on the SGB, though this would not produce WDs with masses that we observe. Meanwhile, for the RGB and TP-AGB donor, the donor's envelope is fully convective, and we see that the mass loss rates peak at much larger values of $\sim 10^{-2}\,M_{\odot} \mbox{yr}^{-1}$ and the filling factor exceeds 1. These point to the onset of instability. While the models do not terminate, the subsequent evolution is likely unreliable -- specifically, the accretor will likely be unable to accept mass at the rate it is being delivered and thermally adjust while remaining in equilibrium, so it will expand itself, leading to CEE. Moreover, once a star greatly fills its Roche lobe, mass loss could occur through the outer Langrangian points (L2/L3) which can carry away more angular momentum and promote unstable mass transfer \citep[e.g.][]{Shu1979ApJ, Reichardt2019MNRAS, Marchant2021A&A}. A caveat to this analysis is that we did not explore the different parameters to tune mass transfer efficiency ($\alpha$, $\beta$, $\delta$, $\gamma$; see \citealt{Soberman1997A&A, Tauris2006csxs}). For simplicity, we assumed that all mass is lost from the vicinity of the accreting star (i.e. $\beta = 1$) and thus we cannot exclude the possibility of stable non-conservative mass transfer in all cases. This may be a topic of interest to explore in future work.

\subsection{Takeaway from the MESA models}

In summary, these considerations suggest that the mass transfer in the progenitors of our WD+MS binaries likely would have become unstable, making them wide PCEBs (though note the caveat above about non-conservative mass transfer). This is in contrast with the predictions of most population synthesis models, which predict no binaries to exist with the periods and masses characteristic of the sample \citep[e.g.][]{Shahaf2024MNRAS}. However, detailed stellar models with realistic envelope structures like those of this work and \citet{2024arXiv240117510B} suggest that it is possible to produce these PCEBs as long as mass transfer begins when the donor is on the TP-AGB where its envelope is very loosely bound. What is still needed is a population model to test whether or not the sheer abundance of the observed systems in this region of the parameter space matches what is expected, given these stellar models and a realistic initial binary population. 

We also mention that there are related alternative theories that could produce wider orbit systems that we have not discussed in this work, with one being the grazing envelope evolution (GEE) where jets from the accretor prevents its complete engulfment in a common envelope (for details, see \citealt{Soker2015ApJ}). GEE has been proposed to explain post-AGB binaries in intermediate orbits $\sim 1\,$AU \citep[e.g.][]{Abu-Backer2018ApJ} comparable to those of the systems in this work.  

\section{Completeness of the sample} \label{sec:completeness}

%There are several considerations when thinking about the completeness of this sample. 

Looking at Figure \ref{fig:P_v_ecc}, there is a lack of systems with lower WD masses that lie in the stable MT region. This is likely at least partly a selection effect. The systems were selected from the \citet{Shahaf2024MNRAS} sample of probable MS + WD binaries based on their AMRF, $\mathcal{A}$, values. Given a primary mass, this requires that they contain a WD that is massive enough such that no single MS companion could produce as large an AMRF as observed. This boundary is shown in Figure 1 of \citet{Shahaf2024MNRAS} (dotted line) and as shown in Figure 11 of \citet{El-Badry2024arXiv}, results in the condition $M_{\rm WD}/M_{\star} > 0.6$.  A stable MT system with a WD mass of $\sim 0.4\,M_{\odot}$ would thus require $M_1 \lesssim 0.7\,M_{\odot}$, and very few of our 31 objects have primary masses this low. 

Thus, less massive WDs can only be detected in systems that host less massive primaries. However, as described in \citep{Shahaf2024MNRAS}, the sample is biased against the least massive MS companions because they are dimmer and  and therefore would not be the photocentric primary if they host a sufficiently hot WD. This is similarly discussed by \citet{Garbutt2024MNRAS} whose sample of WD + FGK binaries were selected on the basis of their UV excess, making them most sensitive to low-mass WD companions and complementary to this sample. Moreover, as described in \citet{Halbwachs2023A&A}, AstroSpectroSB1 solutions include RV measurements which are only made for bright objects ($G_{\rm RVS} < 12$). These factors combined means that at these periods ($\sim 100 - 1000\,$d), there is a bias against finding stable MT systems with the expected WD masses. It may be possible to identify such systems in future work on the basis of their near-zero eccentriciities. 

A detailed discussion of the selection function  will be left for future work. Here we simply emphasize that while WD+MS binaries formed by stable mass transfer are expected to be underrepresented in the sample, most of the objects in the sample are inconsistent with formation through stable mass transfer and are likely products of CEE.

\section{Conclusions} \label{sec:conclusion}

In this work, we presented follow-up spectroscopic observations of 31 MS + WD binaries with orbital periods ranging from $\sim 40 - 300$ d and \texttt{AstroSpectroSB1} solutions from {\it Gaia} DR3. These were selected from the broader sample of candidates selected by \citet{Shahaf2024MNRAS}. By jointly fitting the RVs from our spectra and astrometry, we validated the {\it Gaia} orbits and obtained tightened constraints on their stellar and orbital parameters. Our main findings are as follows:

\begin{enumerate} 
    \item \textit{Presence of a WD companion}: Through our joint fitting, we directly constrained the flux ratio $\mathcal{S} = F_2/F_1$ of our objects. Given these values, we conclude that 26 objects must host dark secondaries to within 1-$\sigma$ errors, in support of the majority indeed hosting WDs. For the remaining objects, while we cannot rule out the possibility of triples given the errors, they are also consistent with dark companions (Section \ref{ssec:flux_ratio}).
    \item \textit{Reliability of {\it Gaia} astrometric solutions}: Comparing the astrometry-only and astrometry + RV solutions, we find that only 2 out of the 31 systems have significant discrepancies in the orbital parameters (Section \ref{ssec:gaia_success}), implying a high success rate for \texttt{AstroSpectroSB1} solutions. 
    \item \textit{Comparison to literature PCEBs}: These objects have $P_{\rm orb} = 40 - 300\,$d. For a given WD mass, this is shorter than predicted for formation through stable MT but longer than the majority of previously known PCEBs (Figure \ref{fig:P_v_M2}), making their mass transfer histories uncertain. 
    \item \textit{Eccentricity}: We find eccentricities that range from $0.02$ to $0.45$, with a median at $\sim 0.07$. This is more circular than MS + MS binaries at comparable periods but more eccentric than most MSP + WD binaries and significantly more eccentric than predicted by the theoretical eccentricity - period relation for stable MT products (\citealt{Phinney1992RSPTA}; Figure \ref{fig:P_v_ecc}). 
    \item \textit{Formation through CEE}: We ran single star MESA models to obtain realistic models of the donor star and trace the binding energy of its envelope. We used the $\alpha$-formalism to calculate final orbital separations for mass transfer beginning at different points on the donor's evolution. We find that if mass transfer begins when the donor is on the TP-AGB, the wide orbits of our systems can form without the need to invoke recombination energy for sufficiently large values of $\alpha_{\rm CE}$. However, by including recombination energy, such orbits can form over a wider range of parameters. We also find that if we only consider the outer region of the envelope (above the final orbital separation of the PCEB) to be ejected, it is significantly easier to form the wide orbits (Section \ref{ssec:cee}).
    \item \textit{Formation through Stable MT}: We also ran several MESA models to test whether MT can remain stable given the mass ratios of the companion and WD progenitor. Both single star models of the donor's response to mass loss as well as binary models of mass transfer from an RGB and AGB donor suggest that mass transfer will become unstable (Section \ref{ssec:stable_mt}). 
\end{enumerate}

\section{Acknowledgements}

We thank Giulia Cinquegrana for advice on the MESA models and Jim Fuller for providing helpful discussions and feedback.  NY and KE acknowledge support from NSF grant AST-2307232.

This work has made use of data from the European Space Agency (ESA) mission
{\it Gaia} (\url{https://www.cosmos.esa.int/gaia}), processed by the {\it Gaia}
Data Processing and Analysis Consortium (DPAC,
\url{https://www.cosmos.esa.int/web/gaia/dpac/consortium}). Funding for the DPAC
has been provided by national institutions, in particular the institutions
participating in the {\it Gaia} Multilateral Agreement.

This job has made use of the Python package GaiaXPy, developed and maintained by members of the Gaia Data Processing and Analysis Consortium (DPAC), and in particular, Coordination Unit 5 (CU5), and the Data Processing Centre located at the Institute of Astronomy, Cambridge, UK (DPCI).

This work made use of the Third Data Release of the GALAH Survey \citep{Buder2021MNRAS}. The GALAH Survey is based on data acquired through the Australian Astronomical Observatory, under programs: A/2013B/13 (The GALAH pilot survey); A/2014A/25, A/2015A/19, A2017A/18 (The GALAH survey phase 1); A2018A/18 (Open clusters with HERMES); A2019A/1 (Hierarchical star formation in Ori OB1); A2019A/15 (The GALAH survey phase 2); A/2015B/19, A/2016A/22, A/2016B/10, A/2017B/16, A/2018B/15 (The HERMES-TESS program); and A/2015A/3, A/2015B/1, A/2015B/19, A/2016A/22, A/2016B/12, A/2017A/14 (The HERMES K2-follow-up program). We acknowledge the traditional owners of the land on which the AAT stands, the Gamilaraay people, and pay our respects to elders past and present. This paper includes data that has been provided by AAO Data Central (datacentral.org.au).

Funding for the Sloan Digital Sky Survey IV has been provided by the Alfred P. Sloan Foundation, the U.S. Department of Energy Office of Science, and the Participating Institutions. 

SDSS-IV acknowledges support and resources from the Center for High Performance Computing  at the University of Utah. The SDSS website is www.sdss4.org.

SDSS-IV is managed by the Astrophysical Research Consortium for the Participating Institutions of the SDSS Collaboration including the Brazilian Participation Group, the Carnegie Institution for Science, Carnegie Mellon University, Center for Astrophysics | Harvard \& Smithsonian, the Chilean Participation Group, the French Participation Group, Instituto de Astrof\'isica de Canarias, The Johns Hopkins University, Kavli Institute for the Physics and Mathematics of the Universe (IPMU) / University of Tokyo, the Korean Participation Group, Lawrence Berkeley National Laboratory, Leibniz Institut f\"ur Astrophysik Potsdam (AIP),  Max-Planck-Institut f\"ur Astronomie (MPIA Heidelberg), Max-Planck-Institut f\"ur Astrophysik (MPA Garching), Max-Planck-Institut f\"ur Extraterrestrische Physik (MPE), National Astronomical Observatories of China, New Mexico State University, New York University, University of Notre Dame, Observat\'ario Nacional / MCTI, The Ohio State University, Pennsylvania State University, Shanghai Astronomical Observatory, United Kingdom Participation Group, Universidad Nacional Aut\'onoma de M\'exico, University of Arizona, University of Colorado Boulder, University of Oxford, University of Portsmouth, University of Utah, University of Virginia, University of Washington, University of Wisconsin, Vanderbilt University, and Yale University. 

\vspace{5mm}
\facilities{Gaia, Max Planck:2.2m}

\software{emcee \citep{Foreman-Mackey2013PASP}}

\appendix

\section{All RVs} \label{sec:all_rvs}

Table \ref{tab:all_rvs} lists all of the follow-up RVs obtained for the 31 objects in our sample. These include RVs computed using the FEROS spectra (Section \ref{ssec:feros}), as well as those from the GALAH DR3 and APOGEE DR17 catalogues (Section \ref{ssec:galah_apogee}).  

\begin{deluxetable}{c c c c c}
\tablecaption{All RVs collected for the 31 objects of this work.}
\tablehead{
\colhead{Name} & \colhead{{\it Gaia} DR3 source ID} & 
\colhead{JD} & \colhead{RV [$\mbox{km s}^{-1}$]} & 
\colhead{Instrument/Catalog} 
} 
\startdata
J2006+1234 & 1803080360967961856 & 2460140.64924 & 0.555 $\pm$ 0.050 & FEROS \\
 J2006+1234 & 1803080360967961856 & 2460224.53102 & 8.655 $\pm$ 0.037 & FEROS \\
 J0145-0208 & 2505477938150534016 & 2460289.60328 & 44.185 $\pm$ 0.027 & FEROS \\
 J0145-0208 & 2505477938150534016 & 2460338.55422 & 27.495 $\pm$ 0.027 & FEROS \\
 J0145-0208 & 2505477938150534016 & 2460141.93724 & 28.215 $\pm$ 0.029 & FEROS \\
 J0145-0208 & 2505477938150534016 & 2460188.73796 & 47.275 $\pm$ 0.028 & FEROS \\
 J0145-0208 & 2505477938150534016 & 2460222.78809 & 59.965 $\pm$ 0.024 & FEROS \\
 J0547-2018 & 2966114550741693440 & 2460287.74003 & 42.185 $\pm$ 0.044 & FEROS \\
 J0547-2018 & 2966114550741693440 & 2460339.66151 & 34.340 $\pm$ 0.046 & FEROS \\
 J0547-2018 & 2966114550741693440 & 2460401.54809 & 50.175 $\pm$ 0.033 & FEROS \\
 J0547-2018 & 2966114550741693440 & 2460225.76536 & 53.735 $\pm$ 0.032 & FEROS \\
 J0602-1747 & 2990664622462566528 & 2460289.73088 & 25.995 $\pm$ 0.025 & FEROS \\
 J0602-1747 & 2990664622462566528 & 2460337.60611 & 17.150 $\pm$ 0.026 & FEROS \\
 J0602-1747 & 2990664622462566528 & 2460401.56500 & 3.845 $\pm$ 0.018 & FEROS \\
 J0602-1747 & 2990664622462566528 & 2460227.86074 & 21.730 $\pm$ 0.024 & FEROS \\
 J1134-2456 & 3533798283973293056 & 2457861.97772 & 14.180 $\pm$ 0.090 & GALAH \\
 J1134-2456 & 3533798283973293056 & 2460336.73239 & 22.975 $\pm$ 0.055 & FEROS \\
 J1134-2456 & 3533798283973293056 & 2460399.76851 & -6.860 $\pm$ 0.038 & FEROS \\
 J1134-2456 & 3533798283973293056 & 2460141.48883 & 12.590 $\pm$ 0.048 & FEROS \\
 J2007-0840 & 4192464229290454912 & 2460140.66734 & -80.580 $\pm$ 0.047 & FEROS \\
 J2007-0840 & 4192464229290454912 & 2460188.65999 & -80.030 $\pm$ 0.070 & FEROS \\
 J2007-0840 & 4192464229290454912 & 2460225.61333 & -88.320 $\pm$ 0.050 & FEROS \\
 J1951-0305 & 4233885203132760320 & 2460141.81006 & -22.970 $\pm$ 0.062 & FEROS \\
 J1951-0305 & 4233885203132760320 & 2460188.61077 & -49.065 $\pm$ 0.034 & FEROS \\
 J1951-0305 & 4233885203132760320 & 2460224.55740 & -52.570 $\pm$ 0.025 & FEROS \\
 J2000-0047 & 4237613470960339456 & 2460141.82301 & -6.495 $\pm$ 0.099 & FEROS \\
 J2000-0047 & 4237613470960339456 & 2460143.83048 & -7.460 $\pm$ 0.068 & FEROS \\
 J2000-0047 & 4237613470960339456 & 2460188.62359 & -35.790 $\pm$ 0.034 & FEROS \\
 J2000-0047 & 4237613470960339456 & 2460226.54698 & -26.585 $\pm$ 0.140 & FEROS \\
 J1553-1117 & 4346259930053416064 & 2460340.87386 & -71.455 $\pm$ 0.027 & FEROS \\
 J1553-1117 & 4346259930053416064 & 2460401.89800 & -41.810 $\pm$ 0.025 & FEROS \\
 J1553-1117 & 4346259930053416064 & 2460141.66734 & -65.380 $\pm$ 0.034 & FEROS \\
 J1553-1117 & 4346259930053416064 & 2460143.70199 & -66.295 $\pm$ 0.028 & FEROS \\
 J1553-1117 & 4346259930053416064 & 2460188.49410 & -64.965 $\pm$ 0.086 & FEROS \\
 J1628+1408 & 4463782295536949632 & 2458634.83817 & 22.697 $\pm$ 0.075 & APOGEE \\
 J1628+1408 & 4463782295536949632 & 2458633.84405 & 22.418 $\pm$ 0.051 & APOGEE \\
 J1628+1408 & 4463782295536949632 & 2460399.86900 & -0.895 $\pm$ 0.036 & FEROS \\
 J1628+1408 & 4463782295536949632 & 2460141.65340 & -0.045 $\pm$ 0.036 & FEROS \\
 J1628+1408 & 4463782295536949632 & 2460188.48152 & 18.350 $\pm$ 0.061 & FEROS \\
 J1834+1525 & 4509912512048265088 & 2460400.88189 & -136.970 $\pm$ 0.039 & FEROS \\
 J1834+1525 & 4509912512048265088 & 2460140.63174 & -151.345 $\pm$ 0.029 & FEROS \\
 J1834+1525 & 4509912512048265088 & 2460143.72760 & -150.980 $\pm$ 0.036 & FEROS \\
 J1834+1525 & 4509912512048265088 & 2460188.56747 & -140.640 $\pm$ 0.050 & FEROS \\
 J1834+1525 & 4509912512048265088 & 2460226.49416 & -135.105 $\pm$ 0.026 & FEROS \\
 J0127-6350 & 4712016939794270720 & 2458007.15118 & 31.593 $\pm$ 0.097 & GALAH \\
 J0127-6350 & 4712016939794270720 & 2460287.56513 & 25.715 $\pm$ 0.042 & FEROS \\
 J0127-6350 & 4712016939794270720 & 2460341.57040 & 40.820 $\pm$ 0.029 & FEROS \\
 J0127-6350 & 4712016939794270720 & 2460299.68352 & 26.780 $\pm$ 0.042 & FEROS \\
\enddata
\end{deluxetable} \label{tab:all_rvs}

\begin{deluxetable}{c c c c c} 
\tablecaption{Table \ref{tab:all_rvs} continued.}
\tablehead{
\colhead{Name} & \colhead{{\it Gaia} DR3 source ID} & 
\colhead{JD} & \colhead{RV [$\mbox{km s}^{-1}$]} & 
\colhead{Instrument/Catalog} 
} 
\startdata
 J0127-6350 & 4712016939794270720 & 2460140.85591 & 55.655 $\pm$ 0.036 & FEROS \\
 J0127-6350 & 4712016939794270720 & 2460187.81023 & 52.555 $\pm$ 0.035 & FEROS \\
 J0127-6350 & 4712016939794270720 & 2460225.69902 & 40.485 $\pm$ 0.036 & FEROS \\
 J0320-6129 & 4722462403337867520 & 2460289.61382 & 28.655 $\pm$ 0.032 & FEROS \\
 J0320-6129 & 4722462403337867520 & 2460340.62545 & 31.510 $\pm$ 0.026 & FEROS \\
 J0320-6129 & 4722462403337867520 & 2460185.78111 & 8.160 $\pm$ 0.026 & FEROS \\
 J0320-6129 & 4722462403337867520 & 2460222.87468 & 3.530 $\pm$ 0.030 & FEROS \\
 J0222-5032 & 4746089877426994048 & 2460285.77515 & 37.190 $\pm$ 0.034 & FEROS \\
 J0222-5032 & 4746089877426994048 & 2460341.58731 & 36.615 $\pm$ 0.047 & FEROS \\
 J0222-5032 & 4746089877426994048 & 2460141.87123 & 41.400 $\pm$ 0.034 & FEROS \\
 J0222-5032 & 4746089877426994048 & 2460185.73478 & 34.920 $\pm$ 0.040 & FEROS \\
 J0222-5032 & 4746089877426994048 & 2460225.71655 & 24.235 $\pm$ 0.032 & FEROS \\
 J0525-4059 & 4807112532250705280 & 2460289.71580 & 53.780 $\pm$ 0.040 & FEROS \\
 J0525-4059 & 4807112532250705280 & 2460340.69988 & 56.735 $\pm$ 0.042 & FEROS \\
 J0525-4059 & 4807112532250705280 & 2460400.50625 & 39.775 $\pm$ 0.041 & FEROS \\
 J0525-4059 & 4807112532250705280 & 2460185.80881 & 27.440 $\pm$ 0.041 & FEROS \\
 J0525-4059 & 4807112532250705280 & 2460224.87441 & 35.680 $\pm$ 0.037 & FEROS \\
 J0502-2803 & 4877335724279916032 & 2460289.70101 & 11.570 $\pm$ 0.027 & FEROS \\
 J0502-2803 & 4877335724279916032 & 2460335.59953 & -7.580 $\pm$ 0.037 & FEROS \\
 J0502-2803 & 4877335724279916032 & 2460227.71491 & -12.235 $\pm$ 0.031 & FEROS \\
 J0050-5739 & 4907589061595812224 & 2460285.63276 & 9.975 $\pm$ 0.029 & FEROS \\
 J0050-5739 & 4907589061595812224 & 2460341.55218 & 14.540 $\pm$ 0.021 & FEROS \\
 J0050-5739 & 4907589061595812224 & 2460140.80234 & 14.245 $\pm$ 0.027 & FEROS \\
 J0050-5739 & 4907589061595812224 & 2460187.76949 & 13.700 $\pm$ 0.023 & FEROS \\
 J0050-5739 & 4907589061595812224 & 2460224.78294 & 17.695 $\pm$ 0.032 & FEROS \\
 J0115-5039 & 4928000193592001920 & 2460285.66595 & 5.485 $\pm$ 0.037 & FEROS \\
 J0115-5039 & 4928000193592001920 & 2460299.66750 & 13.430 $\pm$ 0.019 & FEROS \\
 J0115-5039 & 4928000193592001920 & 2460141.85479 & 29.205 $\pm$ 0.025 & FEROS \\
 J0115-5039 & 4928000193592001920 & 2460185.71925 & 12.350 $\pm$ 0.026 & FEROS \\
 J0115-5039 & 4928000193592001920 & 2460223.67201 & -6.455 $\pm$ 0.049 & FEROS \\
 J0115-5039 & 4928000193592001920 & 2460228.58298 & -7.630 $\pm$ 0.028 & FEROS \\
 J0223-2354 & 5119937334246966656 & 2460287.61944 & 3.530 $\pm$ 0.032 & FEROS \\
 J0223-2354 & 5119937334246966656 & 2460339.56911 & -2.065 $\pm$ 0.038 & FEROS \\
 J0223-2354 & 5119937334246966656 & 2460142.85725 & -3.245 $\pm$ 0.037 & FEROS \\
 J0223-2354 & 5119937334246966656 & 2460187.91411 & 14.860 $\pm$ 0.041 & FEROS \\
 J0223-2354 & 5119937334246966656 & 2460225.73485 & 23.350 $\pm$ 0.036 & FEROS \\
 J0204-2602 & 5120822131869574784 & 2460289.58711 & -7.835 $\pm$ 0.063 & FEROS \\
 J0204-2602 & 5120822131869574784 & 2460340.59518 & 6.785 $\pm$ 0.083 & FEROS \\
 J0204-2602 & 5120822131869574784 & 2460140.92551 & 20.495 $\pm$ 0.059 & FEROS \\
 J0204-2602 & 5120822131869574784 & 2460187.82865 & 12.425 $\pm$ 0.054 & FEROS \\
 J0204-2602 & 5120822131869574784 & 2460226.71387 & 31.905 $\pm$ 0.042 & FEROS \\
 J0621-5218 & 5501078229653801600 & 2457708.17574 & 36.907 $\pm$ 0.083 & GALAH \\
 J0621-5218 & 5501078229653801600 & 2460289.74504 & 32.230 $\pm$ 0.040 & FEROS \\
 J0621-5218 & 5501078229653801600 & 2460335.63448 & 39.050 $\pm$ 0.038 & FEROS \\
 J0621-5218 & 5501078229653801600 & 2460400.52153 & 30.660 $\pm$ 0.041 & FEROS \\
 J0621-5218 & 5501078229653801600 & 2460227.79895 & 31.650 $\pm$ 0.034 & FEROS \\
 J1758-4658 & 5953913841603513984 & 2460398.84694 & 70.100 $\pm$ 0.045 & FEROS \\
 J1758-4658 & 5953913841603513984 & 2460141.68562 & 61.780 $\pm$ 0.054 & FEROS \\
\enddata
\end{deluxetable} \label{tab:all_rvs_2}

\begin{deluxetable}{c c c c c}
\tablecaption{Table \ref{tab:all_rvs} continued.}
\tablehead{
\colhead{Name} & \colhead{{\it Gaia} DR3 source ID} & 
\colhead{JD} & \colhead{RV [$\mbox{km s}^{-1}$]} & 
\colhead{Instrument/Catalogue} 
} 
\startdata
 J1758-4658 & 5953913841603513984 & 2460143.80881 & 62.565 $\pm$ 0.055 & FEROS \\
 J1758-4658 & 5953913841603513984 & 2460188.54970 & 72.575 $\pm$ 0.040 & FEROS \\
 J1758-4658 & 5953913841603513984 & 2460223.54563 & 68.785 $\pm$ 0.057 & FEROS \\
 J1533-4345 & 6001340936385389440 & 2460340.85780 & -43.880 $\pm$ 0.035 & FEROS \\
 J1533-4345 & 6001340936385389440 & 2460401.88275 & -64.910 $\pm$ 0.029 & FEROS \\
 J1533-4345 & 6001340936385389440 & 2460140.53297 & -47.390 $\pm$ 0.030 & FEROS \\
 J1533-4345 & 6001340936385389440 & 2460186.56294 & -41.660 $\pm$ 0.064 & FEROS \\
 J1639-2827 & 6032398738252562688 & 2457124.19103 & -24.522 $\pm$ 0.073 & GALAH \\
 J1639-2827 & 6032398738252562688 & 2460140.56490 & 9.980 $\pm$ 0.032 & FEROS \\
 J1639-2827 & 6032398738252562688 & 2460185.59438 & -22.530 $\pm$ 0.030 & FEROS \\
 J1639-2827 & 6032398738252562688 & 2460227.49316 & -22.410 $\pm$ 0.037 & FEROS \\
 J2234-6856 & 6385252016957373824 & 2457261.11764 & 22.147 $\pm$ 0.092 & GALAH \\
 J2234-6856 & 6385252016957373824 & 2460286.59417 & 5.590 $\pm$ 0.063 & FEROS \\
 J2234-6856 & 6385252016957373824 & 2460140.75390 & 13.255 $\pm$ 0.045 & FEROS \\
 J2234-6856 & 6385252016957373824 & 2460222.67653 & 23.755 $\pm$ 0.059 & FEROS \\
 J2209-6024 & 6409382723774888448 & 2460287.54659 & -34.310 $\pm$ 0.043 & FEROS \\
 J2209-6024 & 6409382723774888448 & 2460286.57739 & -34.315 $\pm$ 0.048 & FEROS \\
 J2209-6024 & 6409382723774888448 & 2460140.78698 & -32.630 $\pm$ 0.037 & FEROS \\
 J2209-6024 & 6409382723774888448 & 2460188.71534 & -8.800 $\pm$ 0.065 & FEROS \\
 J2209-6024 & 6409382723774888448 & 2460222.69333 & -7.195 $\pm$ 0.050 & FEROS \\
 J1956-6409 & 6428833084470488704 & 2460140.58692 & -20.990 $\pm$ 0.025 & FEROS \\
 J1956-6409 & 6428833084470488704 & 2460222.60976 & -17.425 $\pm$ 0.024 & FEROS \\
 J2103-4747 & 6479954396567118976 & 2460287.52910 & -58.590 $\pm$ 0.037 & FEROS \\
 J2103-4747 & 6479954396567118976 & 2460140.73845 & -42.685 $\pm$ 0.026 & FEROS \\
 J2103-4747 & 6479954396567118976 & 2460188.69959 & -55.060 $\pm$ 0.038 & FEROS \\
 J2103-4747 & 6479954396567118976 & 2460222.66053 & -58.580 $\pm$ 0.034 & FEROS \\
 J2103-4747 & 6479954396567118976 & 2460228.54755 & -59.000 $\pm$ 0.029 & FEROS \\
 J1844-6103 & 6632738920293491584 & 2460399.83944 & 110.450 $\pm$ 0.038 & FEROS \\
 J1844-6103 & 6632738920293491584 & 2460141.74950 & 114.900 $\pm$ 0.045 & FEROS \\
 J1844-6103 & 6632738920293491584 & 2460223.56749 & 88.345 $\pm$ 0.052 & FEROS \\
 J1848-5709 & 6637389068504460288 & 2460399.85234 & 47.785 $\pm$ 0.021 & FEROS \\
 J1848-5709 & 6637389068504460288 & 2460140.57660 & 47.705 $\pm$ 0.026 & FEROS \\
 J1848-5709 & 6637389068504460288 & 2460188.58292 & 49.945 $\pm$ 0.031 & FEROS \\
 J1848-5709 & 6637389068504460288 & 2460223.60745 & 57.275 $\pm$ 0.042 & FEROS \\
 J1922-4624 & 6663317786070516736 & 2457207.13280 & -57.091 $\pm$ 0.105 & GALAH \\
 J1922-4624 & 6663317786070516736 & 2460400.86427 & -81.760 $\pm$ 0.022 & FEROS \\
 J1922-4624 & 6663317786070516736 & 2460141.84048 & -76.845 $\pm$ 0.028 & FEROS \\
 J1922-4624 & 6663317786070516736 & 2460188.59658 & -62.295 $\pm$ 0.040 & FEROS \\
 J1922-4624 & 6663317786070516736 & 2460224.58343 & -57.115 $\pm$ 0.028 & FEROS \\
\enddata
\end{deluxetable} \label{tab:all_rvs_3}

\section{SED fitting} \label{sec:sed_fitting_all}

Figures \ref{fig:seds} and \ref{fig:seds_2} shows the observed photometry and best-fit SEDs for all 31 objects. Where available, GALEX NUV/FUV points are plotted, but they were not included in the SED fitting to avoid potential contamination from hot WD. Table \ref{tab:sed_params} summarizes the resulting best-fit parameters. 

\begin{figure}
    \centering
    \includegraphics[width=0.95\textwidth]{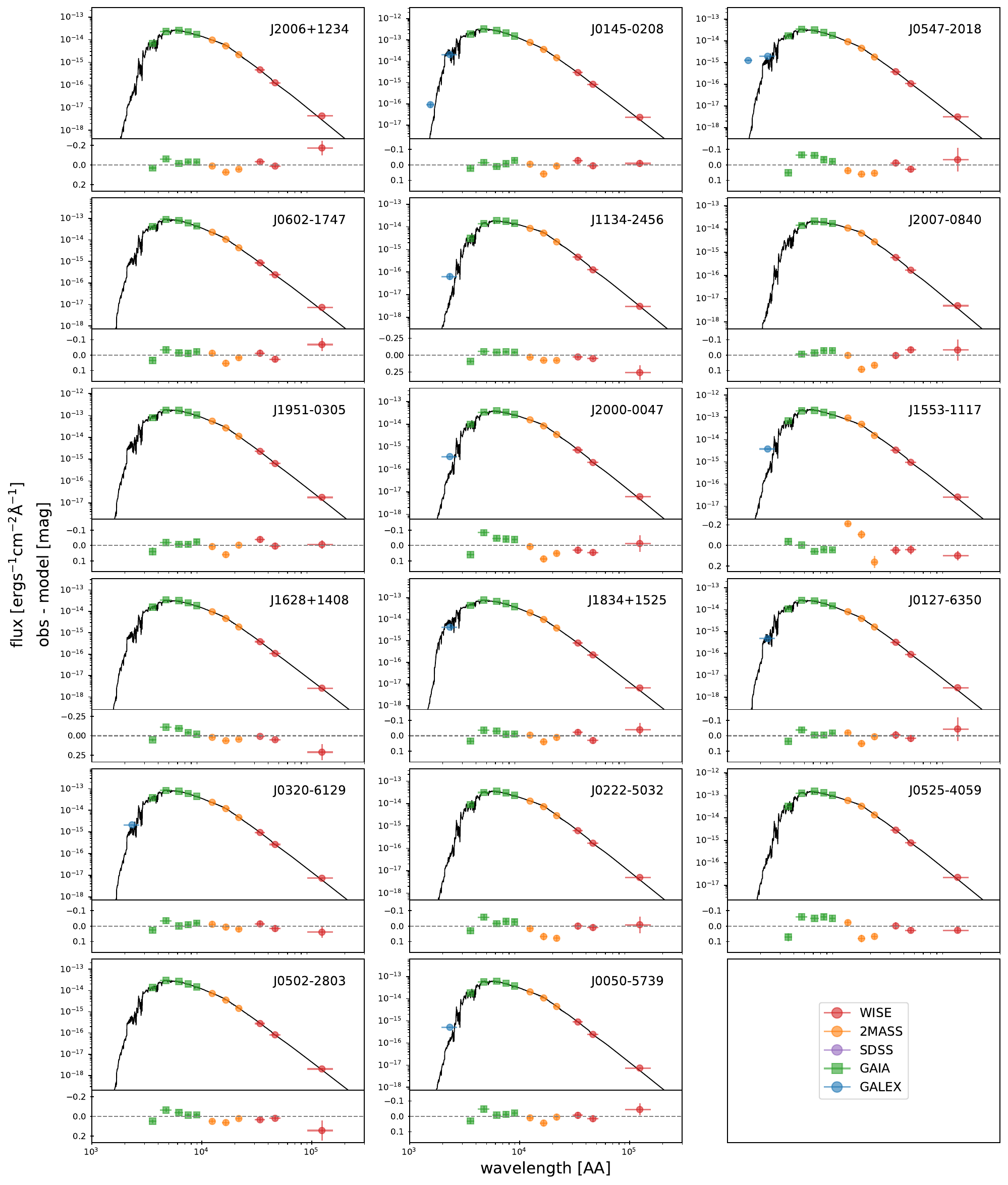}
    \caption{Observed photometry and best-fit SED curves for all 31 objects (continued to next figure). The residuals are plotted on the lower panels, which are typically $\lesssim 0.1$ mag.}
    \label{fig:seds}
\end{figure}

\begin{figure}
    \centering
    \includegraphics[width=0.95\textwidth]{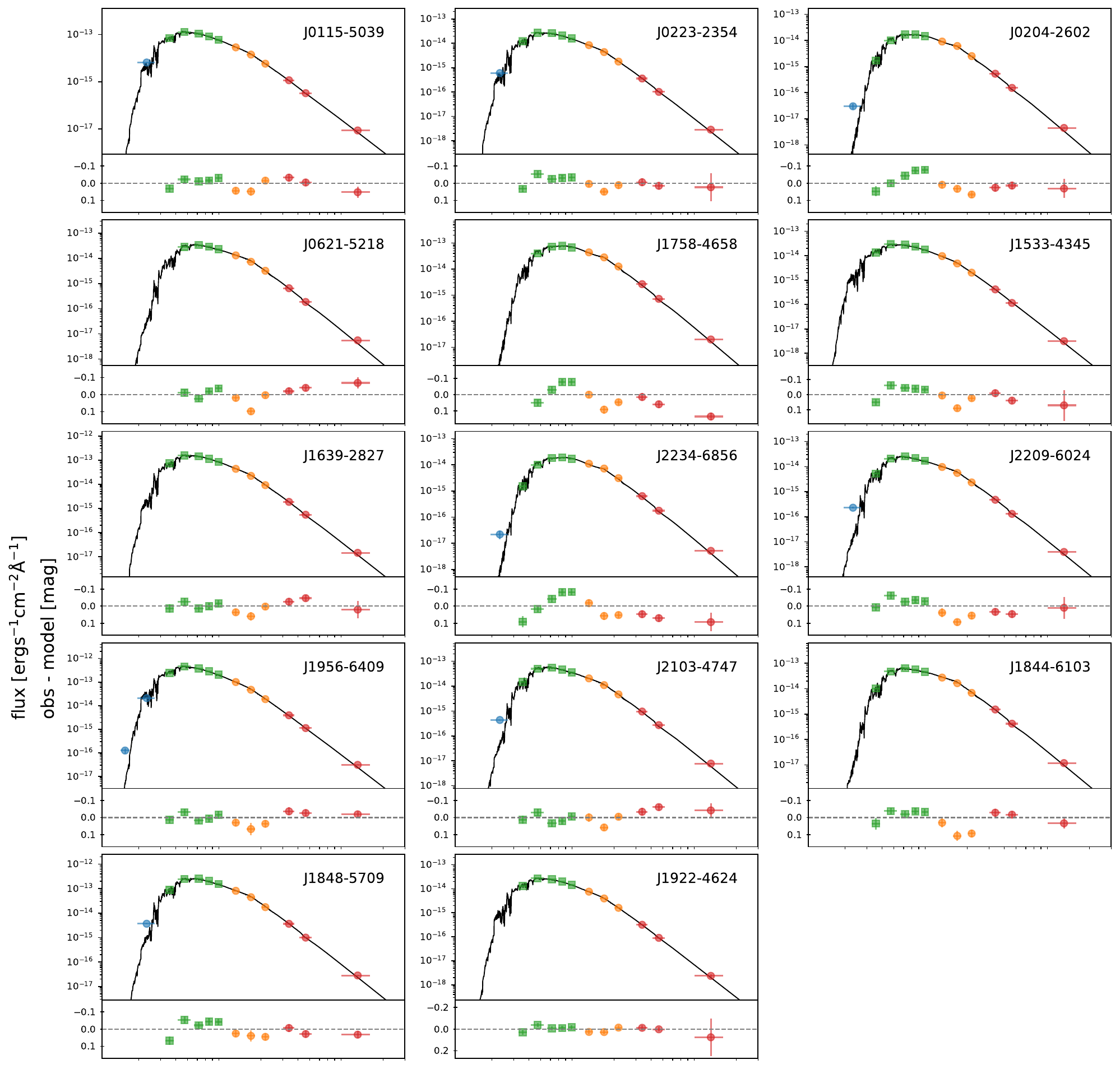}
    \caption{Continued from Figure \ref{fig:seds}.}
    \label{fig:seds_2}
\end{figure}

\begin{deluxetable}{c c c c c c c c c}
\tablecaption{Best-fit parameters from SED fitting of the MS star. From left to right, the columns are: object name, extinction (with typical uncertainties of $\sim 0.02$ mag), star mass, Equivalent Evolutionary Phase, parallax, initial metrallicity of the star, present-day metallicity, effective temperature, and radius.}
\tablehead{
\colhead{Name} & \colhead{E(B-V)} & 
\colhead{$M_1$ [$M_{\odot}$]} & \colhead{EEP} & 
\colhead{$\varpi$ [mas]} & \colhead{[Fe/H]$_{\rm init}$} & 
\colhead{[Fe/H]} & \colhead{T$_{\rm eff}$} & 
\colhead{$R_1$ [$R_{\odot}$]}
} 
\startdata
 J2006+1234 & 0.028 & 0.830 $\pm$ 0.004 & 207.41 $\pm$ 11.67 & 4.68 $\pm$ 0.02 & -0.05 $\pm$ 0.02 & -0.05 $\pm$ 0.02 & 4982 $\pm$ 7 & 0.774 $\pm$ 0.003 \\
 J0145-0208 & 0.000 & 0.939 $\pm$ 0.017 & 334.62 $\pm$ 12.19 & 9.48 $\pm$ 0.02 & -0.12 $\pm$ 0.03 & -0.16 $\pm$ 0.04 & 5763 $\pm$ 17 & 0.891 $\pm$ 0.005 \\
 J0547-2018 & 0.070 & 0.889 $\pm$ 0.005 & 207.57 $\pm$ 12.90 & 3.91 $\pm$ 0.01 & -0.30 $\pm$ 0.03 & -0.30 $\pm$ 0.03 & 5694 $\pm$ 12 & 0.785 $\pm$ 0.003 \\
 J1134-2456 & 0.070 & 0.705 $\pm$ 0.004 & 201.34 $\pm$ 2.96 & 5.60 $\pm$ 0.02 & -0.33 $\pm$ 0.02 & -0.33 $\pm$ 0.02 & 4661 $\pm$ 8 & 0.687 $\pm$ 0.003 \\
 J2007-0840 & 0.083 & 0.803 $\pm$ 0.004 & 204.69 $\pm$ 6.60 & 5.63 $\pm$ 0.02 & 0.12 $\pm$ 0.02 & 0.12 $\pm$ 0.02 & 4601 $\pm$ 10 & 0.773 $\pm$ 0.003 \\
 J1951-0305 & 0.080 & 0.992 $\pm$ 0.021 & 334.87 $\pm$ 18.00 & 8.02 $\pm$ 0.01 & 0.14 $\pm$ 0.03 & 0.11 $\pm$ 0.04 & 5634 $\pm$ 18 & 0.935 $\pm$ 0.006 \\
 J2000-0047 & 0.109 & 0.842 $\pm$ 0.004 & 205.24 $\pm$ 8.02 & 5.90 $\pm$ 0.01 & -0.08 $\pm$ 0.02 & -0.08 $\pm$ 0.02 & 5100 $\pm$ 8 & 0.776 $\pm$ 0.004 \\
 J1553-1117 & 0.000 & 0.792 $\pm$ 0.009 & 369.52 $\pm$ 7.00 & 11.74 $\pm$ 0.02 & -0.02 $\pm$ 0.02 & -0.11 $\pm$ 0.03 & 5007 $\pm$ 13 & 0.855 $\pm$ 0.005 \\
 J1834+1525 & 0.076 & 0.892 $\pm$ 0.009 & 238.11 $\pm$ 31.35 & 5.54 $\pm$ 0.02 & -0.44 $\pm$ 0.03 & -0.45 $\pm$ 0.04 & 5866 $\pm$ 15 & 0.794 $\pm$ 0.004 \\
 J0127-6350 & 0.016 & 0.961 $\pm$ 0.021 & 335.31 $\pm$ 25.11 & 3.19 $\pm$ 0.01 & 0.19 $\pm$ 0.03 & 0.15 $\pm$ 0.05 & 5429 $\pm$ 19 & 0.912 $\pm$ 0.008 \\
 J0320-6129 & 0.012 & 0.932 $\pm$ 0.009 & 255.88 $\pm$ 31.44 & 5.89 $\pm$ 0.01 & -0.02 $\pm$ 0.02 & -0.02 $\pm$ 0.03 & 5479 $\pm$ 12 & 0.839 $\pm$ 0.004 \\
 J0222-5032 & 0.010 & 0.819 $\pm$ 0.004 & 204.28 $\pm$ 6.40 & 5.52 $\pm$ 0.01 & -0.05 $\pm$ 0.02 & -0.05 $\pm$ 0.02 & 4925 $\pm$ 7 & 0.768 $\pm$ 0.003 \\
 J0525-4059 & 0.006 & 0.746 $\pm$ 0.003 & 203.34 $\pm$ 7.45 & 12.98 $\pm$ 0.01 & -0.25 $\pm$ 0.02 & -0.25 $\pm$ 0.02 & 4785 $\pm$ 9 & 0.715 $\pm$ 0.003 \\
 J0502-2803 & 0.060 & 1.087 $\pm$ 0.009 & 248.98 $\pm$ 22.97 & 2.74 $\pm$ 0.01 & 0.20 $\pm$ 0.03 & 0.20 $\pm$ 0.02 & 5830 $\pm$ 11 & 0.967 $\pm$ 0.005 \\
 J0050-5739 & 0.010 & 0.952 $\pm$ 0.007 & 248.34 $\pm$ 29.73 & 5.69 $\pm$ 0.02 & 0.27 $\pm$ 0.02 & 0.26 $\pm$ 0.02 & 5167 $\pm$ 8 & 0.868 $\pm$ 0.004 \\
 J0115-5039 & 0.009 & 0.955 $\pm$ 0.015 & 341.20 $\pm$ 7.91 & 5.74 $\pm$ 0.01 & -0.06 $\pm$ 0.03 & -0.10 $\pm$ 0.04 & 5776 $\pm$ 18 & 0.917 $\pm$ 0.005 \\
 J0223-2354 & 0.030 & 0.829 $\pm$ 0.006 & 221.04 $\pm$ 25.69 & 4.13 $\pm$ 0.02 & -0.30 $\pm$ 0.03 & -0.30 $\pm$ 0.03 & 5348 $\pm$ 11 & 0.753 $\pm$ 0.005 \\
 J0204-2602 & 0.000 & 0.648 $\pm$ 0.003 & 201.73 $\pm$ 5.42 & 6.69 $\pm$ 0.01 & -0.27 $\pm$ 0.02 & -0.27 $\pm$ 0.02 & 4226 $\pm$ 8 & 0.662 $\pm$ 0.003 \\
 J0621-5218 & 0.016 & 0.801 $\pm$ 0.008 & 232.09 $\pm$ 30.43 & 5.71 $\pm$ 0.01 & -0.10 $\pm$ 0.03 & -0.10 $\pm$ 0.04 & 4879 $\pm$ 14 & 0.762 $\pm$ 0.004 \\
 J1758-4658 & 0.010 & 0.639 $\pm$ 0.003 & 200.51 $\pm$ 1.23 & 15.15 $\pm$ 0.02 & -0.21 $\pm$ 0.02 & -0.21 $\pm$ 0.02 & 4151 $\pm$ 8 & 0.659 $\pm$ 0.003 \\
 J1533-4345 & 0.090 & 0.823 $\pm$ 0.005 & 209.19 $\pm$ 16.68 & 4.50 $\pm$ 0.02 & -0.41 $\pm$ 0.02 & -0.42 $\pm$ 0.02 & 5484 $\pm$ 12 & 0.736 $\pm$ 0.004 \\
 J1639-2827 & 0.024 & 0.874 $\pm$ 0.020 & 388.13 $\pm$ 8.98 & 7.32 $\pm$ 0.02 & -0.02 $\pm$ 0.04 & -0.11 $\pm$ 0.05 & 5482 $\pm$ 17 & 0.947 $\pm$ 0.007 \\
 J2234-6856 & 0.013 & 0.654 $\pm$ 0.004 & 201.25 $\pm$ 3.20 & 7.40 $\pm$ 0.02 & -0.16 $\pm$ 0.02 & -0.16 $\pm$ 0.02 & 4154 $\pm$ 6 & 0.670 $\pm$ 0.003 \\
 J2209-6024 & 0.013 & 0.800 $\pm$ 0.004 & 202.20 $\pm$ 2.73 & 5.06 $\pm$ 0.02 & -0.04 $\pm$ 0.01 & -0.04 $\pm$ 0.01 & 4797 $\pm$ 6 & 0.760 $\pm$ 0.003 \\
 J1956-6409 & 0.012 & 1.048 $\pm$ 0.019 & 390.72 $\pm$ 8.24 & 8.53 $\pm$ 0.02 & 0.18 $\pm$ 0.03 & 0.10 $\pm$ 0.04 & 5857 $\pm$ 17 & 1.128 $\pm$ 0.006 \\
 J2103-4747 & 0.016 & 0.864 $\pm$ 0.005 & 390.09 $\pm$ 2.45 & 5.40 $\pm$ 0.02 & 0.28 $\pm$ 0.02 & 0.19 $\pm$ 0.02 & 5031 $\pm$ 7 & 0.951 $\pm$ 0.003 \\
 J1844-6103 & 0.014 & 0.741 $\pm$ 0.004 & 203.64 $\pm$ 7.37 & 9.52 $\pm$ 0.02 & -0.14 $\pm$ 0.02 & -0.14 $\pm$ 0.02 & 4602 $\pm$ 8 & 0.722 $\pm$ 0.003 \\
 J1848-5709 & 0.010 & 0.849 $\pm$ 0.004 & 207.36 $\pm$ 11.84 & 12.93 $\pm$ 0.02 & -0.10 $\pm$ 0.02 & -0.11 $\pm$ 0.02 & 5174 $\pm$ 10 & 0.777 $\pm$ 0.004 \\
 J1922-4624 & 0.034 & 0.882 $\pm$ 0.018 & 310.79 $\pm$ 33.85 & 3.51 $\pm$ 0.03 & -0.18 $\pm$ 0.03 & -0.20 $\pm$ 0.04 & 5516 $\pm$ 15 & 0.816 $\pm$ 0.006 \\
 J0602-1747 & 0.019 & 1.022 $\pm$ 0.014 & 390.51 $\pm$ 7.03 & 4.26 $\pm$ 0.02 & 0.26 $\pm$ 0.03 & 0.18 $\pm$ 0.03 & 5707 $\pm$ 14 & 1.084 $\pm$ 0.007 \\
 J1628+1408 & 0.130 & 1.002 $\pm$ 0.005 & 210.57 $\pm$ 17.47 & 3.50 $\pm$ 0.01 & -0.08 $\pm$ 0.02 & -0.08 $\pm$ 0.02 & 5868 $\pm$ 11 & 0.884 $\pm$ 0.004 \\
\enddata
\end{deluxetable} \label{tab:sed_params}

\section{RV fitting} \label{sec:rv_fitting_all}

Figures \ref{fig:rv_astro_curves} and \ref{fig:rv_astro_curves_2} show RV curves for all 31 objects using parameters from 50 random draws of the posterior resulting from the fitting of just the {\it Gaia} solution. We also overplot our follow-up RVs -- we see that these are generally in agreement with the predicted curves, suggesting the {\it Gaia} solutions are mostly reliable, though there are a few exceptions. The RV residuals from the best-fit model are plotted on the lower panels. We see that for most objects, the residuals are within $\sim 2.5\,\mbox{km s}^{-1}$. Figures \ref{fig:rv_curves} and \ref{fig:rv_curves_2} show RV curves resulting from the joint fitting of {\it Gaia} solutions and our follow-up RVs. The spread in the RV curves are noticeably smaller in this case, illustrating the tightened constraints on the parameters. We also see that the RV residuals are smaller, typically being less than $\sim 0.2\,\mbox{km s}^{-1}$. On Table \ref{tab:rvs}, we summarize the best-fit parameters from the joint fitting for all objects.

\begin{figure}
    \centering
    \includegraphics[width=0.95\textwidth]{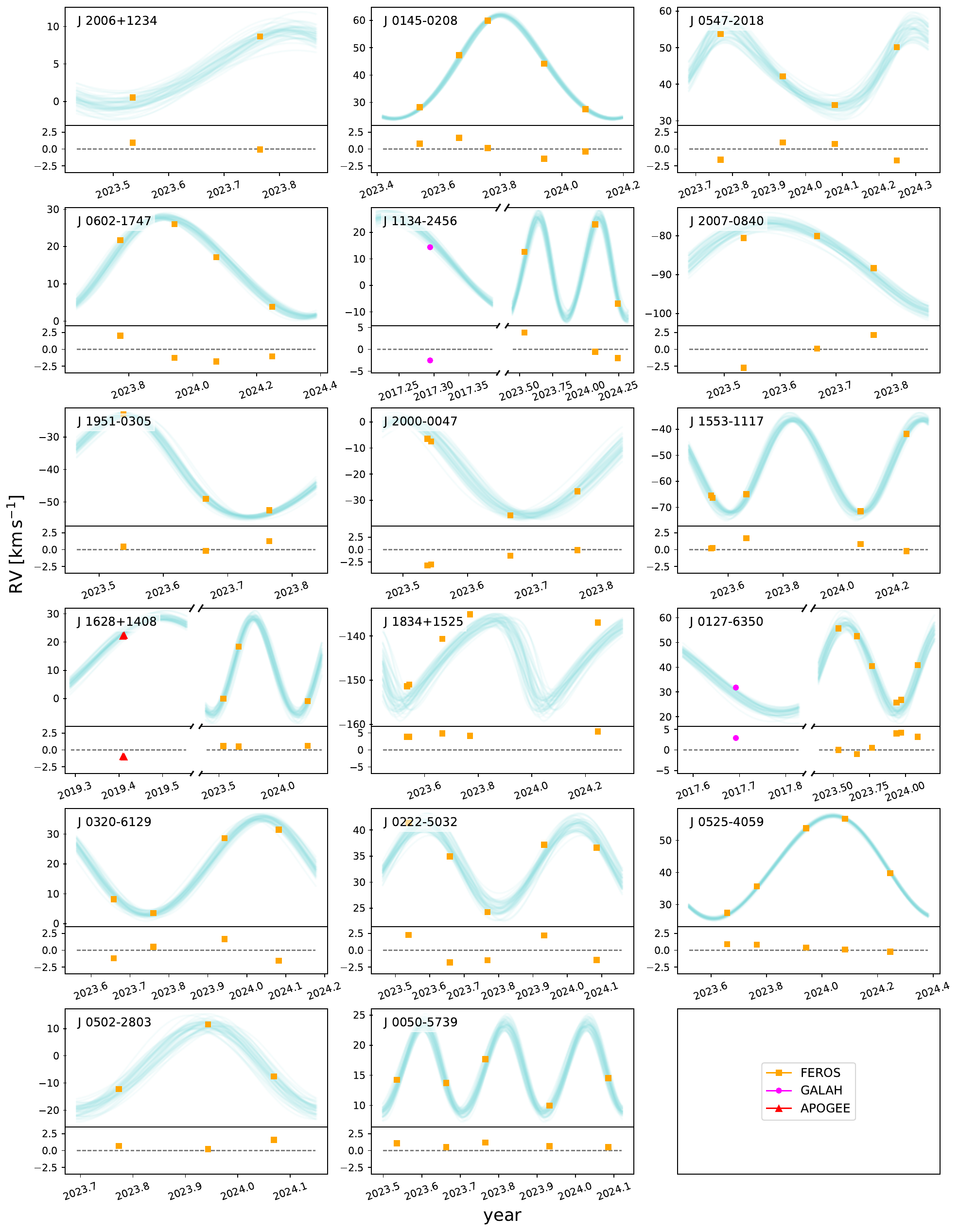}
    \caption{RV curves resulting from the fitting of the Gaia solutions only for all 31 objects (continued to next figure). We plot curves corresponding to parameters from 50 random draws of the posterior to illustrate the uncertainity. We overplot our follow-up RVs, and plot their residuals from the best-fit model on the lower panels of each plot.}
    \label{fig:rv_astro_curves}
\end{figure}

\begin{figure}
    \centering
    \includegraphics[width=0.95\textwidth]{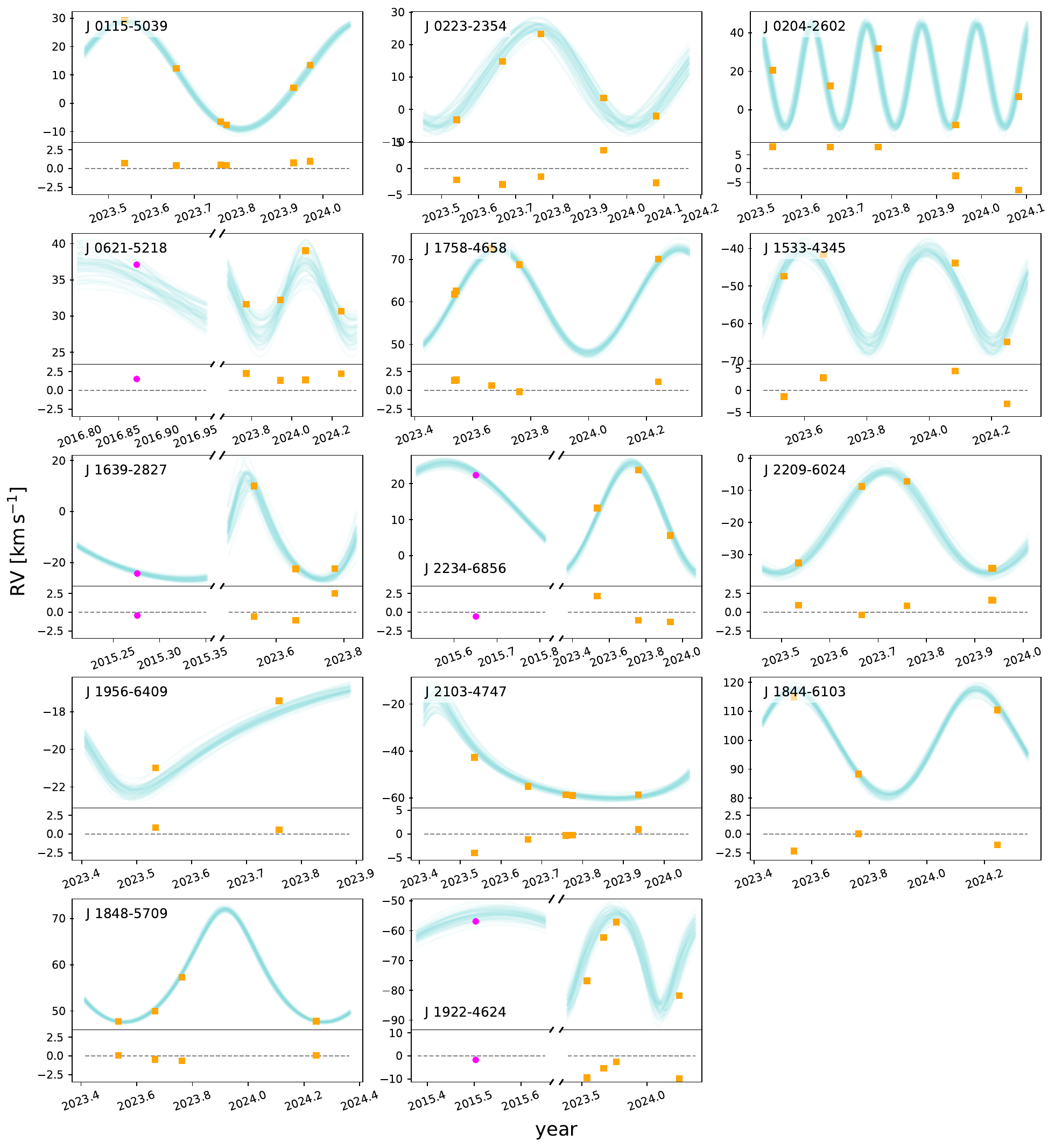}
    \caption{Continued from Figure \ref{fig:rv_astro_curves}. }
    \label{fig:rv_astro_curves_2}
\end{figure}

\begin{figure}
    \centering
    \includegraphics[width=0.95\textwidth]{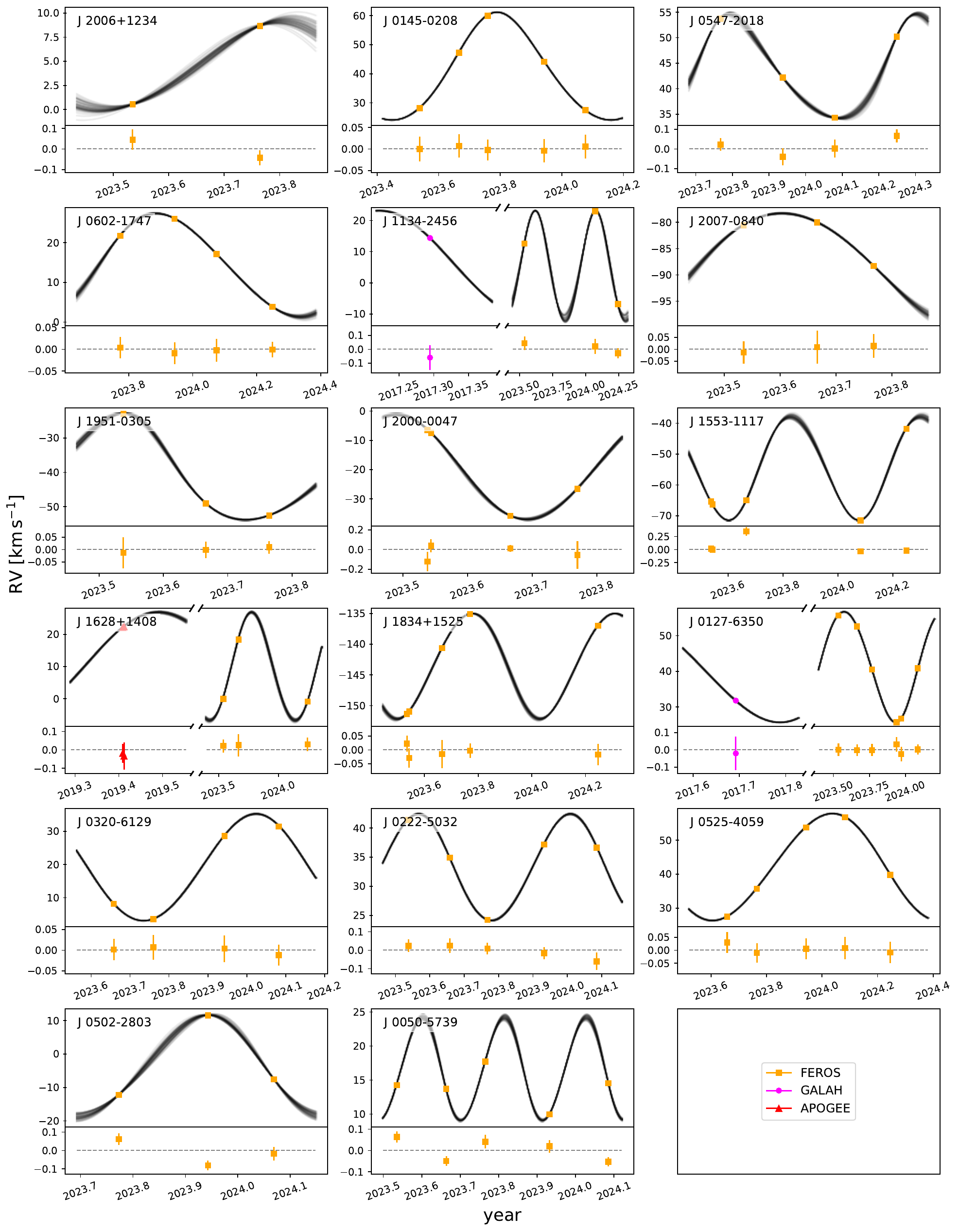}
    \caption{Same as Figure \ref{fig:rv_astro_curves}, but for the joint fitting of both the Gaia solutions and follow-up RVs. We see that there is much less spread in the predicted RV curves, corresponding to improved constraints on the orbits.}
    \label{fig:rv_curves}
\end{figure}

\begin{figure}
    \centering
    \includegraphics[width=0.95\textwidth]{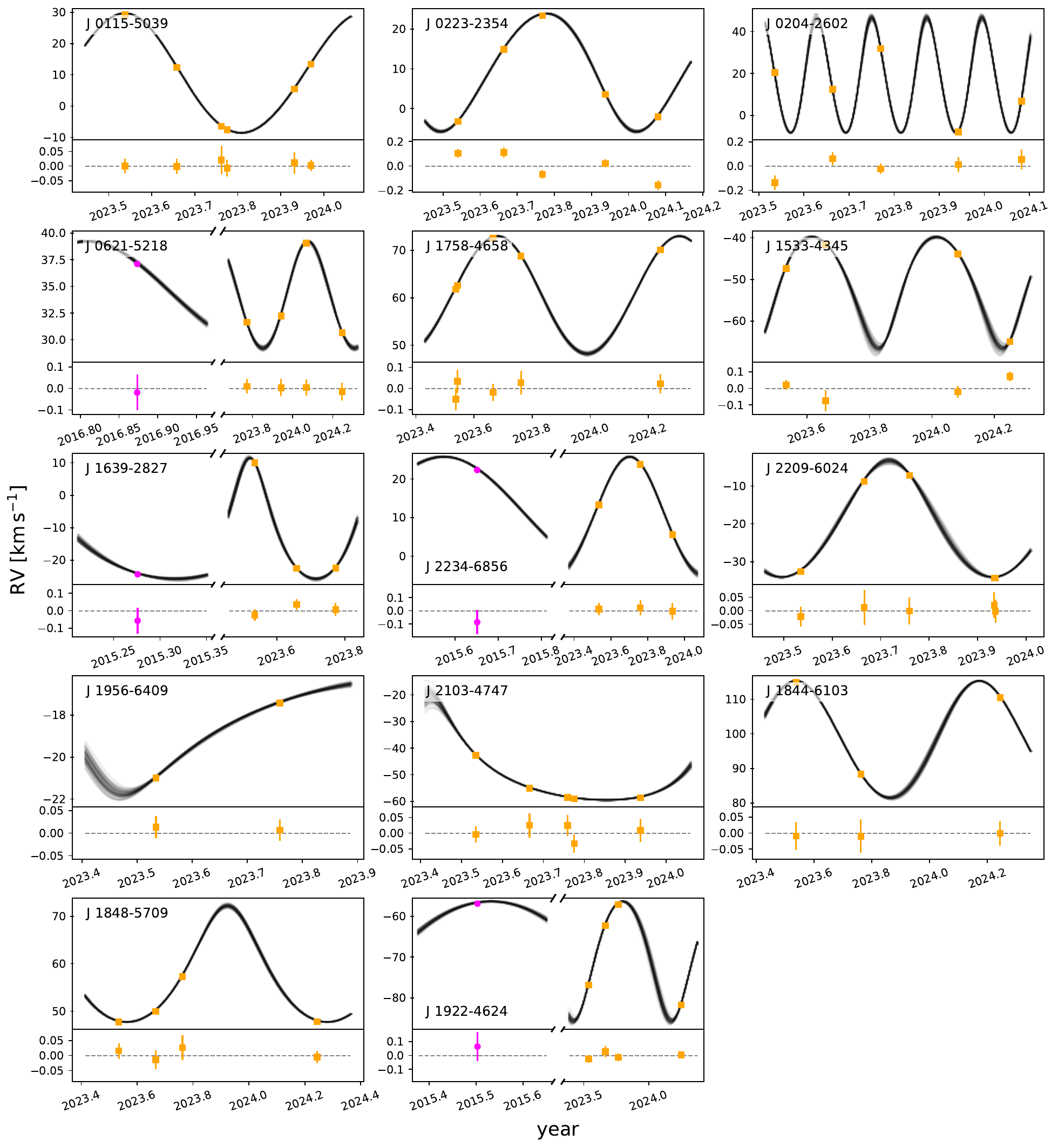}
    \caption{Continued from Figure \ref{fig:rv_curves}.}
    \label{fig:rv_curves_2}
\end{figure}

\begin{longrotatetable}
\begin{deluxetable}{c c c c c c c c c c c c c c c c}
\tablecaption{Best-fit parameters from the joint fitting of the astrometry and RVs. From left to right, the columns are: object name, period, eccentricity, inclination, position angle of the ascending node, argument of periapsis, periastron time, center-of-mass velocity, WD mass, MS star mass, flux ratio.}
\tabletypesize{\scriptsize}
\tablehead{
\colhead{Name} & \colhead{$P_{\rm orb}$ [d]} & 
\colhead{$e$} & \colhead{$i$ [deg]} & 
\colhead{$\Omega$ [deg]} & \colhead{$\omega$ [deg]} & 
\colhead{$t_p$ [JD - 2457389]} & \colhead{$\gamma$ [$\mbox{km s}^{-1}$]} & 
\colhead{$M_{\rm WD}$ [$M_{\odot}$]} & \colhead{$M_{\star}$ [$M_{\odot}$]} & \colhead{$\mathcal{S}$}
} 
\startdata
 J2006+1234 & 219.72 $\pm$ 0.22 & 0.080 $\pm$ 0.017 & 10.99 $\pm$ 1.89 & 72.10 $\pm$ 6.17 & 47.19 $\pm$ 7.93 & 2457407.7 $\pm$ 4.0 & 4.32 $\pm$ 0.25 & 1.03 $\pm$ 0.27 & 0.83 $\pm$ 0.03 & 0.134 $\pm$ 0.112 \\
 J0145-0208 & 259.20 $\pm$ 0.02 & 0.098 $\pm$ 0.001 & 97.28 $\pm$ 0.37 & 136.75 $\pm$ 0.33 & 340.30 $\pm$ 0.66 & 2457371.0 $\pm$ 0.5 & 40.86 $\pm$ 0.01 & 0.80 $\pm$ 0.01 & 0.94 $\pm$ 0.03 & -0.008 $\pm$ 0.002 \\
 J0547-2018 & 187.09 $\pm$ 0.08 & 0.194 $\pm$ 0.026 & 34.99 $\pm$ 1.30 & 321.04 $\pm$ 3.02 & 314.57 $\pm$ 6.04 & 2457411.5 $\pm$ 2.1 & 43.01 $\pm$ 0.17 & 0.61 $\pm$ 0.03 & 0.89 $\pm$ 0.03 & -0.002 $\pm$ 0.020 \\
 J1134-2456 & 166.88 $\pm$ 0.04 & 0.025 $\pm$ 0.016 & 100.79 $\pm$ 0.66 & 70.90 $\pm$ 0.75 & 38.06 $\pm$ 4.90 & 2457351.8 $\pm$ 2.3 & 5.33 $\pm$ 0.14 & 0.52 $\pm$ 0.02 & 0.70 $\pm$ 0.03 & -0.005 $\pm$ 0.013 \\
 J2007-0840 & 212.72 $\pm$ 0.07 & 0.086 $\pm$ 0.013 & 42.83 $\pm$ 1.27 & 145.04 $\pm$ 1.40 & 205.36 $\pm$ 5.85 & 2457311.6 $\pm$ 3.5 & -87.93 $\pm$ 0.21 & 0.51 $\pm$ 0.03 & 0.80 $\pm$ 0.03 & -0.015 $\pm$ 0.017 \\
 J1951-0305 & 157.96 $\pm$ 0.04 & 0.152 $\pm$ 0.013 & 64.45 $\pm$ 0.53 & 45.62 $\pm$ 0.91 & 39.25 $\pm$ 3.39 & 2457311.5 $\pm$ 1.4 & -40.22 $\pm$ 0.12 & 0.58 $\pm$ 0.01 & 0.99 $\pm$ 0.03 & -0.009 $\pm$ 0.005 \\
 J2000-0047 & 150.34 $\pm$ 0.02 & 0.024 $\pm$ 0.015 & 67.60 $\pm$ 0.55 & 90.20 $\pm$ 1.03 & 50.17 $\pm$ 8.49 & 2457438.3 $\pm$ 3.6 & -19.55 $\pm$ 0.11 & 0.62 $\pm$ 0.01 & 0.84 $\pm$ 0.03 & 0.003 $\pm$ 0.011 \\
 J1553-1117 & 172.93 $\pm$ 0.02 & 0.046 $\pm$ 0.012 & 71.82 $\pm$ 0.59 & 164.09 $\pm$ 0.55 & 241.54 $\pm$ 2.48 & 2457426.3 $\pm$ 1.2 & -54.44 $\pm$ 0.15 & 0.56 $\pm$ 0.02 & 0.79 $\pm$ 0.03 & -0.014 $\pm$ 0.005 \\
 J1834+1525 & 193.96 $\pm$ 0.08 & 0.054 $\pm$ 0.011 & 35.13 $\pm$ 2.93 & 116.60 $\pm$ 2.15 & 130.46 $\pm$ 6.26 & 2457388.1 $\pm$ 3.3 & -143.32 $\pm$ 0.09 & 0.50 $\pm$ 0.05 & 0.89 $\pm$ 0.03 & -0.033 $\pm$ 0.037 \\
 J0127-6350 & 249.27 $\pm$ 0.03 & 0.052 $\pm$ 0.001 & 85.39 $\pm$ 0.78 & 206.25 $\pm$ 0.84 & 221.02 $\pm$ 1.70 & 2457320.3 $\pm$ 1.2 & 41.74 $\pm$ 0.02 & 0.62 $\pm$ 0.01 & 0.96 $\pm$ 0.03 & -0.010 $\pm$ 0.005 \\
 J0320-6129 & 207.43 $\pm$ 0.03 & 0.027 $\pm$ 0.004 & 99.37 $\pm$ 0.52 & 115.75 $\pm$ 0.50 & 28.91 $\pm$ 10.03 & 2457431.1 $\pm$ 6.0 & 18.66 $\pm$ 0.03 & 0.60 $\pm$ 0.01 & 0.93 $\pm$ 0.03 & -0.004 $\pm$ 0.008 \\
 J0222-5032 & 160.71 $\pm$ 0.07 & 0.041 $\pm$ 0.006 & -153.14 $\pm$ 1.25 & 306.13 $\pm$ 2.07 & 223.78 $\pm$ 3.93 & 2457439.0 $\pm$ 2.3 & 32.92 $\pm$ 0.02 & 0.67 $\pm$ 0.04 & 0.82 $\pm$ 0.03 & -0.059 $\pm$ 0.033 \\
 J0525-4059 & 297.93 $\pm$ 0.02 & 0.046 $\pm$ 0.002 & 94.77 $\pm$ 0.14 & 76.57 $\pm$ 0.17 & 86.23 $\pm$ 1.97 & 2457412.4 $\pm$ 1.7 & 41.91 $\pm$ 0.02 & 0.60 $\pm$ 0.01 & 0.75 $\pm$ 0.03 & -0.005 $\pm$ 0.003 \\
 J0502-2803 & 174.73 $\pm$ 0.05 & 0.067 $\pm$ 0.028 & 92.90 $\pm$ 0.99 & 73.93 $\pm$ 1.01 & 47.09 $\pm$ 14.82 & 2457340.1 $\pm$ 7.5 & -4.30 $\pm$ 0.07 & 0.56 $\pm$ 0.02 & 1.09 $\pm$ 0.03 & -0.055 $\pm$ 0.017 \\
 J0050-5739 & 77.55 $\pm$ 0.02 & 0.058 $\pm$ 0.013 & 17.39 $\pm$ 2.37 & 45.63 $\pm$ 3.22 & 75.29 $\pm$ 5.92 & 2457388.8 $\pm$ 1.6 & 16.53 $\pm$ 0.09 & 0.71 $\pm$ 0.13 & 0.95 $\pm$ 0.03 & 0.041 $\pm$ 0.065 \\
 J0115-5039 & 201.38 $\pm$ 0.01 & 0.037 $\pm$ 0.001 & 95.56 $\pm$ 0.39 & 249.27 $\pm$ 0.42 & 27.12 $\pm$ 2.57 & 2457336.8 $\pm$ 1.5 & 9.59 $\pm$ 0.01 & 0.75 $\pm$ 0.01 & 0.96 $\pm$ 0.03 & 0.002 $\pm$ 0.004 \\
 J0223-2354 & 192.19 $\pm$ 0.08 & 0.093 $\pm$ 0.008 & -127.94 $\pm$ 0.83 & 319.32 $\pm$ 1.12 & 319.69 $\pm$ 1.34 & 2457417.5 $\pm$ 1.6 & 9.82 $\pm$ 0.02 & 0.66 $\pm$ 0.02 & 0.83 $\pm$ 0.03 & -0.075 $\pm$ 0.013 \\
 J0204-2602 & 44.51 $\pm$ 0.00 & 0.078 $\pm$ 0.009 & 118.45 $\pm$ 1.74 & 309.59 $\pm$ 0.94 & 323.43 $\pm$ 3.04 & 2457366.9 $\pm$ 0.4 & 16.92 $\pm$ 0.24 & 0.59 $\pm$ 0.02 & 0.65 $\pm$ 0.03 & -0.004 $\pm$ 0.013 \\
 J0621-5218 & 165.87 $\pm$ 0.07 & 0.037 $\pm$ 0.013 & 17.39 $\pm$ 2.00 & 45.36 $\pm$ 2.28 & 338.47 $\pm$ 11.38 & 2457344.0 $\pm$ 5.0 & 33.95 $\pm$ 0.02 & 0.51 $\pm$ 0.08 & 0.80 $\pm$ 0.03 & -0.015 $\pm$ 0.047 \\
 J1758-4658 & 228.45 $\pm$ 0.04 & 0.018 $\pm$ 0.006 & 132.13 $\pm$ 0.34 & 313.09 $\pm$ 0.47 & 44.98 $\pm$ 9.19 & 2457480.6 $\pm$ 5.9 & 60.22 $\pm$ 0.09 & 0.53 $\pm$ 0.02 & 0.64 $\pm$ 0.03 & -0.041 $\pm$ 0.010 \\
 J1533-4345 & 144.89 $\pm$ 0.05 & 0.178 $\pm$ 0.007 & 133.42 $\pm$ 1.18 & 302.07 $\pm$ 1.67 & 196.11 $\pm$ 6.50 & 2457355.1 $\pm$ 2.9 & -51.22 $\pm$ 0.19 & 0.55 $\pm$ 0.02 & 0.82 $\pm$ 0.03 & -0.008 $\pm$ 0.017 \\
 J1639-2827 & 139.44 $\pm$ 0.02 & 0.245 $\pm$ 0.015 & 117.56 $\pm$ 0.95 & 124.92 $\pm$ 0.76 & 358.17 $\pm$ 1.68 & 2457346.9 $\pm$ 0.8 & -11.91 $\pm$ 0.17 & 0.65 $\pm$ 0.02 & 0.87 $\pm$ 0.03 & 0.025 $\pm$ 0.013 \\
 J2234-6856 & 296.89 $\pm$ 0.05 & 0.033 $\pm$ 0.009 & 108.03 $\pm$ 0.23 & 345.24 $\pm$ 0.25 & 9.63 $\pm$ 4.05 & 2457240.8 $\pm$ 3.7 & 9.49 $\pm$ 0.14 & 0.58 $\pm$ 0.02 & 0.65 $\pm$ 0.03 & -0.022 $\pm$ 0.010 \\
 J2209-6024 & 158.68 $\pm$ 0.03 & 0.073 $\pm$ 0.024 & 126.91 $\pm$ 0.85 & 334.96 $\pm$ 0.96 & 11.46 $\pm$ 3.30 & 2457355.7 $\pm$ 1.6 & -20.11 $\pm$ 0.18 & 0.61 $\pm$ 0.02 & 0.80 $\pm$ 0.03 & 0.010 $\pm$ 0.010 \\
 J1956-6409 & 280.24 $\pm$ 0.15 & 0.351 $\pm$ 0.008 & 10.98 $\pm$ 1.99 & 15.12 $\pm$ 2.97 & 144.69 $\pm$ 3.23 & 2457304.7 $\pm$ 0.7 & -18.33 $\pm$ 0.08 & 0.56 $\pm$ 0.15 & 1.05 $\pm$ 0.03 & -0.049 $\pm$ 0.064 \\
 J2103-4747 & 268.72 $\pm$ 0.09 & 0.450 $\pm$ 0.015 & 102.87 $\pm$ 0.49 & 138.99 $\pm$ 0.50 & 344.13 $\pm$ 0.61 & 2457408.6 $\pm$ 0.5 & -48.68 $\pm$ 0.37 & 0.75 $\pm$ 0.06 & 0.86 $\pm$ 0.03 & 0.023 $\pm$ 0.019 \\
 J1844-6103 & 231.08 $\pm$ 0.06 & 0.037 $\pm$ 0.010 & 91.65 $\pm$ 0.21 & 219.48 $\pm$ 0.40 & 323.28 $\pm$ 5.96 & 2457347.4 $\pm$ 3.9 & 97.59 $\pm$ 0.23 & 0.59 $\pm$ 0.01 & 0.74 $\pm$ 0.03 & -0.005 $\pm$ 0.005 \\
 J1848-5709 & 261.62 $\pm$ 0.04 & 0.179 $\pm$ 0.006 & 131.27 $\pm$ 0.31 & 94.09 $\pm$ 0.38 & 0.66 $\pm$ 0.67 & 2457405.2 $\pm$ 0.6 & 57.50 $\pm$ 0.09 & 0.62 $\pm$ 0.02 & 0.85 $\pm$ 0.03 & 0.011 $\pm$ 0.005 \\
 J1922-4624 & 274.55 $\pm$ 0.15 & 0.127 $\pm$ 0.012 & 64.99 $\pm$ 1.10 & 39.22 $\pm$ 1.45 & 179.69 $\pm$ 2.79 & 2457354.7 $\pm$ 2.4 & -69.38 $\pm$ 0.40 & 0.64 $\pm$ 0.02 & 0.88 $\pm$ 0.03 & 0.004 $\pm$ 0.016 \\
 J0602-1747 & 296.88 $\pm$ 0.09 & 0.089 $\pm$ 0.013 & 113.54 $\pm$ 0.72 & 45.93 $\pm$ 0.61 & 276.50 $\pm$ 1.31 & 2457239.4 $\pm$ 1.3 & 14.13 $\pm$ 0.11 & 0.61 $\pm$ 0.02 & 1.02 $\pm$ 0.03 & 0.004 $\pm$ 0.007 \\
 J1628+1408 & 260.97 $\pm$ 0.05 & 0.035 $\pm$ 0.013 & 79.97 $\pm$ 0.53 & 51.80 $\pm$ 0.41 & 329.35 $\pm$ 8.74 & 2457337.9 $\pm$ 6.4 & 9.53 $\pm$ 0.15 & 0.74 $\pm$ 0.02 & 1.00 $\pm$ 0.03 & -0.002 $\pm$ 0.010 \\
\enddata
\end{deluxetable} \label{tab:rvs}
\end{longrotatetable}

\bibliographystyle{aasjournal}



\end{document}